\documentclass[usenatbib]{mnras}

\usepackage{bm}
\usepackage{multirow}
\usepackage{rotating}
\usepackage{colortbl}
\usepackage{xcolor}
\usepackage[fleqn]{amsmath}
\usepackage{amssymb}
\usepackage{amsfonts}
\usepackage{verbatim}
\usepackage{scalefnt}
\usepackage[percent]{overpic}
\usepackage[T1]{fontenc} 
\usepackage{aecompl} 
\usepackage{cleveref}
\usepackage[normalem]{ulem} 

\def\msun{~{\rm M}_{\sun}} 
\def\mbh{M_{\rm BH}}
\def\rg{R_{\rm g}}
\def\feddsixteen{f_{\rm Edd, 16}}
\def\mdisc{M_{\rm disc}}
\def\jdisc{J_{\rm disc}}
\def\rdisc{R_{\rm disc}}
\def\lsim{\mathrel{\rlap{\lower 3pt \hbox{$\sim$}} \raise 2.0pt \hbox{$<$}}}
\def\gsim{\mathrel{\rlap{\lower 3pt \hbox{$\sim$}} \raise 2.0pt \hbox{$>$}}}

\newcommand{\comments}[1]{} 
\newcommand\T{\rule{0pt}{2.6ex}}       
\newcommand\B{\rule[-1.2ex]{0pt}{0pt}} 
\defcitealias{Kato_et_al_2008}{K08}
\defcitealias{Kitaki_et_al_2018}{K18}
\defcitealias{Cenci_et_al_2021}{C21}

\newcommand{\soutPC}{\bgroup\markoverwith{\textcolor{cyan}{\rule[0.5ex]{2pt}{1pt}}}\ULon}

\newcommand{\soutEC}{\bgroup\markoverwith{\textcolor{orange}{\rule[0.5ex]{2pt}{1pt}}}\ULon}

\newcommand{\WB}[1]{\textcolor{black}{#1}}
\newcommand{\soutWB}{\bgroup\markoverwith{\textcolor{purple}{\rule[0.5ex]{2pt}{1pt}}}\ULon}

\title[A super-Eddington model for spinning BHs]{A novel sub-grid model for super-Eddington accretion of spinning black holes in galaxy-scale simulations}

\author[W.-B. Kao et al.]
{Wei-Bo Kao,$^{1,2}$\thanks{E-mail: wei-bo.kao@physics.ox.ac.uk}
Pedro R. Capelo,$^{3}$
Elia Cenci,$^{3,4}$
Lucio Mayer,$^{3}$
Alessandro Lupi$^{5}$
and Luca Sala$^{6}$\\
$^{1}$Department of Physics, ETH Zürich, Wolfgang-Pauli-Strasse 27, CH-8093 Zürich, Switzerland\\
$^{2}$Sub-department of Astrophysics, University of Oxford, Keble Road, Oxford OX1 3RH, United Kingdom\\
$^{3}$Department of Astrophysics, University of Zurich, Winterthurerstrasse 190, CH-8057 Z{\"u}rich, Switzerland\\
$^{4}$Department of Astronomy, University of Geneva, Chemin d'Ecogia, CH-1290 Versoix, Switzerland\\
$^{5}$DiSAT, Universit\`{a} degli Studi dell'Insubria, via Valleggio 11, IT-22100 Como, Italy\\
$^{6}$Universit\"{a}ts-Sternwarte, Fakult\"{a}t f\"{u}r Physik, Ludwig-Maximilians-Universit\"{a}t M\"{u}nchen, Scheinerstr. 1, DE-81679 M\"{u}nchen, Germany}

\date{Accepted XXX. Received YYY; in original form ZZZ}

\pubyear{\the\year{}}

\begin{document}
\label{firstpage}
\pagerange{\pageref{firstpage}--\pageref{lastpage}}
\maketitle

\begin{abstract}
Super-Eddington accretion has been proposed to explain the existence of black holes (BHs) with masses exceeding a billion solar masses within the first billion years after the Big Bang. We present a novel accretion disc-based sub-grid model for BH mass and spin evolution in the super-Eddington regime, implemented in the hydrodynamics code \textsc{gizmo}. In our model, motivated by results of radiation-hydrodynamics simulations of accretion discs, the growth of the BH is mediated by a sub-grid accretion disc, comprising an inner photon-trapping region described by simulation-based fitting formulae and an outer thin $\alpha$-disc with three regions. We incorporate a self-consistent spin evolution prescription that transitions between the Bardeen-Petterson effect and inner thick-disc precession, depending on the accretion rate. We perform a suite of idealised simulations of a BH embedded in a gaseous circumnuclear disc and a spherically distributed stellar component to explore the conditions under which super-Eddington accretion can be sustained in the environment of a realistic galactic nucleus. Simulations with misaligned gas inflows onto an initially aligned BH-disc system yield very high Eddington ratios, triggered by the rapid removal of disc angular momentum via inflows. 
These results highlight the importance of angular momentum misalignment in enabling super-Eddington accretion and suggest that such episodes are difficult to trigger unless the system resides in a highly dynamical environment -- a condition more likely to occur in high-redshift galaxies. Our model potentially provides a way to grow moderate-mass BH seeds to the sizes required to explain the bright high-redshift quasars.
\end{abstract}





\section{Introduction}

It is widely accepted that most galaxies host a supermassive black hole (SMBH) at their centre \citep[][]{Kormendy_and_Richstone_1995}. Some SMBHs are responsible for the luminous emission observed in active galactic nuclei (AGN; \citealt{Schmidt_1963}) due to the accretion of gas from their surroundings \citep[][]{Lynden-Bell_1969, Antonucci_1993}. SMBHs also play a crucial role in galaxy formation and evolution through accretion and feedback processes, potentially driving their co-evolution with their host galaxies \citep[][]{King_2003, Di_Matteo_et_al_2005, Fabian_2012, Kormendy_Ho_2013, Heckman_and_Best_2014, King_and_Pounds_2015}. 

Prior to the launch of the James Webb Space Telescope \WB{\citep[JWST;][]{Gardner_et_al_2006}}, observations of AGN at $z \gsim 6$ revealed SMBHs with $M_{\rm BH} > 10^9 \msun$ \citep[][]{Mortlock_et_al_2011, Banados_et_al_2018, Yang_et_al_2020, Wang_et_al_2021, Yang_et_al_2021, Fan_et_al_2023}. With the launch of JWST, such observations have been extended to $z \sim 10$, with detections of SMBHs with masses as high as $10^6 \msun$ \citep[][]{Larson_et_al_2023, Maiolino_et_al_2024}. These discoveries present a significant challenge in explaining how such massive SMBHs already exist so early in cosmic history \citep[see a recent review by][]{Inayoshi_2020}. 

One possible explanation is the formation of heavy seed BHs with high initial masses under peculiar conditions \citep[][]{Dijkstra_et_al_2006, Latif_et_al_2015, Boekholt_et_al_2018, Wise_et_al_2019, Volonteri_et_al_2021, Toyouchi_et_al_2023, Zwick_et_al_2023, Mayer_et_al_2024, Mayer_et_al_2025}. Alternatively, seed BHs could undergo episodes of super-Eddington accretion with rapid mass growth \WB{\citep[][]{Madau_et_al_2014, Volonteri_et_al_2015, Lupi_et_al_2016, Regan_et_al_2019, Hu_et_al_2022_BHGrowth, Massonneau_et_al_2023, Sassano_et_al_2023, Lupi_et_al_2024, Piana_et_al_2024, Shi_et_al_2024, Chiu_et_al_2025, Gordon_et_al_2025, Husko_et_al_2025, Prole_et_al_2025, Quadri_et_al_2025, Sanati_et_al_2025}}.

Candidates for super-Eddington accretion events have been observed in the local Universe in tidal disruption events \citep[TDEs;][]{Lin_et_al_2017, Dai_et_al_2018}, ultra-luminous X-ray sources \citep[ULXs;][]{Bachetti_et_al_2014,Motch_et_al_2014}, and AGN \citep[][]{Jin_et_al_2012, Du_et_al_2015, Du_et_al_2018, Regan_et_al_2019, Liu_et_al_2021, Abuter_et_al_2024, Marziani_et_al_2025}. Recently, \citet{Abuter_et_al_2024} \WB{and \WB{\citet{GRAVITY+2025}}} dynamically measured the mass of 
super-Eddington-accreting SMBH\WB{s} at $z\sim 2$ \WB{and $z \sim 4$, respectively.} 
JWST \WB{has} also identified a potential super-Eddington-accreting SMBH at $z \sim 4$ \citep[][]{Suh_et_al_2024}\WB{, and \citet{Ighina_et_al_2025} used X-ray observations to report a super-Eddington-accreting SMBH candidate at $z \sim 6$.} Moreover, \citet{Leung_et_al_2024} and \citet{Lupi_et_al_2024_JWST_superEdd} suggest that JWST may observe an extensive population of super-Eddington-accreting SMBHs at $z > 4$. 

\WB{Little Red Dots \citep[LRDs;][]{Matthee_et_al_2024} are exotic sources recently discovered with JWST. They exhibit extremely broad emission lines (FWHM > 1000 km~s$^{-1}$; e.g. \citealt{Greene_et_al_2024, Kocevski_et_al_2025}) and weak X-ray emission \citep[e.g.][]{Ananna_et_al_2024, Yue_et_al_2024, Maiolino_et_al_2025_xray}, that could arise from super-Eddington accretion onto SMBHs (e.g. \citealt{Lambrides_et_al_2024, Pacucci_and_Narayan_2024, Inayoshi_et_al_2025, Madau_2025}). However, alternative physical scenarios have been proposed to account for their full spectral properties, including reddened AGN \citep[e.g.][]{Li_et_al_2025, Volonteri_et_al_2025}, dusty star-formation \citep[e.g.][]{Baggen_et_al_2024, Barro_et_al_2024}, newborn heavy BH seeds \citep[e.g.][]{Cenci_and_Habouzit_2025, Jeon_et_al_2025}, their progenitor supermassive stars \citep[e.g.][]{Zwick_et_al_2025}, or gas-enshrouded AGN \citep[e.g.][]{Inayoshi_and_Maiolino_2025, Lin_et_al_2025, Rusakov_et_al_2025}.}

The thin $\alpha$-disc model is commonly used to describe the structure of accretion flows around BHs (\citealt{Shakura_Sunyaev_1973, Frank_et_al_2002, Kato_et_al_2008}, hereafter \citetalias{Kato_et_al_2008}). However, super-Eddington accretion flows cannot be adequately described by the commonly used standard thin $\alpha$-disc model, since advection significantly alters the structure of the disc. Instead, they are better described by the slim disc model \citep[][]{Abramowicz_et_al_1988}, a one-dimensional (1D) accretion disc model that accounts for the advection of the photons. Approximate analytical solutions have been derived by \citet{Wang_and_Zhou_1999} and \citet{Watarai_2006}, whereas numerical relativistic solutions have been obtained by \citet{Sadowski_et_al_2009, Sadowski_2011}. 

Further simulations have provided deeper insight into this process \WB{(e.g. \citealt{Ohsuga_et_al_2005,Jiang_et_al_2014, Inayoshi_et_al_2016,Sadowski_and_Narayan_2016, Kitaki_et_al_2018}, hereafter \citetalias{Kitaki_et_al_2018}; \citealt{Jiang_et_al_2019, Kitaki_et_al_2021, Hu_et_al_2022_RHDsimulaion, Yoshioka_et_al_2024,Zhang_et_al_2025})}. These studies have elucidated how extreme physical conditions inherent to super-Eddington flows -- such as high radiation pressure, complex magnetic fields, strong outflows, and turbulent accretion -- affect the accretion process \citep[][]{Mayer_2019,Jiang_and_Dai_2024}. 

The mass of an SMBH is not its only fundamental parameter, its spin being also crucial to its evolution. The spin determines the location of the innermost stable circular orbit (ISCO; \citealt{Bardeen_et_al_1972}), which modulates the rate of BH growth by altering the radiative efficiency and the structures of the disc \citep[][]{Sadowski_et_al_2009, Lopez_Armengol_et_al_2021, Inayoshi_and_Ichikawa_2024}.  A rapidly spinning SMBH is also responsible for the production of relativistic jets \citep[][]{Blandford_and_Znajek_1977}. Furthermore, both the direction and magnitude of the spin influence the feedback exerted on the BH surroundings, impacting the accretion process and the evolution of the host galaxy \citep[][]{Silk_and_Rees_1998, McKinney_et_al_2012, King_and_Pounds_2015, Sala_et_al_2021, Massonneau_et_al_2023_spin, Ricarte_et_al_2023, Bollati_et_al_2024}.

The growth of SMBHs during galactic evolution is a complex and multi-scale process spanning over ten orders of magnitude in spatial scale, from the event horizon of the SMBH to the scales of entire galaxies and the cosmic web. Resolving BH physics across this vast spatial range in cosmological simulations remains computationally unfeasible. Progress has been made by focusing on a limited range of spatial or temporal scales \citep[e.g.][]{Guo_et_al_2023, Guo_et_al_2024, Hopkins_2024}.

Another widely adopted approach is the accretion disc particle method, in which the SMBH is represented as an unresolved composite SMBH-accretion disc particle (e.g. \citealt{Power_et_al_2011, Dubois_et_al_2014, Fiacconi_et_al_2018, Cenci_et_al_2021}, hereafter \citetalias{Cenci_et_al_2021}; \citealt{Tartenas_and_Zubovas_2022, Massonneau_et_al_2023_spin, Massonneau_et_al_2023, Koudmani_et_al_2024, Husko_et_al_2025}). In this framework, the SMBH is evolved by tracking the mass and angular momentum flows onto the accretion disc.

In many of these models, the BH mass accretion rate is estimated using the Bondi-Hoyle-Lyttleton prescription \citep[BHL;][]{Hoyle_and_Lyttleton_1939, Bondi_and_hoyle_1944, Bondi_1952}. However, this approach neglects the influence of the gas angular momentum, which strongly regulates accretion. To address this limitation, \citet{Fiacconi_et_al_2018} introduced a steady-state accretion disc-mediated accretion rate, which was further refined by \citetalias{Cenci_et_al_2021} and \citet{Koudmani_et_al_2024}. By contrast, \citet{Tartenas_and_Zubovas_2022} introduced the viscous evolution of the disc, allowing for a more general treatment in scenarios wherein the inflow rate changes rapidly, finding that accounting for the disc viscous evolution leads to different BH growth histories and BH feedback effects compared to the BHL prescription.

In this paper, we develop the first sub-grid model of an SMBH-accretion disc particle that is capable of capturing the evolution of both the SMBH mass and spin allowing for super-Eddington accretion. Our model builds upon and extends the framework introduced by \citetalias{Cenci_et_al_2021} (see also \citealt{Fiacconi_et_al_2018}) and is implemented in the publicly available code \textsc{gizmo} \citep[][]{Hopkins_2015}. 

The remainder of the paper is structured as follows. Section~\ref{sec:super_edd_model} provides a detailed theoretical description of this model. Section~\ref{sec:numerical_setup} outlines its numerical implementation and simulation setup. Section~\ref{sec:results} presents our results. In Section~\ref{sec:discussion}, we discuss the relevance of super-Eddington accretion and the caveats of our model. Finally, we summarise our findings in Section~\ref{sec:conclusions}. 


\section{The super-Eddington model}\label{sec:super_edd_model}

An astrophysical BH is characterised by its mass, $M_{\text{BH}} = 10^6 M_{\text{BH},6} \msun$, and angular momentum, $\bm{J}_{\rm BH} = J_{\rm BH} \,\bm j_{\rm BH}$, where  $\bm{J}_{\rm BH}$ and $J_{\rm BH}$ denote the BH angular momentum vector and magnitude, respectively.\footnote{We neglect the BH charge, as it is generally believed that electrically charged BHs cannot exist stably in astrophysical environments \citep[e.g.][]{Gibbons_1975,Blandford_and_Znajek_1977}.} The (dimensionless) spin parameter of the BH is defined as $a_{\rm BH} = c J_{\text{BH}}/(G M_{\text{BH}}^2)$, where $G$ is the gravitational constant, and $c$ is the speed of light in vacuum. Since the maximum angular momentum of a BH is $G M_{\rm BH}^2/c$ \citep[][]{Kerr_1963}, the theoretical range of $a_{\rm BH}$ is between 0 (for a non-spinning BH) and 1 (for a maximally spinning BH).\footnote{We note that other authors \citep[e.g.][]{Kerr_1963} define the BH spin as $J_{\rm BH}/(M_{\rm BH}c)$ (with units of length), so that its maximum value is given by $GM_{\rm BH}/c^2$.}

An accreting BH releases energy according to

\begin{equation}
    \WB{\mathcal L_{\rm BH}} = \eta \dot{M}_{\rm BH, accr} c^2~,
    \label{eq:LBH}
\end{equation}

\noindent where $\WB{\mathcal L_{\rm BH}}$ is the BH (bolometric) luminosity originating from the conversion of gravitational potential energy into radiation as gas is accreted, $\dot{M}_{\rm BH, accr}$ is the BH mass accretion rate, and $\eta = 0.1\eta_{0.1}$ is the radiative efficiency, that is how much of the rest energy of the accreting gas is converted into radiation. This luminosity has a theoretical upper limit, given by the \citet{Eddington_1916} luminosity, $\WB{\mathcal L_{\rm Edd}}$, which defines the maximum accretion power a celestial object undergoing a spherically symmetric accretion process (such as an accreting star or BH) can achieve when the outward pressure of radiation, generated in the accretion flow, counteracts the inward pull of gravity:

\begin{equation}
    \WB{\mathcal L_{\rm Edd}} = \frac{4 \pi G M_{\rm BH} \mu_{\rm e} m_{\rm p} c}{\sigma_{\rm T}}~,
\end{equation}

\noindent where $m_{\rm p}$ is the proton mass, $\mu_{\rm e}$ is the mean molecular weight per electron, and $\sigma_{\rm T}$ is the \WB{\citet{Thomson_1906} electron} scattering cross-section.\footnote{\WB{We note that a more accurate definition would be given by $4 \pi GM_{\rm BH} c/\kappa$, where $\kappa$ is the opacity of the gas. The two definitions coincide only when $\kappa$ is equal to the (low-energy limit; \citealt{Klein_Nishima_1929}) electron scattering opacity, i.e. $\kappa = \kappa_{\rm es} = \sigma_{\rm T}/(\mu_{\rm e}m_{\rm p})$.\label{foot:Thomson}}} We assume \WB{a fully ionized gas with a hydrogen mass fraction $X = 1$, so that} $\mu_{\rm e} = 1$ 
(as in \citetalias{Kato_et_al_2008}). 

Following Equation~\eqref{eq:LBH}, the Eddington mass accretion rate is then $\dot M_{\rm Edd, \eta} = \WB{\mathcal L_{\rm Edd}}/(\eta c^2)$ and we also introduce the corresponding Eddington ratio as $f_{\rm Edd, \eta} = \dot M_{\rm BH,accr} / \dot M_{\rm Edd, \eta}$. The radiative efficiency is not a constant and can vary as a function of spin and accretion rate \citep[][]{Bardeen_et_al_1972,Sadowski_et_al_2009,Madau_et_al_2014}. However, a characteristic value is usually assumed for the definition of Eddington mass accretion rate, so that the latter quantity does not depend on the varying radiative efficiency (although it still obviously depends on the varying BH mass). Therefore, we also define the Eddington mass accretion rate and Eddington ratio for a fixed value of $\eta$, for ease of comparison with other literature, picking $\eta = 1/16$ \citep[as in][]{Sadowski_et_al_2009, Madau_et_al_2014} and obtaining $\dot M_{\rm Edd, 16} = 16 \WB{\mathcal L_{\rm Edd}}/ c^2$ and $f_{\rm Edd, 16} = \dot M_{\rm BH,accr} / \dot M_{\rm Edd, 16}$.\footnote{Other authors define the Eddington mass accretion rate as $\dot M_{\rm Edd, 10} = 10 \WB{\mathcal L_{\rm Edd}}/ c^2$ \citep[e.g.][]{Jiang_et_al_2019, Hu_et_al_2022_RHDsimulaion, Massonneau_et_al_2023} or $\dot M_{\rm Edd, 1} = \WB{\mathcal L_{\rm Edd}}/ c^2$ (e.g. \citetalias{Kato_et_al_2008}; \citealt{Kitaki_et_al_2021, Liu_et_al_2021}), with corresponding definitions of the Eddington ratio as $f_{\rm Edd, 10} = \dot M_{\rm BH,accr} / \dot M_{\rm Edd, 10}$ and $f_{\rm Edd, 1} = \dot M_{\rm BH,accr} / \dot M_{\rm Edd, 1}$ (the latter denoted as $\dot m$ in some works). The relation between the different definitions of the Eddington ratio is $f_{\rm Edd,16} = 10 f_{\rm Edd,10}/16 = f_{\rm Edd,1}/16 = f_{\rm Edd,\eta}/(16 \eta) = 10 f_{\rm Edd,\eta}/(16 \eta_{0.1})$. Note that $f_{\rm Edd}$ in \citet{Perego_et_al_2009, Dubois_et_al_2014, Fiacconi_et_al_2018, Bustamante_et_al_2019}; \citetalias{Cenci_et_al_2021}; \citet{Sala_et_al_2021} is equal to our $f_{\rm Edd,\eta}$.\label{foot:fEdd}}

We mediate accretion onto the BH via an accretion disc with physical properties that reflect the state of the BH and the surrounding gas reservoir. The accretion disc size is $R_{\rm disc}$, the total mass is $M_{\rm disc}$, and the total angular momentum is denoted as $\bm{J}_{\rm disc} = J_{\rm disc} \,\bm j_{\rm disc}$, with $J_{\rm disc}$ and $\bm j_{\rm disc}$ being its magnitude and unit vector, respectively. To model the accretion disc structure, we define the disc surface density and specific angular momentum at a given radius $R$ (in cylindrical coordinates) as, respectively, $\Sigma(R)$ and $\bm L(R) = L(R) \, \bm l(R)$, where $L(R)$ is the magnitude and $\bm l(R)$ is the unit vector. The disc temperature, volume density, total pressure, and scale height are denoted as $T(R)$, $\rho(R)$, $P(R)$, and $H(R)$, respectively. Here, $H$ is defined as $H = c_{\rm s}/\Omega_{\rm K}$, where $c_{\rm s}$ is the sound speed of the gas and $\Omega_{\rm K} = \sqrt{G\mbh/R^3}$ is its Keplerian angular velocity. 

In the accretion disc, the mass inflow is driven by the radial viscosity, $\nu_1$, as it is responsible for transporting angular momentum \WB{outwards}, thereby driving the inflow. Following \citetalias{Kato_et_al_2008}, the viscosity prescription is given by the $r\phi$-component of the shear stress tensor, $t_{r \phi} = -\alpha P$, where $\alpha = 0.1\alpha_{0.1}$ is a dimensionless constant 
\WB{to describe} the viscosity: this is equivalent to $\nu_1 = 2\alpha c_{\rm s} H/3$. We note the presence of an extra factor of $2/3$ in this expression compared to the original formulation by \citeauthor{Shakura_Sunyaev_1973} (\citeyear{Shakura_Sunyaev_1973}; see Appendix~\ref{app:difference_alpha_disc} for further details). 

In our model, we consider the situation in which the BH might not be aligned with the disc (i.e. $\bm J_{\rm disc}$ is not parallel to $\bm J_{\rm BH}$) and investigate the evolution of the BH in this scenario. In this case, the spinning BH generates a frame-dragging effect on the misaligned disc, known as the Lense-Thirring effect \citep[][]{Lense_and_Thirring_1918}. Consequently, when a disc is tilted relative to the spinning BH, it naturally becomes warped (i.e. the tilt angle of the disc varies with radius). We note that, in this model, we assume that $\bm j_{\rm disc}$ corresponds to the direction of the outer part of the warped accretion disc \citep[as in][]{Perego_et_al_2009, Fiacconi_et_al_2018}. The justification for this assumption is given in Section~\ref{sec:LT_low}. For a thin disc, the vertical viscosity, $\nu_2$, acts to damp the warps and tends to align each disc ring with its neighbours.\footnote{The third viscosity, $\nu_3$, which represents a torque that induces precession when a disc ring is misaligned with its neighbours, is often neglected, as it is typically negligible compared to other viscosities in a thin Keplerian disc \citep[][]{Ogilvie_et_al_1999}.} This is related to $\nu_1$ by the relation

\begin{equation}
  \frac{\nu_2}{\nu_1} = \frac{\xi}{2 \alpha^2}~, 
  \label{eq:nu_2}
\end{equation}

\noindent where $\xi$ is a dimensionless constant of order unity \citep[][]{Papaloizou_and_Pringle_1983, Lodato_and_Pringle_2007}.

\subsection{Model overview}\label{sec:model_overview}

\begin{table*}
    \centering
    \caption{Overview of our sub-grid model.}
    \label{tab:overview}
    \begin{tabular}{p{4.0cm} p{12.5cm}}   
    \hline
    \rule{0pt}{3ex}\textbf{Components} & \textbf{Relevant equations and quantities} \T \B \\
    \hline
    \rule{0pt}{3ex}Disc structure (Sec.~\ref{sec:accretion_disc_structure}) 
    &Use four regions to describe the accretion disc: 
    \[\text{Structure} : 
    \begin{cases} 
        \text{Photon-trapping region} \text{ ($L$ and $\Sigma$ follow Eqs~\ref{eq:sigma_trap} and \ref{eq:L_trap})}\quad   \text{if } R_{\rm ISCO} \leq R < R_{\rm trap} \\
        \text{Region~(a) of thin $\alpha$-disc}  \text{ (Eqs~\ref{eq:sigma_a} and \ref{eq:L_Keplerian})}\quad  \text{if } \max (R_{\rm ISCO}, R_{\rm trap}) \leq R < R_{\rm ab}\\
        \text{Region~(b) of thin $\alpha$-disc} \text{ (Eqs~\ref{eq:sigma_b} and \ref{eq:L_Keplerian})} \quad  \text{if } R_{\rm ab} \leq R < R_{\rm bc}  \\
        \text{Region~(c) of thin $\alpha$-disc} \text{ (Eqs~\ref{eq:sigma_c} and \ref{eq:L_Keplerian})}\quad  \text{if } R_{\rm bc} \leq R  
    \end{cases}\]\\
    &$R_{\rm ISCO}, R_{\rm trap}, R_{\rm ab},$ and $R_{\rm bc}$ are given by Eqs~\eqref{eq:r_isco}, \eqref{eq:r_trap}, \eqref{eq:r_ab}, and \eqref{eq:r_bc}, respectively. \T \B \\
    \hline
    \rule{0pt}{3ex}Mass accretion rate (Sec.~\ref{sec:mass_accretion}) 
    &$\feddsixteen$ is calculated using the Newton-Raphson method by solving for it based on the integrated disc properties, specifically $\mdisc$ and $\jdisc$. The maximum value for $\feddsixteen$, $f_{\rm Edd,16,max}$, is given by Eq.~\eqref{eq:f_edd_max}.  \\
    \hline
    \rule{0pt}{3ex}Angular momentum (Sec.~\ref{sec:angular_momentum}) 
    \rule{0pt}{1.3ex}&${\bm{\dot J}}_{\text{BH}} = {\bm{\dot J}}_{\rm BH, acc} + {\bm{\dot J}}_{\rm BH, LT}$ \\
    &${\bm{\dot J}}_{\rm BH, acc}$ is given by Eq.~\eqref{eq:J_BH_accretion}.
    \[ 
    {\bm{\dot J}}_{\rm BH, LT} : 
        \begin{cases}
            \text{Bardeen-Petterson effect (Eq.~\ref{eq:LT_lowedd})} & \text{if } \feddsixteen \leq \hat f_{\rm Edd, 16}\\
            \text{Inner precessing thick\WB{-}disc (Eq.~\ref{eq:LT_highedd})} & \text{if } \feddsixteen > \hat f_{\rm Edd, 16}
        \end{cases}
\]Disc and BH (counter)-align instantaneously if $R_{\rm disc} < R_{\rm warp}$ (Eq.~\ref{eq:r_warp}).
    \\
    \\
    \hline
    \rule{0pt}{3ex}Self-gravitating mass (Sec.~\ref{sec:self_gravitating}) 
    &We limit $M_{\rm disc} \leq M_{\rm sg}$, where $M_{\rm sg} = M_{\rm disc} (R=R_{\rm sg})$. $R_{\rm sg}$ is given by Eq.~\eqref{eq:r_sg_abc}. \WB{When $\mdisc = M_{\rm sg}$, we additionally require $J_{\rm disc} \leq J_{\rm sg}$, where $J_{\rm sg} = J_{\rm disc} (R=R_{\rm sg})$.} \T \B \\  
    \hline
    \rule{0pt}{3ex}Radiative efficiency (Sec.~\ref{sec:radiative_efficiency}) & 
    Eqs~\eqref{eq:radiative_efficiency}--\eqref{eq:radiative_efficiency_C} are utilised to compute the radiative efficiency. \T \B \\
    \hline
    \rule{0pt}{3ex}\WB{Maximum $a_{\rm BH}$ (Sec.~\ref{sec:max_a_BH})} & 
    \WB{$a_{\rm BH} \leq \min(0.998,  a_{\rm BH,max})$, where $a_{\rm BH,max}$ is defined as the value at which the dimensionless spin-up parameter $s=0$ (Eq.~\ref{eq:spin_up_parameter}).}\T \B \\
    \hline
    \end{tabular}
\end{table*}

In our model, a BH particle represents a sub-grid system consisting of a BH surrounded by its accretion disc. The unresolved accretion disc is described by its global properties: $M_{\rm disc}$ and $\bm{J}_{\rm disc}$. Our model extends from previous accretion disc-based BH growth sub-grid models (\citealt{Power_et_al_2011, Dubois_et_al_2014, Fiacconi_et_al_2018}; \citetalias{Cenci_et_al_2021}; \citealt{Massonneau_et_al_2023, Koudmani_et_al_2024}). The mass and angular momentum of the BH and accretion disc are evolved considering accretion from the accretion disc onto the BH, external inflows from the large-scale gas onto the disc, and relativistic torques exerted by the spinning BH onto the disc. 

Most previous sub-grid models describe the structure of the accretion disc using only the outer region of the thin $\alpha$-disc model [i.e. region~(c), introduced in Section~\ref{sec:accretion_disc_structure}]. However, the thin $\alpha$-disc model [even when considering all regions: (a), (b), and (c)] is valid only when the disc is optically thick and radiative cooling is efficient. Consequently, it applies only at intermediate Eddington ratios, $0.01 \lesssim  f_{\rm Edd, 16} \lesssim 0.2$ \citep[][]{Laor_and_Netzer_1989, Koratkar_and_Blaes_1999, Narayan_and_McClintock_2008}.

\citet{Koudmani_et_al_2024} extended this popular sub-grid model by incorporating an analytical advection-dominated inflow-outflow solution \citep[ADIOS;][]{Blandford_and_Begelman_1999, Blandford_and_Begelman_2004} to account for lower mass accretion rates ($\feddsixteen \lesssim 0.01$). 

In our model, we focus instead on the super-Eddington regime and employ simulation results from \citetalias{Kitaki_et_al_2018}, alongside a more detailed thin $\alpha$-disc model from \citetalias{Kato_et_al_2008} [that includes also the inner regions (a) and (b), introduced in Section~\ref{sec:accretion_disc_structure}], to develop an accretion disc model also applicable to super-Eddington accretion. 

Our model builds upon that of \citetalias{Cenci_et_al_2021}, which itself extends from that of \citet{Fiacconi_et_al_2018}. Table~\ref{tab:overview} illustrates a summary of our accretion disc-based BH growth sub-grid model for super-Eddington accretion. It outlines the methods used to compute the disc structure (Section~\ref{sec:accretion_disc_structure}), BH mass accretion rate (Section~\ref{sec:mass_accretion}), angular momentum evolution (Section~\ref{sec:angular_momentum}), self-gravitating conditions (Section~\ref{sec:self_gravitating}), 
radiative efficiency (Section~\ref{sec:radiative_efficiency})\WB{, and a novel upper limit for the BH spin (Section~\ref{sec:max_a_BH})}. The differences between our model and that of \citetalias{Cenci_et_al_2021} are detailed in Appendix~\ref{app:difference_between_cenci_2021}. 

\subsection{Accretion disc structure}\label{sec:accretion_disc_structure}

We begin defining the relevant length-scales of the accretion disc, moving from the BH outwards. We first set the radius of the ISCO, $R_{\rm ISCO}$, as the inner boundary of the accretion disc. Following \citet{Bardeen_et_al_1972},  $R_{\rm ISCO}$ is calculated as

\begin{equation}
    \frac{R_{\rm ISCO}}{R_{\rm g}} = 3 + Z_2 \mp \sqrt{(3 - Z_1)(3 + Z_1 + 2Z_2)}~,
    \label{eq:r_isco}
\end{equation}

\noindent where $R_{\rm g} = GM_{\rm BH}/c^2$ is the gravitational radius of the BH, the upper and lower signs indicate orbits that are prograde and retrograde relative to the BH spin, respectively, and $Z_1$ and $Z_2$ are two functions of $a_{\rm BH}$:

\begin{align}
    Z_1(a_{\rm BH}) &= 1 + (1 - a_{\rm BH}^2)^{1/3} \left[ (1 + a_{\rm BH})^{1/3} + (1 - a_{\rm BH})^{1/3} \right] \, , \\
    Z_2(a_{\rm BH}) &= \sqrt{3a_{\rm BH}^2 + Z_1^2(a_{\rm BH})} \, .
\end{align}

Photon-trapping plays a crucial \WB{role} for super-Eddington accretion \citep[][]{Begelman_1978}. In the inner parts of the disc, photons can become trapped in the radial flow and are unable to escape from the disc surface, eventually being accreted by the BH. This occurs when the photon diffusion time-scale, $t_{\rm diff}$, exceeds the accretion time-scale in the (cylindrical) radial direction, $t_{{\rm acc,} R}$. Following \citetalias{Kato_et_al_2008}, we define $t_{\rm diff} = 3H \tau/c$, where $\tau$ is the vertical optical depth in the disc, and $t_{{\rm acc,} R} = R/|v_R|$, where $v_R$ is the radial velocity. Here, $\tau = \kappa_{\rm es} \Sigma/2$, where $\kappa_{\rm es} = 0.20\, (1+X) \ \rm{cm}^2$~g$^{-1}$ is the \WB{(low-energy limit)} electron scattering opacity \WB{(see Footnote~\ref{foot:Thomson}), assuming that the gas is fully ionized [which implies $\mu_{\rm e} = 2/(1+X)$]}. We assume a hydrogen mass fraction $X=1$ \WB{(i.e. $\mu_{\rm e} = 1$)} to calculate $\kappa_{\rm es}$ \citepalias[as in][]{Kato_et_al_2008}, the exact value of which does not significantly affect the results \citep[][]{Frank_et_al_2002}. The photon-trapping radius, $R_{\rm trap}$, is defined as the radius where the photon diffusion time-scale equals the accretion time-scale. To derive this, we employ the continuity equation to obtain

\begin{equation}
    \frac{R_{\rm trap}}{R_{\rm g}} = 48 \, \feddsixteen \left(\frac H R\right)~.
    \label{eq:r_trap}
\end{equation}

Based on results from two-dimensional (2D) radiation hydrodynamics (RHD) simulations of super-Eddington accretion flows in \citet{Kitaki_et_al_2021}, we set $H/R = 1$. Photons within $R_{\rm trap}$ are accreted onto the BH instead of escaping from the accretion disc. If $\feddsixteen \lesssim 0.1$, $R_{\rm trap}$ would become smaller than $R_{\rm ISCO}$. In this case, the photon-trapping region cannot exist stably and we assume that it disappears if $R_{\rm trap} < R_{\rm ISCO}$.

For $R < R_{\rm trap}$, the assumptions underlying the thin $\alpha$-disc break down, as the flow is radiatively inefficient and the advection of the photon entropy becomes significant. The slim disc model accounts for the photon advection and constructs a 1D, stable, and steady accretion disc model \citep[][]{Abramowicz_et_al_1988}. Although approximate analytical solutions have been derived \citep[][]{Wang_and_Zhou_1999, Watarai_2006}, the multidimensional motion and outflows driven by strong radiation pressure are not considered in the slim disc model. 

\citetalias{Kitaki_et_al_2018} performed 2D RHD simulations of super-Eddington accretion flows with outflows around a non-spinning BH, assuming a constant $\alpha_{0.1}= 1$. They were the first to obtain fitting formulae for the structure of super-Eddington accretion discs covering a wide range of values of $\dot M_{\rm BH, acc}$ and $\mbh$. We employ these fitting formulae to establish the inner structure of the accretion disc for $R < R_{\rm trap}$. 

For $R>R_{\rm trap}$, we assume that the accretion disc structure follows the  solutions of the thin $\alpha$-disc model. This assumption is supported by both analytical and simulation results \citep[][]{Watarai_2006,Kitaki_et_al_2021}. The pressure of the accretion disc can be written as $P = P_{\rm rad} + P_{\rm gas}$, where $P_{\rm rad}$ and $P_{\rm gas}$ are the radiation and gas pressure, respectively. In this work, we assume that the gas temperature is above $10^4 \ \rm K$, so that the disc opacity, $\kappa$, is dominated by two sources: electron scattering opacity, $\kappa_{\rm es}$, and free-free absorption opacity, $\kappa_{\rm ff} \propto T^{-7/2}$, but we refer to Section \ref{sec:caveat_opacity} for a discussion on the validity of this assumption. The thin $\alpha$-disc model is further divided into three regions based on the dominant sources of opacity and pressure (\citealt{Shakura_Sunyaev_1973}; \citetalias{Kato_et_al_2008}):

\begin{itemize}

\item Region~(a) -- the inner region. This region has the highest temperature amongst the three: radiation pressure dominates ($P \sim P_{\rm rad}$) and electron scattering is the primary opacity source ($\kappa \sim \kappa_{\rm es}$).  This region exists between $\max (R_{\rm ISCO}, R_{\rm trap}) \leq R < R_{\rm ab}$, where $R_{\rm ISCO}$, $R_{\rm trap}$, and $R_{\rm ab}$ are defined by Equations~\eqref{eq:r_isco}, \eqref{eq:r_trap}, and \eqref{eq:r_ab}, respectively.
    
\item Region~(b) -- the middle region. Gas pressure dominates ($P \sim P_{\rm gas}$) and electron scattering remains the primary source of opacity ($\kappa \sim \kappa_{\rm es}$). This region exists between $R_{\rm ab} \leq R < R_{\rm bc}$, where $R_{\rm bc}$ is defined by Equation~\eqref{eq:r_bc}.
    
\item Region~(c) -- the outer region. Gas pressure dominates ($P \sim P_{\rm gas}$) and free-free absorption becomes the primary source of opacity ($\kappa \sim \kappa_{\rm ff}$). This region extends beyond $R \geq R_{\rm bc}$.
    
\end{itemize}

Following \citetalias{Kato_et_al_2008}, the transitions between the different regions of the thin $\alpha$-disc occur at the characteristic radii\footnote{We note that $R_{\rm bc}$ in \citetalias{Kato_et_al_2008} is nearly identical to that given by \citet{Shakura_Sunyaev_1973}, but $R_{\rm ab}$ is approximately three times larger than that of \citet{Shakura_Sunyaev_1973}. The differences arise from slight variations in definitions (see Appendix~\ref{app:difference_alpha_disc} for further details).} 

\begin{equation}
    \frac{R_{\rm ab}}{R_{\rm g}} = 1.12 \times 10^3 \, M_{\text{BH},6}^{2/21} \, \alpha_{0.1} ^{2/21} \,\feddsixteen^{16/21}~,
    \label{eq:r_ab}
\end{equation}

\begin{equation}
    \frac{R_{\rm bc}}{R_{\rm g}} = 3.15 \times 10^4 \, \feddsixteen^{2/3}~.
    \label{eq:r_bc}
\end{equation}

Unlike in \citet{Koudmani_et_al_2024}, we neglect an additional transition radius, $R_{\rm tran}/ R_{\rm g} \sim 6 \left (1.25 \times 10^{-2}/f_{\rm Edd, 16}\right)^2$, below which the thin $\alpha$-disc model breaks down \citep[][]{Liu_et_al_1999,Yuan_et_al_2018}. For $R<R_{\rm tran}$, the disc becomes radiatively inefficient due to low gas density and to extended cooling time. This region can instead be better described by the ADIOS solution or by an advection-dominated accretion flow \citep[ADAF;][]{Narayan_and_Yi_1994,Narayan_and_Yi_1995a,Natayan_and_Yi_1995b,Abramowicz_et_al_1995,Chen_et_al_1995}. The value of $R_{\rm tran}$ is inversely correlated with $f_{\rm Edd, 16}$, with the ADAF and ADIOS models becoming more relevant at lower mass accretion rates. We neglected the impact of $R_{\rm tran}$ in our model, since it remains smaller than $R_{\rm ISCO}$ for $f_{\rm Edd, 16} \gtrsim 0.01$. Situations involving low accretion rates, for which the ADAF or ADIOS models become significant, are beyond the scope of this study. 

\begin{figure}
    \centering
    \includegraphics[width=0.9\linewidth]{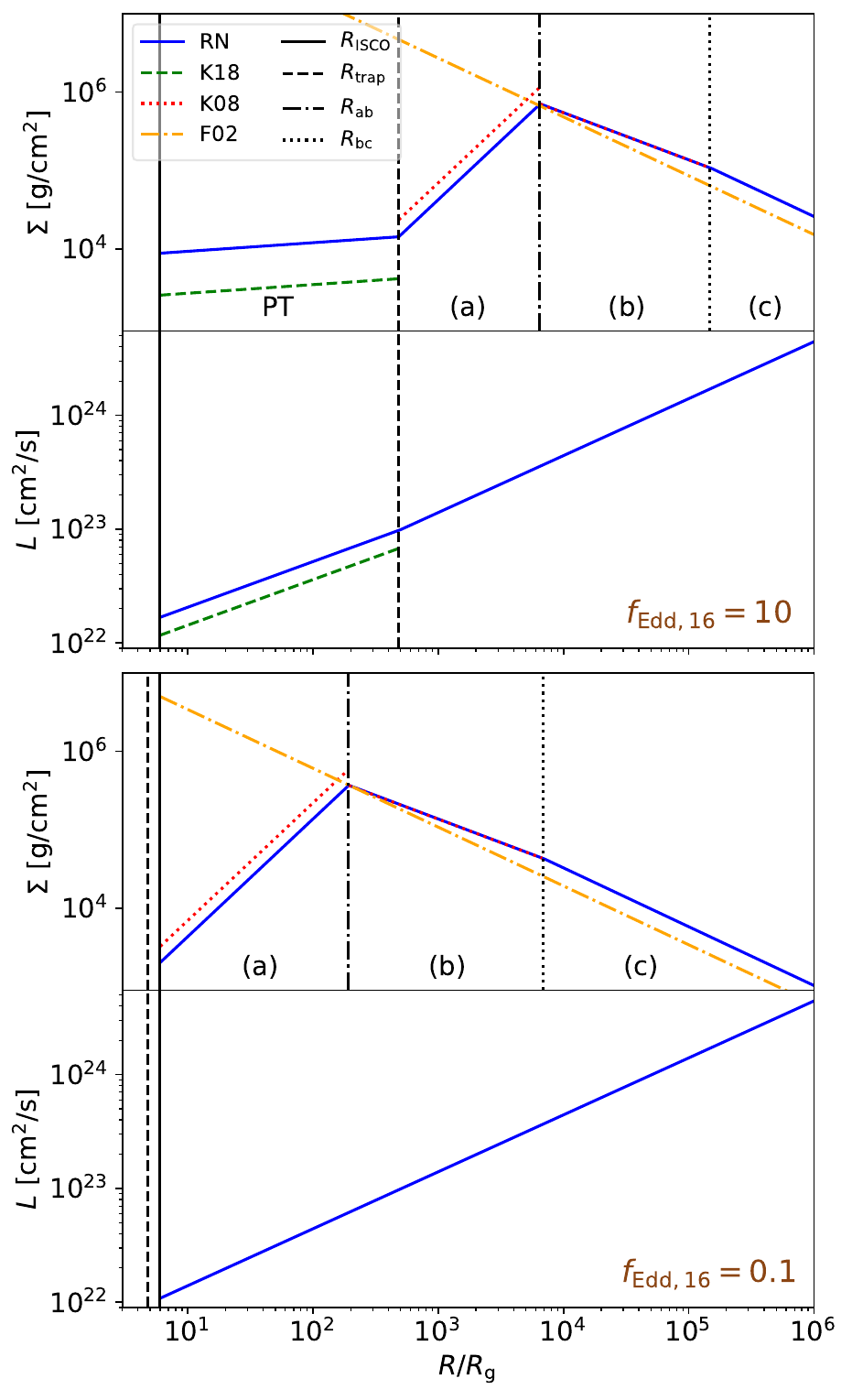}  
    \caption{Surface density profile (first and third panels) and specific angular momentum profile (second and fourth panels) for accretion discs with $\feddsixteen = 10$ and 0.1, assuming $M_{\rm BH,6}=1$, $\alpha_{0.1} = 1$, and $a_{\rm BH} = 0$. The blue solid lines represent the profile constructed using our model, renormalised to ensure continuity (RN in the figure). The green dashed lines correspond to the fitting formulae from \citetalias{Kitaki_et_al_2018}, i.e. Equations~\eqref{eq:sigma_trap} and \eqref{eq:L_trap}. The red dotted lines show the surface density profile of regions~(a) and (b) from \citetalias{Kato_et_al_2008}, i.e. Equations~\eqref{eq:sigma_a} and \eqref{eq:sigma_b}. We also plot the surface density profile from \citet{Frank_et_al_2002}, which includes only region~(c) of the thin $\alpha$-disc and has been commonly used in previous sub-grid models. The black vertical lines in the first and third panels indicate $R_{\rm ISCO}$ (solid), $R_{\rm trap}$ (dashed), $R_{\rm ab}$ (dashed-dotted), and $R_{\rm bc}$ (dotted). The black vertical lines in the second and fourth panels indicate $R_{\rm ISCO}$ and $R_{\rm trap}$. For $\feddsixteen = 0.1$, $R_{\rm trap} < R_{\rm ISCO}$, meaning that the photon-trapping region does not exist. Texts ``PT'', ``(a)'', ``(b)'', and ``(c)'' indicate the photon-trapping region and regions (a), (b), and (c) of the disc, respectively. The value of $R_{\rm tran}/\rg$, which is $10^{-5}$ for $\feddsixteen = 10$ and $10^{-1}$ for $\feddsixteen = 0.1$, is not shown in this figure due to its small value. $R_{\rm sg}/\rg$ (for $ Q_{\rm min} = 1$) is equal to $4 \times 10^4$ [in region~(b)] and $2.9 \times 10^5$ [in region~(c)] for $\feddsixteen = 10$ and 0.1, respectively. 
    }
    \label{fig:profile_continuous}
\end{figure}

Unlike $R_{\rm tran}$, the radii $R_{\rm trap}$, $R_{\rm ab}$, and $R_{\rm bc}$ are all positively correlated with the mass accretion rate, because a higher mass accretion rate increases $T$, enhancing the importance of radiation pressure and electron scattering opacity. As a result, at intermediate mass accretion rates, only region~(c) of the thin $\alpha$-disc remains dominant, as the other regions shrink significantly in comparison. Consequently, most accretion disc-based BH growth sub-grid models consider only region~(c) when describing the disc structure \citep[e.g. see footnote~1 in][]{Fiacconi_et_al_2018}. However, for super-Eddington accretion, the contribution from the inner regions may become non-negligible (see the top two panels of Figure~\ref{fig:profile_continuous}), necessitating the inclusion of all three regions of the thin $\alpha$-disc in our model. 

Finally, we define the self-gravitating radius, $R_{\rm sg}$, beyond which the accretion disc is assumed to fragment due to self-gravity. Accordingly, we impose the condition that $R_{\rm disc} \leq R_{\rm sg}$. $R_{\rm sg}$ is determined using the \citet{Toomre_1964} parameter $Q(R) = {\Omega_{\rm K} c_{\rm s}}/{(\pi G \Sigma})$. We define $R_{\rm sg}$ as the radius where $Q(R_{\rm sg}) = Q_{\rm min}$, with $Q_{\rm min} \sim \mathcal O(1)$ is a constant that can be set in the code and determines the maximum size of the disc. The self-gravitating mass is defined as $M_{\rm sg}$ = $\mdisc(\rdisc = R_{\rm sg})$. A detailed calculation of $R_{\rm sg}$ and $M_{\rm sg}$ is provided in Section~\ref{sec:self_gravitating}.  

To model the inner accretion disc structure, we adopt the fitting formulae from simulation results of \citetalias{Kitaki_et_al_2018} to compute the surface gas density, $\Sigma_{\rm trap}$, and specific angular momentum, $L_{\rm trap}$, in the photon-trapping region of the accretion disc, where $R < R_{\rm trap}$: 

\begin{equation}
    \Sigma_{\rm trap} = \frac {\dot{M}_{\rm BH,accr}} {2 \pi R v_R}= 222 \, \feddsixteen^{0.98} \left(\frac{R}{R_{\rm g}}\right)^{0.11} \ \rm{g \ cm^{-2}}~,
    \label{eq:sigma_trap}
\end{equation}

\begin{equation}
\begin{split}
    L_{\rm trap} &= 1.27 \, \feddsixteen^{0.01} \left(\frac{G \mbh}{c} \right)  \left(\frac{R}{\rg}\right)^{0.4} \\
    &= 5.622 \times 10^{21} \, M_{\rm BH, 6} \, \feddsixteen^{0.01} \left(\frac{R}{\rg}\right)^{0.4} \ \rm{cm^2}~\rm{s^{-1}}~. \label{eq:L_trap}
\end{split}
\end{equation}

\cite{Watarai_2006} derived an analytical formula for the slim disc model, predicting a surface density profile of $\Sigma \propto R^{-0.5}$ in the photon-trapping region. The slope is negative in \citet{Watarai_2006}, whereas it is positive in \citetalias{Kitaki_et_al_2018}. The discrepancy may be attributed to the presence of large-scale circulation and outflows observed in the 2D simulations of \citetalias{Kitaki_et_al_2018}, which are not included in the analytical model.

\citet{Sadowski_2011} calculated 1D numerical solutions for the relativistic slim disc model. For $f_{\rm Edd, 16} = 10$ and $R<200 \,\rg < R_{\rm trap}$, their results indicate a slightly larger specific angular momentum and a steeper surface density profile compared to those of \citetalias{Kitaki_et_al_2018}. The discrepancy may be attributed to the inclusion of general relativistic (GR) effects considered in \citet{Sadowski_2011}. Furthermore, \citetalias{Kitaki_et_al_2018} consider a 2D flow with outflows, which may further contribute to the differences. 

Following \citetalias{Kato_et_al_2008}, we obtain the asymptotic limits of the surface density profile for the three regions of the thin $\alpha$-disc around a non-spinning BH by assuming that the pressure is dominated by either $P_{\rm gas}$ or $P_{\rm rad}$ and the opacity $\kappa$ is dominated by either $\kappa_{\rm es}$ or $\kappa_{\rm ff}$: 

\begin{equation}
    \Sigma_{\rm a} = 21.9 \, \alpha_{0.1}^{-1} \, \feddsixteen^{-1} \, \left(\frac{R}{\rg}\right)^{3/2} \, f^{-1} \ \rm{g \ cm^{-2}}~,
    \label{eq:sigma_a}
\end{equation}

\begin{equation}
    \Sigma_{\rm b} = 3.45 \times 10^7 \, \alpha_{0.1}^{-4/5} \, M_{\rm BH,6}^{1/5} \, \feddsixteen^{3/5} \, \left(\frac{R}{\rg}\right)^{-3/5}\, f^{3/5} \ \rm{g \ cm^{-2}}~,
    \label{eq:sigma_b}
\end{equation}

\begin{equation}
    \Sigma_{\rm c} = 1.67 \times 10^8 \, \alpha_{0.1}^{-4/5} \, M_{\rm BH,6}^{1/5} \, \feddsixteen^{7/10} \, \left(\frac{R}{\rg}\right)^{-3/4}\, f^{7/10} \ \rm{g \ cm^{-2}}~,
    \label{eq:sigma_c}
\end{equation}

\noindent where $f = 1 - \sqrt{6 \rg/R}$, and we assume $f = 1$ for simplicity, since its value differs from unity only very close to the ISCO radius. $\Sigma_{\rm a}$, $\Sigma_{\rm b}$, and $\Sigma_{\rm c}$ are the surface (gas) densities of regions~(a), (b), and (c), respectively. Keplerian angular momentum is assumed in all three regions of the thin $\alpha$-disc: 

\begin{equation}
\begin{split}
    L_{\rm K} &=  \left(\frac{G \mbh}{c} \right) \left(\frac{R}{\rg}\right)^{0.5} \\
    &= 4.427 \times 10^{21} \, M_{\rm BH, 6} \left(\frac{R}{\rg}\right)^{0.5} \ \rm{cm^2~s^{-1}}~. \label{eq:L_Keplerian}
\end{split}
\end{equation}

To construct the surface density profile of the entire accretion disc, we first compute the surface density using Equation~\eqref{eq:sigma_c} for region~(c) of the thin $\alpha$-disc, where 
\WB{$R \ge R_{\rm bc}$}. In all inner regions, we renormalise the surface density profile for each region to ensure continuity at the boundaries. This approach results in a continuous surface density profile without abrupt transitions between different regions. When the mass accretion rate is low, the structure of the accretion disc is dominated by region~(c), as the other regions significantly decrease in size (Equation~\ref{eq:r_bc}). Therefore, during renormalisation, we preserve the exact results in region~(c) rather than in other regions. This ensures that the surface density profile remains accurate at low mass accretion rates and maintain consistency with other accretion disc-based sub-grid models that consider only region~(c) of the $\alpha$-disc. We apply the same method even when $R_{\rm disc} < R_{\rm bc}$, such that the construction always starts from region~(c), with a cut-off applied at $R_{\rm disc}$.\footnote{We have tested that applying this renormalisation does not affect the BH mass evolution. Moreover, $R_{\rm disc}$ is always greater than $R_{\rm ab}$ (see Section~\ref{sec:mass_accretion} for the cause) and the renormalisation of region (b) is minimal (see Figure~\ref{fig:profile_continuous}).} We apply a similar procedure for the specific angular momentum profile and apply the renormalisation also to the photon-trapping region. Hereafter, unless otherwise stated, all surface density and specific angular momentum profiles used in the calculations are the renormalised ones.

\begin{figure}
    \centering
    \includegraphics[width=0.95\linewidth]{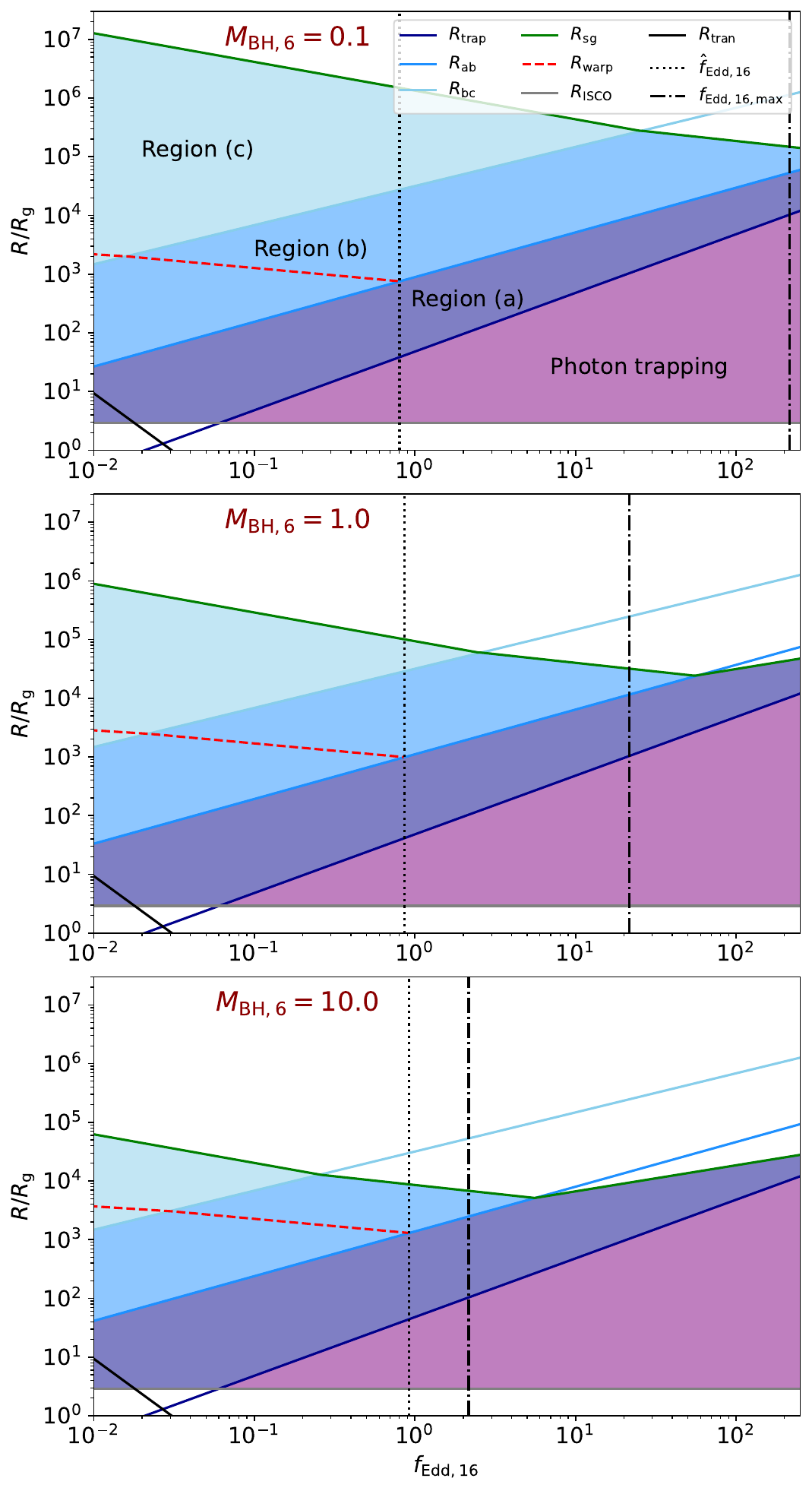}
    \caption{Characteristic radii as a function of $f_{\rm Edd, 16}$ for different values of $\mbh$: $R_{\rm trap}$ (dark blue line), $R_{\rm ab}$ (medium blue line), $R_{\rm bc}$ (light blue line), $R_{\rm sg}$ (green line), $R_{\rm ISCO}$ (grey line), and $R_{\rm tran}$ (solid black line). The warp radius ($R_{\rm warp}$, red dashed line) is defined in Section~\ref{sec:angular_momentum}. 
    The top, middle, and bottom panels correspond to $M_{\rm BH,6} = 0.1, 1$, and 10, respectively, with $\alpha_{0.1} = 1$,  $Q_{\rm min} = 1$, and $a_{\rm BH} = 0.8$ (prograde disc). 
    The shaded regions indicate the parameter space corresponding to different disc regions. The vertical black dotted lines represent $\hat f_{\rm Edd, 16}$, the critical value of $\feddsixteen$ distinguishing the two torque models (see Section~\ref{sec:angular_momentum} for more details). The vertical black dashed-dotted lines represent $f_{\rm Edd, 16, max}$ (defined in Section~\ref{sec:mass_accretion}), assuming $\mdisc = M_{\rm sg}$. 
    } 
\label{fig:characteristic_radius}
\end{figure}

In the top two panels of Figure~\ref{fig:profile_continuous}, we show the (renormalised) surface density and specific angular momentum profiles for $\feddsixteen = 10$ (i.e. $f_{\rm Edd,\eta}/\eta_{0.1} = 16$ in the equations of \citetalias{Cenci_et_al_2021}), with $\alpha_{0.1} = 1$, $M_{\rm BH,6}=1$, and $a_{\rm BH} = 0$. The panels also include the surface density and specific angular momentum profiles without renormalisation (i.e. the exact Equations~\ref{eq:sigma_trap}, \ref{eq:L_trap}, \ref{eq:sigma_a}, and \ref{eq:sigma_b}). For comparison, we also plot the surface density profile of region~(c) of the thin $\alpha$-disc from \citet{Frank_et_al_2002}, which is commonly used to describe the thin $\alpha$-disc in sub-grid models (e.g. \citealt{Perego_et_al_2009, Fiacconi_et_al_2018}; \citetalias{Cenci_et_al_2021}; \citealt{Koudmani_et_al_2024}). However, in our model, we refer to \citetalias{Kato_et_al_2008} for the thin $\alpha$-disc model, as it provides detailed equations for all three regions of the thin $\alpha$-disc. Notably, the surface density in region~(c) of \citetalias{Kato_et_al_2008} is approximately 1.7 times higher than that in \citet{Frank_et_al_2002}. This discrepancy arises from differences in definitions and opacity values adopted in the two references (see Appendix~\ref{app:difference_alpha_disc} for further details, wherein we also compare these two models to the original model by \citealt{Shakura_Sunyaev_1973}).

While the differences between the renormalised surface density and the original surface density equations in regions~(a) and (b) of the thin $\alpha$-disc model are relatively small, they become non-negligible in the photon-trapping region. For example, the surface density can differ by a factor of a few, as seen in the top two panels of this figure. One possible explanation for this discrepancy is that the effective $\alpha$ might change in this region, due to the strong photon advection \citep[e.g.][]{Jiang_et_al_2019}. Another is the inclusion of outflows in the simulations of \citetalias{Kitaki_et_al_2018}. By contrast, the discrepancy in specific angular momentum is much smaller compared to the surface density profile. 

In the bottom two panels of the same figure, we present the case of a lower Eddington ratio, $\feddsixteen = 0.1$ (i.e. $f_{\rm Edd,\eta}/\eta_{0.1} = 0.16$). In this scenario, $R_{\rm trap} < R_{\rm ISCO}$, indicating that the photon-trapping region is absent. Additionally, regions~(a) and (b) shift significantly inwards within the disc, as lower accretion rates lead to weaker radiation pressure and a reduced dominance of electron scattering opacity.

Note that, in general, our renormalisation does not significantly affect the total mass and angular momentum of the disc, as these are dominated by the outer parts of the accretion disc -- typically corresponding to region~(b) or region~(c) -- due to their much larger area (proportional to $R^2$).

In reality, both the surface density profile and the specific angular momentum profile should not only be continuous but also smooth. Although transitions between different regions should ideally be smooth rather than exhibiting discontinuous changes in slopes, we neglect this issue due to the difficulty in determining the exact values of $\Sigma$ and $L$ (and related quantities) at these transitions. For instance, at $R \sim R_{\rm ab}$, both $P_{\rm gas}$ and $P_{\rm rad}$ contribute significantly to the total pressure, making it challenging to derive a simple analytical solution. Instead, we ensure continuity in the profiles while acknowledging that a fully smooth transition would require a more detailed treatment of the disc structure \citep[e.g.][]{Hure_et_al_1994a, Derdzinski_and_Mayer_2023, Gangardt_et_al_2024}.  

Figure~\ref{fig:characteristic_radius} illustrates the characteristic radii as a function of $\feddsixteen$, for a few different values of BH mass, with each shaded region indicating the parameter space corresponding to a specific disc region. For example, consider a BH with $M_{\rm BH, 6} = 1$: when $\feddsixteen \gtrsim 2.5$, region~(c) of the accretion disc vanishes, since $R_{\rm bc} > R_{\rm sg}$. This supports our claim that at higher mass accretion rates, the contributions of the inner regions of the accretion disc become increasingly significant.

\subsection{Mass accretion}\label{sec:mass_accretion}

At each time step, the growth rates of $M_{\rm BH}$ and $M_{\rm disc}$, denoted as $\dot{M}_{\text{BH,growth}}$ and $\dot M_{\text{disc}}$, respectively, are calculated using the following equations: 

\begin{align}
    \dot{M}_{\text{BH,growth}} &= (1 - \eta) \, \dot{M}_{\text{BH,accr}}~,     \label{eq:M_BH_growth} \\
    \dot M_{\text{disc}}  &= \dot{M}_{\text{in}} - \dot{M}_{\text{BH,accr}} - \dot M_{\rm out}~,
\end{align}

\noindent where the term $(1-\eta)$ represents the fraction of mass that is effectively accreted onto the BH due to radiative losses.\footnote{In the common case wherein only the BH is modelled (i.e. with no accretion disc), the BH growth rate should be $\dot{M}_{\text{BH,growth}} = (1 - \eta_{\rm m}) \, \dot{M}_{\text{BH,accr}}$, where $\eta_{\rm m} \geq \eta$ is the fraction of mass-energy released by accretion in any form \citep[e.g.][]{Volonteri_et_al_2015, Capelo_et_al_2023}. In our model, $\eta_{\rm m} = \eta$, as any mass loss due to feedback is described by $\dot M_{\rm out}$.} The mass outflow rate due to BH feedback, $\dot{M}_{\rm out}$, is set to zero in this work, as we do not consider BH feedback. The mass accretion rate onto the disc, $\dot{M}_{\rm{in}}$, is determined by the gas dynamics on resolved scales.

Instead of using the BHL prescription for $\dot{M}_{\text{BH,accr}}$ \citep[as in, e.g.][]{Dubois_et_al_2014, Bustamante_et_al_2019, Massonneau_et_al_2023, Sala_et_al_2024, Husko_et_al_2025}, which neglects the influence of angular momentum transport on accretion disc scales, we update $\dot{M}_{\text{BH,accr}}$ using an accretion disc-mediated accretion rate \citep[][]{Fiacconi_et_al_2018}. Since the accretion disc is unresolved in the simulation, $\dot{M}_{\text{BH,accr}}$ is determined based on integrated properties of the disc, $M_{\rm disc}$ and $J_{\rm disc}$, within a sub-grid model. For instance, equation~(5) in \citetalias{Cenci_et_al_2021} (see also equations~2 and A5 in \citealt{Fiacconi_et_al_2018}) describes the relation between $M_{\rm disc}$, $J_{\rm disc}$, and $f_{\rm Edd, \eta}$. However, these relations are derived under the assumption that the whole accretion disc follows the region~(c) solution of the thin $\alpha$-disc model, which is not consistent with the multi-region disc structure considered in this work. 

Within our more complex disc model, it is not feasible any longer to derive a simple analytic equation relating $\feddsixteen$ to $\mdisc$ and $\jdisc$, as was done in previous works. Instead, we first calculate $\mdisc$ and $\jdisc$ via numerical integration using the following equations:

\begin{align}
    \mdisc &= 2\pi \int^{\rdisc}_{R_{\rm ISCO}} \Sigma(R, \feddsixteen) \, R \, \rm{d} R ~,   \label{eq:M_disc_integral} \\   
    \jdisc &= 2\pi \int^{\rdisc}_{R_{\rm ISCO}} \Sigma(R, \feddsixteen) \, L(R, \feddsixteen) \, R \, \rm{d} R ~. \label{eq:J_disc_integral}
\end{align}

For simplicity, we assume $R_{\rm ISCO}=0$ when performing the integration. These integrations allow us to determine $M_{\rm disc}$ and $J_{\rm disc}$ as functions of $\feddsixteen$ and $R_{\rm disc}$. We then apply the Newton-Raphson method to invert these two functions numerically and compute $\feddsixteen$ from given values of $M_{\rm disc}$ and $J_{\rm disc}$. The initial guess for the Newton-Raphson method is taken as the $\feddsixteen$ and $\rdisc$ values from the previous time step. 

However, the Newton-Raphson method fails to compute $\feddsixteen$ when multiple or no solutions exist for a given pair of $M_{\rm disc}$ and $J_{\rm disc}$ values. As shown in Appendix~\ref{app:Newton_Raphson_method}, degeneracies arise when region~(a) dominates the disc structure instead of regions~(b) or (c), leading to multiple or no valid solutions of $\feddsixteen$ for the same $(M_{\rm disc}, \, J_{\rm disc})$. To address this, we define an upper limit for the Eddington ratio, given by

\begin{equation}
    f_{\rm Edd,16,max} = 21.6 \, M_{\rm BH,6}^{-1} \alpha_{0.1}^{4/10} Q_{\rm min}^{-7/10} \left(\frac{M_{\rm disc}}{M_{\rm sg}}\right)^{3/5}~.
    \label{eq:f_edd_max}
\end{equation}

\noindent To prevent degeneracies \WB{or the absence of a solution in our numerical approach}, we impose the constraint $\feddsixteen \leq f_{\rm Edd,16,max}$.

Many previous sub-grid models that do not account for super-Eddington accretion impose an upper limit on the Eddington ratio, capping it below a value close to unity to prevent super-Eddington accretion (e.g. \citealt{Power_et_al_2011, Dubois_et_al_2014, Fiacconi_et_al_2018, Bustamante_et_al_2019}; \citetalias{Cenci_et_al_2021}). In contrast, our model introduces a new cap at $f_{\rm Edd, 16, max}$, which is approximately 20 for $M_{\rm BH, 6}= 1$ (and $M_{\rm disc} = M_{\rm sg}$). However, we caution that for more massive BHs ($\mbh \gtrsim 10^8 \msun$), $f_{\rm Edd, 16, max} \lesssim 1$. Additionally, previous sub-grid models that consider only region~(c) of the thin $\alpha$-disc solution might become inaccurate when $\feddsixteen > f_{\rm Edd, 16, max}$, as the disc should instead be dominated by region~(a), which has significantly different structural properties (see Figure~\ref{fig:profile_continuous}). 

\subsection{Angular momentum evolution}\label{sec:angular_momentum}

In this section, we describe the evolution of the BH angular momentum due to gas accretion at $R_{\rm ISCO}$ (see Section~\ref{sec:angular_momentum_accretion}) and the Lense-Thirring effect. The evolution of the angular momentum of the BH is given by

\begin{equation}
    {\bm{\dot J}}_{\text{BH}} = {\bm{\dot J}}_{\rm BH, acc} + {\bm{\dot J}}_{\rm BH, LT} \WB{- {\bm{\dot J}}_{\rm BH, jet}}~, \label{eq:update_jbh}
\end{equation}

\noindent where ${\bm{\dot J}}_{\rm BH, acc}$ represents the change in angular momentum due to accretion\WB{,} 
${\bm{\dot J}}_{\rm BH, LT}$ is the Lense-Thirring torque\WB{, and ${\bm{\dot J}}_{\rm BH, jet}$ accounts for any spin-down effect due to, e.g. jets \citep[e.g.][]{Blandford_and_Znajek_1977,Ricarte_et_al_2023}, which we set to zero in this work}. We adopt two different models to compute the Lense-Thirring torque, for low and high mass accretion rates (see Section~\ref{sec:LT_torque}). 

Due to the conservation of angular momentum, $\bm{J}_{\rm disc}$ evolves according to 

\begin{equation}
    \bm{\dot J}_{\rm disc} = - \bm{\dot J}_{\rm BH} + \bm{\dot J}_{\rm in} \WB{-\bm{\dot J}_{\rm out}}~,
    \label{eq:update_jdisc}
\end{equation}

\noindent where $\bm{\dot J}_{\rm in}$ is the angular momentum inflow onto the accretion disc from resolved scales \WB{and $\bm{\dot J}_{\rm out}$ is the angular momentum outflow due to BH feedback effects \citep[see, e.g.][]{Bollati_et_al_2024}, which is set to zero in this work}.

\subsubsection{Accretion}\label{sec:angular_momentum_accretion}

The specific angular momentum at $R_{\rm ISCO}$, denoted as $\Lambda_{\rm ISCO}$, depends on $a_{\rm BH}$ and $M_{\rm BH}$. Following \citet{Bardeen_et_al_1972}, it is computed using 

\begin{equation}
    \Lambda_{\text{ISCO}}(a_{\rm BH}) = \pm \frac{GM_{\rm BH}}{c\lambda} \frac{\lambda^2 \mp 2a_{\rm BH} \sqrt{\lambda} + a_{\rm BH}^2}{\left( \lambda - 3 \pm 2a_{\rm BH} / \sqrt{\lambda} \right)^{1/2}}~,
    \label{eq:lambda_isco}
\end{equation}

\noindent  where $\lambda = R_{\rm ISCO} / \rg$ (given by Equation~\ref{eq:r_isco}), and the upper and lower signs correspond to a prograde and retrograde disc, respectively. The evolution of the BH angular momentum due to accretion is given by
 
\begin{equation}
    {\bm{\dot J}}_{\rm BH, acc} = \WB{\dot{M}_{\rm{BH, accr}}} \, \Lambda_{\rm{ISCO}} \, \bm{j}_{\rm{BH}}~.
    \label{eq:J_BH_accretion}
\end{equation}

Following \citetalias{Cenci_et_al_2021}, the disc \WB{is assumed to be depleted and its mass removed} if $\jdisc/\mdisc < \Lambda_{\rm ISCO}$. In such cases,it is refilled stochastically through inflows from the surrounding gas \WB{(for example, when there is no gas inflow from the surroundings, or in very special configurations; e.g. \citealt{Fiacconi_et_al_2018}; \citetalias{Cenci_et_al_2021}).} However, in the simulations presented in this paper, such depletion never occurs. 

\subsubsection{The Lense-Thirring effect}\label{sec:LT_torque}

The spinning BH induces Lense-Thirring precession on the surrounding accretion disc. The precession frequency, $\omega_{\rm LT}$, is given by \citep[][]{Lense_and_Thirring_1918}

\begin{equation}
    \omega_{\rm LT} = \frac{2 G J_{\rm BH}}{c^2 R^3} ~.
    \label{eq:omega_LT}
\end{equation}

The response of the disc to Lense-Thirring precession depends on the parameters $\alpha$ and $H/R$. For a thin disc ($\alpha > H/R$), the system lies in the diffusive regime, in which the warp is communicated through the vertical viscosity, $\nu_2$, related to $\nu_1$ via Equation~\eqref{eq:nu_2}. It is important to note that $\xi$ is a function of $\alpha$ \citep[][]{Lodato_and_Pringle_2007}. For larger values of $\alpha$, $\xi$ can increase to around 1. We adopt $\xi = 0.7$ in this study, as it is more appropriate for $\alpha_{0.1} = 1$ \citepalias[as in][]{Cenci_et_al_2021}. In this regime, the precessional motion is damped by the vertical viscosity.

When $\alpha < H/R$, the disc enters the wave-like regime, wherein bending waves effectively communicate the warp because the sound-crossing time-scale becomes shorter than the vertical viscous time-scale, rendering the effect of vertical viscosity negligible \citep[see the reviews by][]{Ingram_and_Motta_2019, Fragile_and_Liska_2024}. In this regime, if the outer boundary of the disc is misaligned with the BH, the entire region remains misaligned and undergoes solid-body precession. Beyond the bending wave radius, $R_{\rm bw} = 6 (H/R)^{-4/5} a_{\rm BH}^{2/5} \, \rg$, bending waves propagate smoothly, maintaining a nearly constant tilt angle. However, for $R <R_{\rm bw}$, the wavelength of these waves depends strongly on radius, leading to significant variations in the tilt angle. GR magnetohydrodynamics (GRMHD) simulations suggest that this effect can induce a substantial decrease in density within this region \citep[][]{Fragile_et_al_2009, Ingram_et_al_2009}. 

In our accretion disc model, the photon-trapping region lies in the wave-like regime, since it is geometrically thick ($H/R = 1$), whereas all three regions of the thin $\alpha$-disc are in the diffusive regime.

\begin{figure*}
    \centering
    \includegraphics[width=0.9\linewidth]{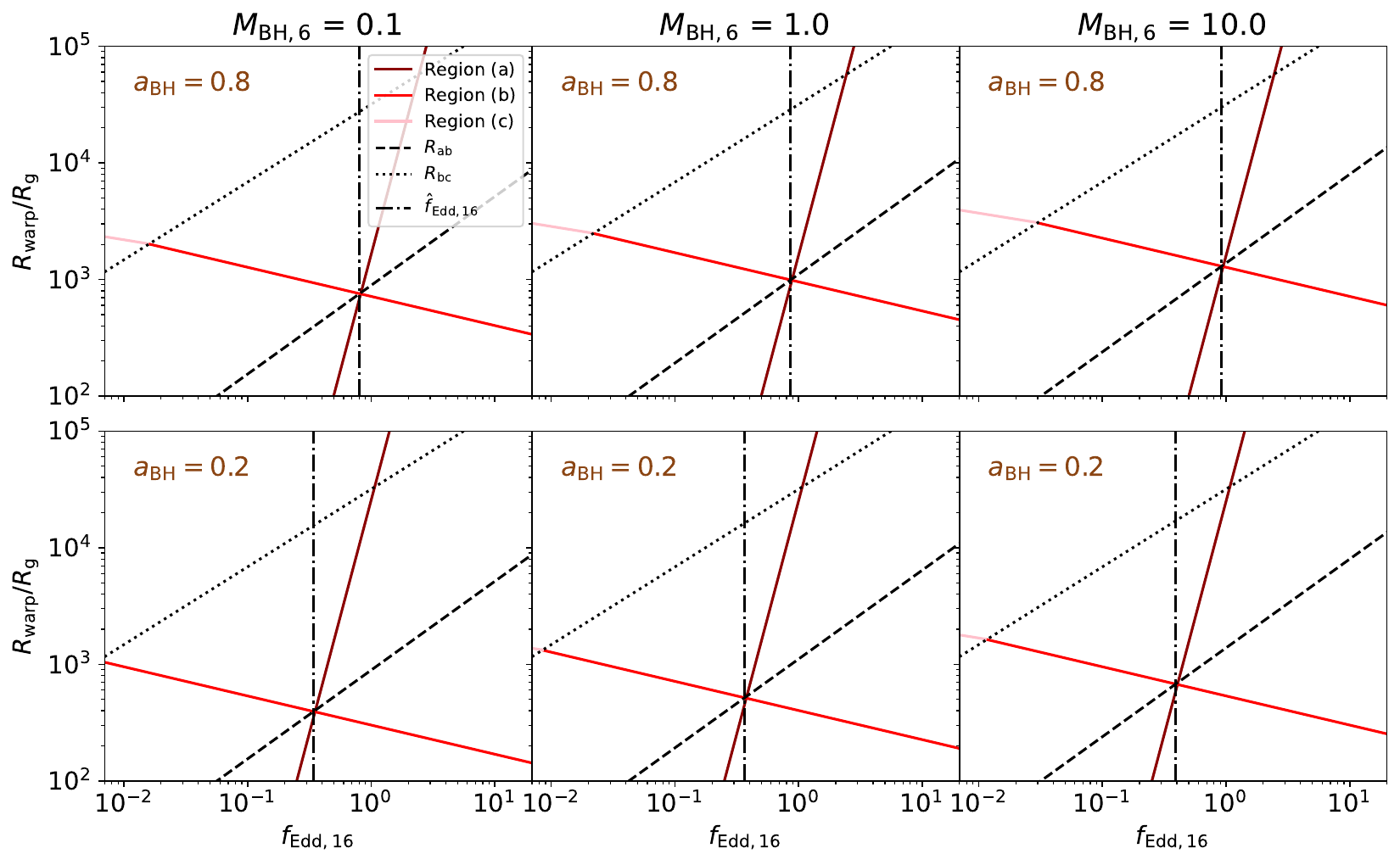} 
    \caption{The warp radius, $R_{\rm warp}$, as a function of $f_{\rm Edd,16}$ for $a_{\rm BH}=0.8$ (top row) and 0.2 (bottom row) and $M_{\rm BH, 6}$ = 0.1 (left-hand column), 1 (central column), and 10 (right-hand column), with $\alpha_{0.1} = 1$ and $\xi=0.7$. For each region, $R_{\rm warp}$ is calculated by assuming a power-law surface density profile, using \WB{Equations~\eqref{eq:r_warp_a}, \eqref{eq:r_warp_b}, and \eqref{eq:r_warp_c}}, properly readjusted after the renormalisation of the surface density and specific angular momentum profiles. Dark red, red, and pink lines correspond to regions~(a), (b), and (c), respectively. Black lines indicate $R_{\rm bc}$ (dotted), $R_{\rm ab}$ (dashed), and $\hat{f}_{\rm Edd, 16}$ (dash-dotted). When $f_{\rm Edd, 16} < \hat f_{\rm Edd, 16}$, $R_{\rm warp} = R_{\rm warp, b}$ if $R_{\rm warp, b} < R_{\rm bc}$; otherwise, $R_{\rm warp} = R_{\rm warp, c}$. When $f_{\rm Edd, 16} > \hat f_{\rm Edd, 16}$, $R_{\rm warp, a} > R_{\rm ab}$ and $R_{\rm warp, b} < R_{\rm ab}$. In this case, the Bardeen-Petterson configuration cannot be reached. For $f_{\rm Edd, 16} > \hat f_{\rm Edd, 16}$, we extrapolate $R_{\rm warp, b}$ and $R_{\rm warp, a}$ beyond their regions of validity to better illustrate this in the figure.
    \label{fig:r_warp}
    }
\end{figure*}

Within the diffusive regime, we qualitatively estimate the shape of the disc by considering the balance between the Lense-Thirring torque and vertical shear. Following equations~(1) and~(9) in \citet{Martin_et_al_2007}, the torque induced by the Lense-Thirring precession on the disc at radius $R$ is given by $\bm \tau_{\rm LT} = -\omega_{\rm LT} \Sigma \bm{j}_{\rm BH} \times \bm{L}$, whereas the vertical shear exerted by the vertical viscosity is $\bm \tau_{\rm visc} = \frac 1 R \frac{\partial}{\partial R} \left( \frac{1}{2} R \nu_2  \Sigma  L \frac{\partial \bm l}{\partial R} \right)$. The equilibrium state of the warped disc is reached when $\bm \tau_{\rm visc} = \bm \tau_{\rm LT}$. To characterise the relative strength of the vertical shear and the Lense-Thirring torque, we introduce the dimensionless parameter $\mathcal{A}$:

\begin{equation}
    \mathcal A = \frac{\frac{1} {R^2} \frac{1}{2} R \nu_2 \Sigma \frac{L}{R}}{{\omega_{\rm LT}  \Sigma {L}} } = \frac{\nu_2}{2\omega_{\rm LT}R^2} = \frac{c^2}{4 G J_{\rm BH}}  \nu_2(R)R~.
\end{equation}

If $\nu_2(R)$ is a continuous and monotonically increasing function, then $\mathcal{A} > 1$ in the outer part of the disc and $\mathcal{A} < 1$ in the inner part. This implies that the inner region of the disc is governed by the Lense-Thirring torque and aligns with the BH, while the outer region remains misaligned due to the relatively weaker Lense-Thirring torque. This configuration is also known as the Bardeen-Petterson effect \citep[][]{Bardeen_and_Petterson_1975, Papaloizou_and_Pringle_1983, Nealon_et_al_2015, Liska_et_al_2019}. A smooth transition occurs around the warp radius, $R_{\rm warp}$, which is defined by the condition $\mathcal A\left(R=R_{\rm warp}\right) = 1$.\footnote{This definition is equivalent to requiring that the Lense-Thirring precession period is comparable to the vertical warp diffusion time-scale: $\omega_{\rm LT}^{-1} (R_{\rm warp}) \sim R_{\rm warp}^2/\nu_2(R_{\rm warp})$.}

For simplicity, we first assume a power\WB{-}law expression for $\nu_1$: $\nu_1 = C R^{\beta}$. Using this assumption, we derive $R_{\rm warp}$ based on its definition:

\begin{equation}
    R_{\rm warp} = \frac{4GJ_{\rm BH}}{c^2\nu_{2}(R_{\rm warp})} = \left( \frac{8G^2 M_{\rm BH}^2 a_{\rm BH} \alpha^2}{\xi C c^3} \right)^{1/(1+\beta)}~.
    \label{eq:r_warp}
\end{equation}

\noindent We use this equation to reformulate $\mathcal{A}$ in terms of $R_{\rm warp}$:

\begin{equation}
    \mathcal A= \left(\frac{R}{R_{\rm warp}}\right) \left(\frac{\nu_2}{\nu_{2}(R_{\rm warp})}\right) = \left(\frac{R}{R_{\rm warp}}\right)^{1+\beta}.
    \label{ratio_LT_viscosity_torque}
\end{equation}

For instance, consider an accretion disc composed only of region~(c) of the thin $\alpha$-disc (i.e. $\beta=3/4$). The corresponding warp radius, $R_{\rm warp, c}$, and the dimensionless parameter, $\mathcal{A}_{\rm c}$, can be determined using Equations~\eqref{eq:r_warp} and~\eqref{ratio_LT_viscosity_torque}. The same method is applicable also to regions~(a) and (b):\footnote{
\WB{When} $\beta = -3/2$, 
$\mathcal A (R)$ is a decreasing function, meaning that $\mathcal A > 1$ in the inner disc \WB{($R<R_{\rm warp,a}$; which exists only if $R_{\rm warp,a} > R_{\rm ISCO}$)}, preventing alignment with the BH. Nonetheless, we still calculate $R_{\rm warp, a}$ and use it to describe $\mathcal{A}_{\rm a}(R)$ for simplicity. We also note that this approach is not applicable to the photon-trapping region, as it lies in the wave-like regime. Finally, we note that Equations~\eqref{eq:r_warp_a}--\eqref{eq:r_warp_c} are not renormalised, but the computations in the code and the results we show are based on their renormalised version. The same applies to Equations~\eqref{eq:r_sg_a}--\eqref{eq:r_sg_c}.}

\begin{align}
    \frac{R_{\rm warp, a}}{\rg} &= 9.18 \times 10^2 \, \xi^{2}  \alpha_{0.1}^{-2} \, a_{\rm BH}^{-2} \, \feddsixteen^{4}~, \label{eq:r_warp_a}\\
    \frac{R_{\rm warp, b}}{\rg} &= 8.78 \times 10^2 \, \xi^{-5/8} M_{\rm BH,6}^{1/8} \, \alpha_{0.1}^{3/4} \, a_{\rm BH}^{5/8} \, \feddsixteen^{-1/4}~, \label{eq:r_warp_b}\\ 
    \frac{R_{\rm warp, c}}{\rg} &= 1.19 \times 10^3 \, \xi^{-4/7} M_{\rm BH,6}^{4/35} \, \alpha_{0.1}^{24/35} \, a_{\rm BH}^{4/7} \, \feddsixteen^{-6/35}~, \label{eq:r_warp_c}\\ 
    \mathcal A_{\rm a} &= \left(\frac{R}{R_{\rm warp, a}}\right)^{-1/2}~, \label{eq:A_a}\\ 
    \mathcal A_{\rm b} &= \left(\frac{R}{R_{\rm warp, b}}\right)^{8/5}~, \label{eq:A_b}\\
    \mathcal A_{\rm c} &= \left(\frac{R}{R_{\rm warp, c}}\right)^{7/4}~. \label{eq:A_c} 
\end{align}

We then use the above equations to characterise the alignment behaviour of the accretion disc in our model. In region~(a), $\mathcal{A}(R)$ decreases with increasing $R$, whereas in regions~(b) and (c), $\mathcal{A}(R)$ increases with increasing $R$. Since $\mathcal{A}(R)$ is a continuous function 
\WB{of radius}, its minimum value occurs at $R=R_{\rm ab}$. If $\min(\mathcal{A}) = \mathcal{A}(R_{\rm ab}) > 1$, the vertical shear generated by viscosity is always more significant than the Lense-Thirring torque throughout the disc. Therefore, the disc does not reach the Bardeen-Petterson configuration and remains misaligned with the BH. This condition is satisfied when $R_{\rm warp, a} > R_{\rm ab}$ and $R_{\rm warp, b} < R_{\rm ab}$. Conversely, if $\min(\mathcal{A}) = \mathcal{A}(R_{\rm ab}) < 1$, we assume that the innermost part of region~(b) aligns with the BH, since 
\WB{$\mathcal{A} < 1$} there. Consequently, region~(b) supplies aligned gas inflow to region~(a) and the photon-trapping region, causing them to align with the BH as well.

Figure~\ref{fig:r_warp} illustrates $R_{\rm warp}$ (after renormalisation of the surface density and the specific angular momentum) in different regions as a function of $\feddsixteen$ for a few values of $\mbh$ and $a_{\rm BH}$ (note that this result is the same for prograde/retrograde orbits, since the Lense-Thirring torque has the same magnitude in both cases). We define a critical Eddington ratio, $\hat f_{\rm Edd, 16}$, such that $R_{\rm warp, b}(\hat f_{\rm Edd, 16}) = R_{\rm ab}(\hat f_{\rm Edd, 16})$. For $\feddsixteen < \hat f_{\rm Edd, 16}$, 
\WB{$R_{\rm warp , b} > R_{\rm ab}$}, which implies $\min(\mathcal{A}) < 1$. This condition allows the system to reach the Bardeen-Petterson configuration. If $\feddsixteen \ll \hat f_{\rm Edd, 16}$, then $R_{\rm warp, b} > R_{\rm bc}$, meaning that $R_{\rm warp, b}$ cannot lie within region~(b). In this case, we assume $R_{\rm warp} = R_{\rm warp, c}$. Otherwise, if $\feddsixteen$ is larger, such that $R_{\rm warp, b}$ lies within region~(b), $R_{\rm warp} = R_{\rm warp, b}$. It is evident that when $\feddsixteen > \hat f_{\rm Edd, 16}$, $R_{\rm warp, a} > R_{\rm ab}$ and $R_{\rm warp, b} < R_{\rm ab}$, leading to $\min(\mathcal{A}) > 1$. In this regime, the system cannot achieve the Bardeen-Petterson configuration, and the Lense-Thirring torque is instead generated by precession in the misaligned wave-like photon-trapping region.

Using Equations~\eqref{eq:r_ab} and~\eqref{eq:r_warp_b}, we obtain $\hat f_{\rm Edd, 16} = 0.79 \, \xi^{-21/34}\, M_{\rm BH,6}^{1/34} \, a_{\rm BH}^{21/34} \, \alpha_{0.1}^{11/17}$. However, when $a_{\rm BH} \ll 1$, $\hat f_{\rm Edd, 16} \ll 1$, leading to $R_{\rm trap}$ < $R_{\rm ISCO}$. In this case, the photon-trapping region disappears. From Equations~\eqref{eq:r_isco} and~\eqref{eq:r_trap}, the condition $R_{\rm trap} < R_{\rm ISCO}$ occurs when $\feddsixteen < R_{\rm ISCO}/(48\rg) = \lambda/48 \sim 0.125$. Here, we assume $R_{\rm ISCO} \sim 6 \rg$, since $a_{\rm BH} \ll 1$. Consequently, the photon-trapping region disappears when $\feddsixteen < 0.125$. If $\hat f_{\rm Edd, 16} < \feddsixteen < 0.125$, the photon-trapping region vanishes, but the disc remains misaligned with the BH. This implies that the torque between the BH and the disc would be extremely weak. For simplicity, we assume that the Lense-Thirring torque in this regime still follows the Bardeen-Petterson configuration. Given that $a_{\rm BH} \ll 1$, the Lense-Thirring torque is weak and it would not significantly affect the results. Thus, we redefine the critical value for $\feddsixteen$ as

\begin{equation}
    \hat f_{\rm Edd, 16} = \max \left(0.125, \, 0.79 \, \xi^{-21/34}\, M_{\rm BH,6}^{1/34} \, a_{\rm BH}^{21/34} \, \alpha_{0.1}^{11/17}\right) ~.
    \label{eq:f_edd_crit}
\end{equation}

\begin{figure}
    \centering
    \includegraphics[width=1.0\linewidth]{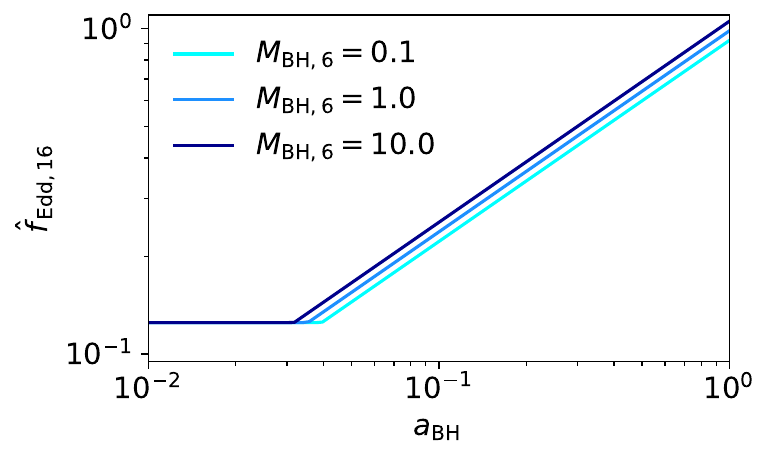}  
    \caption{Critical Eddington ratio, $\hat f_{\rm Edd, 16}$ (Equation~\ref{eq:f_edd_crit}), as a function of the BH spin parameter, $a_{\rm BH}$, with $\alpha_{0.1} = 1$ and $\xi = 0.7$. 
    The different curves correspond to $M_{\rm BH,6} = 0.1$ (cyan), 1 (blue), and 10 (dark blue). 
    }
    \label{fig:f_edd_crit}
\end{figure}

Figure~\ref{fig:f_edd_crit} presents $\hat f_{\rm Edd, 16}$ as a function of $a_{\rm BH}$  
for $M_{\rm BH, 6} = 0.1, 1$, and 10. It is evident that the dependence of $\hat f_{\rm Edd, 16}$ on $\mbh$ is minimal, with $a_{\rm BH}$ being the primary factor influencing its value. For a rapidly spinning BH ($a_{\rm BH} \gtrsim 0.5$), we find that $\hat f_{\rm Edd, 16} \approx 1$.

When $\feddsixteen \leq \hat f_{\rm Edd, 16}$, we assume that the torque is exerted via the Bardeen-Petterson effect, leading to alignment between the inner disc and the BH spin (see Section~\ref{sec:LT_low}). Conversely, when $\feddsixteen > \hat f_{\rm Edd, 16}$, the disc remains misaligned with the BH, and the Lense-Thirring torque is instead driven by the inner precessing thick\WB{-}disc (see Section~\ref{sec:LT_high}).

\subsubsection{Lense-Thirring effect for low mass accretion rates}\label{sec:LT_low}

In this section, we compute the Lense-Thirring effect for $\feddsixteen \leq \hat f_{\rm Edd, 16}$, when the Bardeen-Petterson configuration is achieved. The gravito-magnetic torque exerted by the Lense-Thirring effect in a misaligned disc is derived in \citet{Fiacconi_et_al_2018}, based on the formulation of \citet{Martin_et_al_2007}:

\begin{equation}
    {\bm{\dot J}}_{\rm BH, LT} = - \frac{\bm{J}_{\text{BH}}}{t_{\rm{gm}}} \times \left\{ \sin \left( \frac{\pi}{7} \right) \, \bm{j}_{\text{disc}} + \cos \left( \frac{\pi}{7} \right) (\bm{j}_{\text{BH}} \times \bm{j}_{\text{disc}}) \right\}~, 
    \label{eq:LT_lowedd}
\end{equation}

\noindent where $t_{\rm gm}$ represents the characteristic time-scale over which the gravito-magnetic interaction significantly alters the BH spin direction. We re-derive the time-scale using expressions from \citet{Fiacconi_et_al_2018}, assuming that the disc structure follows region~(c) of the thin $\alpha$-disc as described in \citetalias{Kato_et_al_2008}:

\begin{equation}
    t_{\text{gm}} \simeq 9.5 \times 10^{-2} \, \xi^{-5/7} \, \alpha_{0.1}^{58/35} \, M_{\text{BH},6}^{-2/35} \, a_{\rm BH}^{5/7} \, \feddsixteen^{-32/35} \, \text{Myr}~.
    \label{eq:t_gm}
\end{equation}

We note that the derivations of ${\bm{\dot J}}_{\rm BH, LT}$ and $t_{\rm gm}$ are both based on the assumption that the accretion disc structure is well-described by region~(c). In our model, when $f_{\rm Edd, 16} \leq \hat f_{\rm Edd, 16}$, region~(a) aligns with the BH and does not contribute to the torque. Additionally, the structure of region~(b) closely resembles that of region~(c) (see Figure~\ref{fig:profile_continuous}), as the only difference between them is the dominant opacity source. Consequently, we adopt the same assumption in our calculation.

As discussed in Section~\ref{sec:super_edd_model}, we assume that $\bm{J}_{\rm disc}$ is aligned with the outer section of the warped disc, as this region contains the majority of the disc angular momentum. Specifically, the outer section of the warped disc corresponds to 
\WB{$R > R_{\rm warp}$}. This assumption holds as long as $R_{\rm disc} \gg R_{\rm warp}$. If this condition is not met, the angular momentum contribution from the inner section, which is aligned with the BH, would be comparable to that of the outer section. In this case, $\bm{J}_{\rm disc}$ would not align with the outer section. Moreover, the condition $R_{\rm disc} \gg R_{\rm warp}$ is an implicit assumption in the derivation of  Equation~\eqref{eq:LT_lowedd} \citep[][]{Martin_et_al_2007}. Even though the condition $R_{\rm disc} \gg R_{\rm warp}$ is always verified in the simulations conducted in this study (but may not hold for $\mbh \gtrsim 10^8 \msun$; see Figure~\ref{fig:characteristic_radius}), we modify the angular momentum model when $R_{\rm disc} < R_{\rm warp}$. In this regime, the strong Lense-Thirring torque significantly reduces the time-scale for (counter)-alignment between the disc and the BH. Following \citet{Dotti_et_al_2013}, \citet{Fiacconi_et_al_2018}, and \citetalias{Cenci_et_al_2021}, we assume that the disc and the BH instantaneously (counter)-align with each other. $\bm{J}_{\rm BH}$ aligns with the total angular momentum of the BH-disc system, defined as $\bm{J}_{\rm tot} = \bm{J}_{\rm disc} + \bm{J}_{\rm BH}$, and $\bm J_{\rm disc}$ aligns with $\bm{J}_{\rm tot}$ if 

\begin{equation}
    \bm{j}_{\mathrm{BH}} \cdot \bm{j}_{\mathrm{disc}} \geq -\frac{J_{\mathrm{disc}}}{2J_{\mathrm{BH}}}~.
    \label{eq:counter_align_King_2005}
\end{equation}

\noindent Otherwise, $\bm J_{\rm disc}$ would end up counter-aligned with $\bm{J}_{\rm tot}$ \citep[][]{King_et_al_2005}. 

\subsubsection{Lense-Thirring effect for high mass accretion rates}\label{sec:LT_high}

\begin{figure*}
    \centering
    \includegraphics[width=1.0\linewidth]{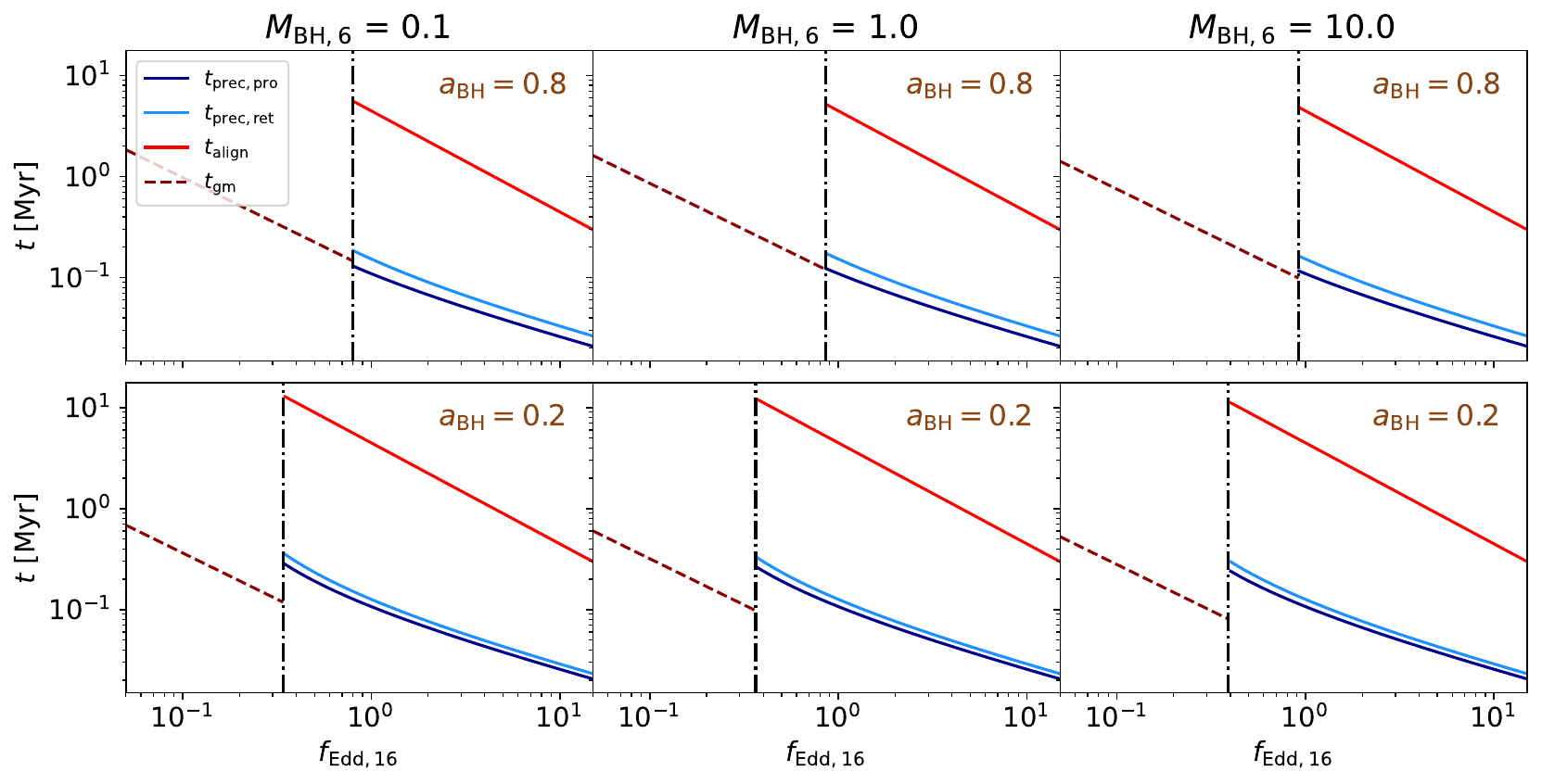}  
    \caption{Time-scales $t_{\rm align}$ (red line), $t_{\rm gm}$ (dashed dark red line), and $t_{\rm prec}$ for prograde ($t_{\rm prec, pro}$; dark blue line) and retrograde ($t_{\rm prec, ret}$; light blue line) discs as a function of $\feddsixteen$, displayed for $a_{\rm BH} = 0.8$ (top row) and $0.2$ (bottom row), and for $M_{\rm BH, 6} = 0.1$ (left-hand column), 1 (central column), and 10 (right-hand column), with $\alpha_{0.1} = 1$ and $\xi = 0.7$. 
    The vertical black dash-dotted line marks $\hat{f}_{\rm Edd, 16}$. We find that $t_{\rm prec} \sim t_{\rm gm}$, whereas $t_{\rm align} \gg t_{\rm gm}$ at $\feddsixteen = \hat f_{\rm Edd, 16}$.
    \label{fig:precession_alignment_time}}
\end{figure*}

In this section, we calculate the Lense-Thirring effect for $\feddsixteen > \hat f_{\rm Edd, 16}$. In this regime, the accretion disc remains misaligned with the BH, and the thick disc in the photon-trapping region, which lies in the wave-like regime, undergoes precession due to the Lense-Thirring effect. This scenario is analogous to the truncated\WB{-}disc model described in \citet{Koudmani_et_al_2024}.

Following \citet{Ingram_and_Motta_2019}, the angular frequency of precession for the inner thick disc, $\omega_{\text{prec}}$, is given by 

\begin{equation}
    \omega_{\text{prec}} = \frac{\int_{R_{\text{in}}}^{R_{\text{trap}}} \omega_{\rm LT}(R) \Sigma(R) L(R) R \, {\rm d} R}{\int_{R_{\text{in}}}^{R_{\text{trap}}} \Sigma(R) L(R) R \, {\rm d} R}~.
\end{equation}

As noted by \citet{Ingram_et_al_2009} and \citet{Koudmani_et_al_2024}, it is crucial to account for the bending wave radius $R_{\rm bw}$, as the density decreases significantly for $R < R_{\rm bw}$. Therefore, we define $R_{\rm in} = \max(R_{\rm ISCO}, R_{\rm bw})$. In our calculations, we assume $H/R=1$ when calculating $R_{\rm bw}$ \citep[as shown in][]{Kitaki_et_al_2021}. Assuming $\Sigma(R) L(R) \propto R^s$, we obtain:

\begin{equation}
    \begin{split}
        \omega_{\rm prec} &= 8 c a_{\rm BH} \rg^2 \frac{\int_{R_{\text{in}}}^{R_{\text{trap}}} R^{s+1-3} \,\text d R}{\int_{R_{\text{in}}}^{R_{\text{trap}}}  R^{s+1} \, \text d R} \\
        &= \frac{8 c a_{\rm BH}}{\rg} \frac{s+2}{s-1} \frac{(R_{\rm trap}/\rg)^{s-1} - (R_{\rm in}/\rg)^{s-1}}{(R_{\rm trap}/\rg)^{s+2} - (R_{\rm in}/\rg)^{s+2}}~.
    \end{split}
\end{equation}

Similarly to what \WB{is} done by \citet{Koudmani_et_al_2024}, the Lense-Thirring torque for a precessing disc is calculated as follows: 

\begin{equation}
    {\bm{\dot{J}}}_{\rm BH, LT} = - \bm{J}_{\rm BH} \times \left\{ t_{\rm prec}^{-1} \, \bm{j}_{\rm disc} + t_{\rm align}^{-1} \left(\bm{j}_{\rm BH} \times \bm{j}_{\rm disc} \right) \right\}
    ~,
    \label{eq:LT_highedd}
\end{equation}

\noindent where the first term describes the precession between the BH and the disc, which does not alter the magnitude of $J_{\rm disc}$ and $J_{\rm BH}$, and the second term represents the additional alignment torque between the BH and the disc, as found in both simulations and analytic models \citep[][]{Volonteri_et_al_2005, Liska_et_al_2018}. Here, $t_{\rm align}$ is the alignment time-scale, defined as 

\begin{equation}
    t_{\rm align} = \frac{1}{2\pi} \frac{M_{\rm BH}}{\dot M_{\rm BH}} = \frac {4.47}{f_{\rm Edd,16}} \rm ~ Myr~,
\end{equation}

\noindent and $t_{\rm prec}$ is the precession time-scale for the BH:

\begin{equation}
    t_{\rm prec} = \frac{1}{\omega_{\rm prec} } \frac {J_{\rm BH}}{J_{\rm disc, trap}}~,
    \label{eq:t_prec}
\end{equation}

\noindent where $J_{\rm disc, trap}$ is the angular momentum of the disc within the photon-trapping region.

Figure~\ref{fig:precession_alignment_time} presents $t_{\rm prec}$, $t_{\rm acc}$, and $t_{\rm gm}$ as functions of $\feddsixteen$, for a few values of $\mbh$ and $a_{\rm BH}$. It is evident that these time-scales show minimal dependence on the BH mass, while their dependence on the BH spin is more pronounced. For the Lense-Thirring effect at low mass accretion rates (i.e. $\feddsixteen < \hat f_{\rm Edd, 16}$), both the alignment and precession time-scales are of the order of $t_{\rm gm}$ (Equation~\ref{eq:LT_lowedd}). However, for $\feddsixteen > \hat f_{\rm Edd, 16}$, the precession time-scale for both prograde and retrograde discs becomes much shorter than the alignment time-scale. It is also evident that $t_{\rm prec}$ is comparable to $t_{\rm gm}$, while $t_{\rm align} \gg t_{\rm gm}$ at 
\WB{$\feddsixteen > \hat f_{\rm Edd, 16}$}. This can be quantified by computing the ratio ${t_{\rm{gm}}}/{t_{\rm align}}$:

\begin{equation}
    \frac{t_{\rm{gm}}}{t_{\rm align}} = 2.1 \times 10^{-2} \, \xi^{-5/7} \,\alpha_{0.1}^{58/35} \, M_{\text{BH},6}^{-2/35} \, a_{\rm BH}^{5/7} \, \WB{f_{\rm Edd,16}^{3/35}}~.
    \label{eq:t_gm_over_t_align}
\end{equation}

\noindent This result demonstrates that $t_{\rm{align}}$ is one to two orders of magnitude larger than $t_{\rm gm}$, indicating that, at high mass accretion rates, the alignment process between the BH and the disc occurs much more slowly.\footnote{If $\hat f_{\rm Edd, 16} \gg 10^4$ or $M_{\rm BH, 6} \ll 10^{-6}$, then $t_{\rm gm}$ would become comparable to $t_{\rm align}$ \WB{at $\feddsixteen = \hat f_{\rm Edd, 16}$}. However, this lies well outside the parameter space considered in this study.}

\subsection{Self-gravitating mass}\label{sec:self_gravitating}

In this section, we discuss how we calculate $R_{\rm sg}$ and $M_{\rm sg}$, as well as the motivation for imposing the constraint $\rdisc < R_{\rm sg}$.

In each region of the thin $\alpha$-disc, we utilise the sound speed equation, derived from $c_{\rm s} = \sqrt{P/\rho}$ in \citetalias{Kato_et_al_2008}, to calculate $R_{\rm sg}$. We define $R_{\rm sg, a}$, $R_{\rm sg, b}$, and $R_{\rm sg, c}$ as the self-gravitating radii for regions~(a), (b), and (c), respectively, assuming that the entire disc is described by a single region: 

\begin{align}
     \frac{R_{\rm sg, a}}{\rg} &= 3.45 \times 10^3 \, \alpha_{0.1}^{2/9} \, M_{\rm BH,6}^{-2/9} \, \feddsixteen^{4/9} \, Q_{\rm min}^{2/9}~, \label{eq:r_sg_a}\\
     \frac{R_{\rm sg, b}}{\rg} &= 7.66 \times 10^4 \, \alpha_{0.1}^{14/27} \, M_{\rm BH,6}^{-26/27} \, \feddsixteen^{-8/27} \, Q_{\rm min}^{-20/27}~, \label{eq:r_sg_b}\\
     \frac{R_{\rm sg, c}}{\rg} &= 9.54 \times 10^4 \, \alpha_{0.1}^{28/45} \, M_{\rm BH,6}^{-52/45} \, \feddsixteen^{-22/45} \, Q_{\rm min}^{-8/9}~. \label{eq:r_sg_c}
\end{align}

We define the self-gravitating radius for our disc model as follows:

\begin{equation}
    R_{\rm sg} = 
    \begin{cases}
        R_{\rm sg, a} &\text{if } \max (R_{\rm ISCO}, R_{\rm trap}) \leq R_{\rm sg, a} < R_{\rm ab}~,\\
        R_{\rm sg, b} &\text{if } R_{\rm ab} \leq R_{\rm sg, b} < R_{\rm bc} ~,\\
        R_{\rm sg, c} &\text{if } R_{\rm bc} \leq R_{\rm sg, c} ~.
        \label{eq:r_sg_abc}
    \end{cases}
\end{equation}  

Using $R_{\rm sg}$, we compute $M_{\rm sg}$ by setting $\rdisc = R_{\rm sg}$ in the surface density integration from Equation~\eqref{eq:M_disc_integral}. 

If the accretion disc becomes self-gravitating (i.e. $M_{\rm disc} > M_{\rm sg}$), gravitational instabilities would develop in its outer regions. These instabilities could lead to disc fragmentation and potentially trigger star formation \citep[][]{Shlosman_and_Begelman_1987, Deng_et_al_2017, Chen_et_al_2023, Derdzinski_and_Mayer_2023} or generate non-axisymmetric structures such as spiral arms and clumps \citep[][]{Durisen_et_al_2007}. The exact details of these processes remain uncertain. However, gravitational instabilities would significantly alter the disc structure, as the transport of mass and angular momentum can no longer be adequately described by the thin $\alpha$-disc model and may lead to disc truncation \citep[][]{Goodman_2003, Sirko_and_Goodman_2003, Thompson_et_al_2005, Rafikov_2015}. 

Following \citet{Perego_et_al_2009}, \citet{Dubois_et_al_2014}, \citet{Fiacconi_et_al_2018}, and \citetalias{Cenci_et_al_2021}, we avoid these complexities by assuming that the disc 
\WB{size is limited by $R_{\rm sg}$}, ensuring that the disc does not enter a self-gravitating state. This imposes a constraint on $\dot{M}_{\rm in}$ such that $M_{\rm disc} \leq M_{\rm sg}$ (in the case that $M_{\rm disc}$ is initially larger than $M_{\rm sg}$, then $\dot{M}_{\rm in}$ is set to zero until $M_{\rm disc}$ decreases, due to accretion onto the BH, down to the value of $M_{\rm sg}$). By preventing gravitational instabilities from developing in the outer disc regions, this approach maintains the validity of the thin $\alpha$-disc framework. The limitations of this assumption are discussed in Section~\ref{sec:caveat_accretion_model}. \WB{At variance with previous studies (e.g. \citealt{Fiacconi_et_al_2018}; \citetalias{Cenci_et_al_2021}; \citealt{Koudmani_et_al_2024}), when $\mdisc = M_{\rm sg}$, we additionally require that $\jdisc \leq J_{\rm sg}$,\footnote{\WB{We allow $\jdisc < J_{\rm sg}$, since a misaligned gas inflow onto the accretion disc can decrease the disc angular momentum and thus lead to a lower $\jdisc$. See Figure~\ref{fig:result_misalign_disc_gas} for an example.}} where $J_{\rm sg}$ represents the disc angular momentum corresponding to $\rdisc = R_{\rm sg}$. We impose this additional condition to ensure that the disc remains constrained by the self-gravity limit and that the inflow of angular momentum does not violate this constraint.} 

\subsection{Radiative efficiency}\label{sec:radiative_efficiency}

Due to the strong photon-trapping effect in super-Eddington flows, photons cannot freely escape the accretion disc in the innermost region. Consequently, the radiative efficiency is reduced compared to the efficiency determined solely using the Kerr metric as in \citet{Bardeen_et_al_1972}. To smoothly transition between the sub-Eddington and super-Eddington regimes, we adopt the results from \citet{Madau_et_al_2014}, 
which are derived from fitting numerical solutions of the relativistic slim accretion disc equations presented in \citet{Sadowski_et_al_2009}: 

\begin{equation}
    \frac{\eta}{1/16} = A(a_{\rm BH}) \left[\frac{0.985}{1 + B(a_{\rm BH}) f_{\rm Edd, 16}} + \frac{0.015}{1 + C(a_{\rm BH}) f_{\rm Edd, 16}} \right]~,
    \label{eq:radiative_efficiency}
\end{equation}

\noindent where the coefficients $A$, $B$, and $C$ are given by, respectively,

\begin{align}
    A(a_{\rm BH}) &= (0.9663 \mp 0.9292 \, a_{\rm BH})^{-0.5639}~, \label{eq:radiative_efficiency_A} \\
    B(a_{\rm BH}) &= (4.627 \mp 4.445 \, a_{\rm BH})^{-0.5524}~, \label{eq:radiative_efficiency_B} \\
    C(a_{\rm BH}) &= (827.3 \mp 718.1 \, a_{\rm BH})^{-0.7060}~, \label{eq:radiative_efficiency_C}  
\end{align}

\noindent where the minus and plus signs correspond to a prograde and retrograde disc, respectively. 

We note that \citet{Madau_et_al_2014} originally calibrated this fitting function for prograde-only accretion flows (i.e. Equations~\ref{eq:radiative_efficiency_A}--\ref{eq:radiative_efficiency_C} only had the minus sign), as the results from \citet{Sadowski_et_al_2009} were specified only for those cases. We extended the fitting function to cover also retrograde accretion flows and verified that this extension provides reasonable results: Figure~\ref{fig:eta} illustrates the value of $\eta$ as a function of $a_{\rm BH}$ for different values of $\feddsixteen$.

We note with caution that the results of \citet{Madau_et_al_2014} are derived from numerical solutions of the relativistic slim accretion disc in \citet{Sadowski_et_al_2009}, which exhibit a different disc structure compared to \citetalias{Kitaki_et_al_2018}. It is not straightforward to properly compare the two models. However, we can partially validate Equations~\eqref{eq:radiative_efficiency}--\eqref{eq:radiative_efficiency_C} using simulation results from \citet{Kitaki_et_al_2021}, which are in close agreement with those of \citetalias{Kitaki_et_al_2018}. Their simulations model a non-spinning BH accreting at a rate of $\dot M_{\rm BH, accr} \sim 180 \WB{\mathcal L_{\rm Edd}}/c^2$ (i.e. $\feddsixteen \sim 11.25$). The (bolometric) luminosity observed by a distant observer is $\WB{\mathcal L} \sim 2.5 \WB{\mathcal L_{\rm Edd}}$, implying a radiative efficiency $\eta = \WB{\mathcal L}/(\dot M_{\rm BH,accr} c^2) \sim 0.0139$, which is consistent with $\eta \sim 0.012$ predicted by Equations~\eqref{eq:radiative_efficiency}--\eqref{eq:radiative_efficiency_C} when using $a_{\rm BH} = 0$ and $f_{\rm Edd,16} = 11.25$. These findings suggest that the radiative efficiency adopted in our model remains in reasonable agreement with this expression, even if the underlying disc structure differs.\footnote{We note that \citet{Sadowski_2011} quotes a slightly higher value for the radiative efficiency, of the order of 0.025 \citep[see also][]{Watarai_et_al_2000,Kubota_Done_2019}. However, the proper comparison here is with the results of \citet{Sadowski_et_al_2009}, since those were used for the improved fitting functions by \citet{Madau_et_al_2014}.} 

\subsection{\WB{Maximum value of the black hole spin parameter}}\label{sec:max_a_BH}

\WB{During accretion, the well-known theoretical upper limit of $a_{\rm BH}$ is 0.998, due to the torque exerted by retrograde photons \citep[][]{Thorne_1974}. If we further consider the change of $a_{\rm BH}$ during prograde accretion, another upper limit arises in the case of high accretion rates. At time $t$, the BH angular momentum's magnitude is $J_{\rm BH} = \left({GM_{\rm BH}^2}/{c}\right) a_{\rm BH}$. At time $t+\Delta t$, it becomes $J_{\rm BH}' = \left({GM_{\rm BH}'^2}/{c}\right) a'_{\rm BH}$, where $a'_{\rm BH}$ and $M_{\rm BH}'$ denote the BH spin parameter and mass at $t + \Delta t$, respectively. Here, we assume $\Delta t$ is infinitesimal and neglect higher-order effects.}

\WB{Considering mass accretion onto the BH (Equation~\ref{eq:M_BH_growth}), we derive $M_{\rm BH}' = M_{\rm BH}+(1-\eta)\Delta M_{\rm BH}$, where $\Delta \mbh = \dot M_{\rm BH, accr} \, \Delta t$ is the total mass accreted during this time interval. The corresponding change in angular momentum is given by $J_{\rm BH}' = J_{\rm BH} + \Delta M_{\rm BH} \, \Lambda_{\rm ISCO}$ (Equation~\ref{eq:J_BH_accretion}). Expanding to first order, the variation in the BH spin parameter, $\Delta a_{\rm BH} = a'_{\rm BH} - a_{\rm BH}$, can be expressed as}

\begin{equation}
    \begin{split}
        \frac{\Delta a_{\rm BH}}{a_{\rm BH}} &\approx \frac{J_{\rm BH}' - J_{\rm BH}}{J_{\rm BH}} - 2\frac{M_{\rm BH}' - \mbh}{M_{\rm BH}} \\
        &= \left(\frac{\Lambda_{\rm ISCO}}{a_{\rm BH}} \left(\frac{c}{GM_{\rm BH}}\right) - 2 + 2\eta \right) \, \frac{\Delta M_{\rm BH}}{M_{\rm BH}} ~.
    \end{split}
\end{equation}

\WB{Following \citet{Shapiro_2005, Massonneau_et_al_2023_spin}, we define the dimensionless spin-up parameter as}

\begin{equation}
    \begin{split}
        s &\equiv \frac{{\rm d}a_{\rm BH}}{{\rm d}t} \frac{M_{\rm BH}}{\dot M_{\rm BH, accr}} = \frac{\Delta a_{\rm BH} \, M_{\rm BH}}{\Delta M_{\rm BH}} \\
        &= \Lambda_{\rm ISCO} \left(\frac{c}{GM_{\rm BH}}\right) + \left(2\eta - 2 \right)a_{\rm BH} ~.
        \label{eq:spin_up_parameter}
    \end{split}
\end{equation}

\noindent \WB{This expression is equivalent to equation~(14) in \citet{Shapiro_2005} for the thin-disc model. When the radiative efficiency is determined by the Kerr metric \citep[][]{Bardeen_et_al_1972}, $s$ remains positive for $a_{\rm BH} < 0.998$ (see, e.g. \citealt{Shapiro_2005, Dubois_et_al_2021, Massonneau_et_al_2023}).}

\WB{In contrast, the reduced radiative efficiency caused by photon trapping in the super-Eddington accretion regime (Equation~\ref{eq:radiative_efficiency}) changes the value of $s$ to negative values for high mass accretion rates and rapidly spinning BHs. For instance, when $\feddsixteen = 1$, $s < 0$ if $a_{\rm BH} > 0.973$.}

\begin{figure}
    \centering
    \includegraphics[width=0.9\linewidth]{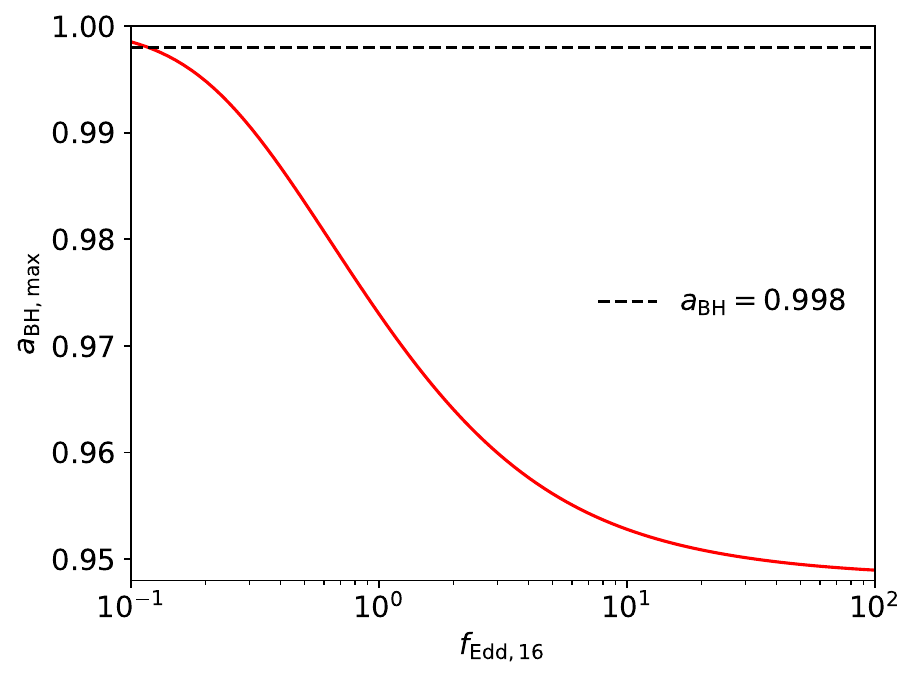}
    \caption{\WB{$a_{\rm BH, max}$ as a function of $\feddsixteen$. $a_{\rm BH, max}$ is defined as the value at which the dimensionless spin-up parameter $s$ (Equation~\ref{eq:spin_up_parameter}) equals zero. Its value is smaller than 0.998 when $\feddsixteen \gtrsim 0.1$ (i.e. $f_{\rm Edd,1} \gtrsim 1.6$) because the reduced radiative efficiency at higher mass accretion rates, caused by the photon trapping effect, decreases the spin-up parameter and lowers the maximum value of BH spin parameter.}}
    \label{fig:a_BH_max}
\end{figure}

\WB{We therefore define the maximum spin parameter $a_{\rm BH, max}$ as the value satisfying $s(a_{\rm BH,max})=0$. For $a_{\rm BH} > a_{\rm BH,max}$, $s$ crosses zero and becomes negative due to the reduced $\Lambda_{\rm ISCO}$ (Equation~\ref{eq:lambda_isco}), which halts the increase of $a_{\rm BH}$ through accretion. The dependence of $a_{\rm BH,max}$ on $\feddsixteen$ is shown in Figure~\ref{fig:a_BH_max}. We find that $a_{\rm BH,max} < 0.998$ when $\feddsixteen \gtrsim 0.1$ \WB{(i.e. $f_{\rm Edd,1} \gtrsim 1.6$)}, and that it asymptotically approaches $a_{\rm BH,max} \sim 0.95$ for $\feddsixteen \gg 1$.}

\WB{Furthermore, the Lense-Thirring torque does not change the magnitude of $a_{\rm BH}$ because it acts perpendicular to the BH spin direction. Consequently, the only term that influences the value of $a_{\rm BH}$ is accretion, and thus $a_{\rm BH}$ cannot exceed $a_{\rm BH,max}$. We cap $a_{\rm BH}$ in this model to prevent numerical errors, using the following equation:}\footnote{\WB{We note that jets powered by the Blandford-Znajek mechanism \citep[][]{Blandford_and_Znajek_1977} could rapidly spin down the BH through jet launching \citep[e.g.][]{McKinney_et_al_2012, Narayan_et_al_2022, Ricarte_et_al_2023}. The derivation in this section would no longer be valid if this effect were taken into consideration (i.e. if ${\bm{\dot J}}_{\rm BH, jet} \neq 0$ in Equation~\ref{eq:update_jdisc}).}}

\begin{equation}
    a_{\rm BH} \leq \min(0.998,  a_{\rm BH,max})~.
\end{equation}


\section{Numerical setup}\label{sec:numerical_setup}

This sub-grid model, as detailed in Section~\ref{sec:super_edd_model}, is built upon the model and code in \citetalias{Cenci_et_al_2021} and has been integrated into the publicly available $N$-body, mesh-less hydrodynamics code \textsc{gizmo}\footnote{\href{http://www.tapir.caltech.edu/~phopkins/Site/GIZMO.html}{http://www.tapir.caltech.edu/\textasciitilde
phopkins/Site/GIZMO.html}} \citep[][]{Hopkins_2015}. The numerical setup follows the methodology outlined in \citetalias{Cenci_et_al_2021}. We provide here a brief summary for completeness. 

The initial conditions consist of a single BH particle, which represents a sub-resolution ``BH+accretion disc'' system, embedded within a spherically distributed stellar structure and surrounded by a gaseous circumnuclear disc (CND).

The spherical stellar component is modelled using the \citet{Hernquist_1990} profile as a function of (spherical) radius $r$:

\begin{equation}
    \rho_{\rm b}(r) = \frac{M_{\rm b}}{2\pi} \frac{r_{\rm b}}{r (r + r_{\rm b})^3}~,
\end{equation}

\noindent where the total mass of the bulge is $M_{\rm b} = 5 \times 10^8 \msun$ and its scale radius is $r_{\rm b} = 100~\text{pc}$.

The CND is a rotationally supported disc in vertical hydrostatic equilibrium. Its surface density profile is given by the exponential disc as follows:

\begin{equation}
    \Sigma_{\text{CND}} (R) = \frac{M_{\text{CND}}}{2\pi R^2_{\text{CND}}} \exp\left(-\frac{R}{R_{\text{CND}}}\right)~,
    \label{eq:CND_surface_density}
\end{equation}

\noindent where the total mass of the disc is $M_{\text{CND}} = 10^8 \msun$ and its scale radius is $R_{\text{CND}} = 50$~pc.

\begin{table*}
    \centering
    \caption{Summary of simulations with different parameters. From left to right, we list the name of each simulation, the initial BH mass, $M_{\rm BH, 0}$, the initial Eddington ratio, $f_{\rm Edd, 16, 0}$, the circularisation radius ratio of the inflow gas, $W_{\rm circ}$, the initial BH spin magnitude, $a_{\rm BH, 0}$, the initial angle between the angular momentum of the BH and that of the disc, $\theta_{\rm BH-disc, 0}$, the minimum Toomre parameter, $Q_{\rm min}$, and the initial angle between the angular momentum of the gas and that of the disc, 
    \WB{$\theta_{\rm gas-disc, 0}$}. \WB{In all runs, we set $M_{\rm disc, 0} = M_{\rm sg}$ and compute $J_{\rm disc, 0}$ accordingly.} The suffixes VL, L, H, and VH denote runs wherein a specific parameter is set to very-low, low, high, and very-high values, respectively.
    \WB{All but the last four} rows show the runs with initial gas-disc alignment (illustrated by the top diagram), whereas the bottom 
    \WB{four} rows show the runs with initial BH-disc alignment (bottom diagram).
    }
    \begin{tabular}{lcccccccc}
        \hline 
        \rule{0pt}{2.5ex}{Simulation name} & $M_{\rm BH, 0}/\rm M_{\sun}$ & $f_{\rm Edd, 16, 0}$ & $W_{\rm circ}$ & $a_{\rm BH, 0}$ & $\theta_{\rm BH-disc,0}$ & $Q_{\rm min}$ & $\theta_{\rm gas-disc, 0}$ & \T \B \\
        \hline 
        \rule{0pt}{3ex}Fiducial & $10^6$ & $1$ & $0.1$ & 0.8 & $5\pi/6$ & 1 & 0 \\
        \rule{0pt}{3ex}Edd-H   & $10^6$ & \textbf{5} & $0.1$ & 0.8 & $5\pi/6$ & 1 & 0 & \multirow{3}{*}{\parbox[c]{14em}{\includegraphics[width=1.6in]{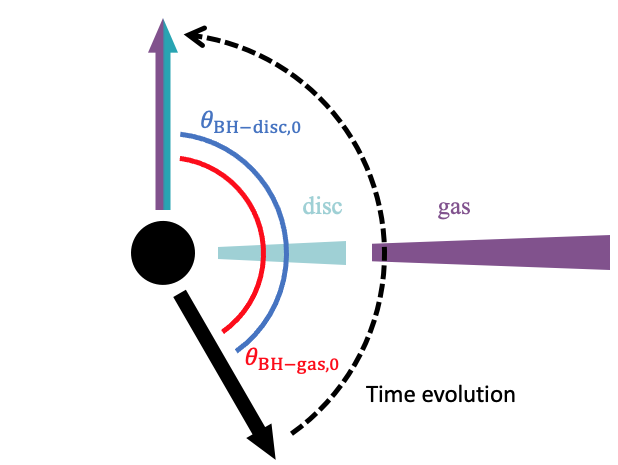}}}\\
        Edd-L   & $10^6$ & \textbf{0.6} & $0.1$ & 0.8 & $5\pi/6$ & 1 & 0\\
        Edd-VL   & $10^6$ & \textbf{0.1} & $0.1$ & 0.8 & $5\pi/6$ & 1 & 0\\
        \rule{0pt}{3ex}Wcirc-VH & $10^6$ & $1$ & \textbf{0.9} & 0.8 & $5\pi/6$ & 1 & 0\\
        Wcirc-H & $10^6$ & $1$ & \textbf{0.5} & 0.8 & $5\pi/6$ & 1 & 0\\
        Wcirc-L & $10^6$ & $1$ & \textbf{0.05} & 0.8 & $5\pi/6$ & 1 & 0\\ 
        Wcirc-VL & $10^6$ & $1$ & \textbf{0.01} & 0.8 & $5\pi/6$ & 1 & 0\\ 
        \rule{0pt}{3ex}aBH-H & $10^6$ & $1$ & $0.1$ & \textbf{\WB{0.95}} & $5\pi/6$ & 1 & 0\\
        aBH-L & $10^6$ & $1$ & $0.1$ & \textbf{0.5} & $5\pi/6$ & 1 & 0\\
        aBH-VL & $10^6$ & $1$ & $0.1$ & \textbf{0.2} & $5\pi/6$ & 1 & 0\\
        \cline{9-9}
        \rule{0pt}{3ex}theta-L & $10^6$ & $1$ & $0.1$ & 0.8 & $\bm{2\pi/3}$ & 1 & \multicolumn{1}{c!{\vrule width 0.5pt}}{0} \\
        theta-VL & $10^6$ & $1$ & $0.1$ & 0.8 & $\bm{\pi/2}$ & 1 & \multicolumn{1}{c!{\vrule width 0.5pt}}{0} & \multirow{3}{*}{\parbox[c]{14em}{\includegraphics[width=1.6in]{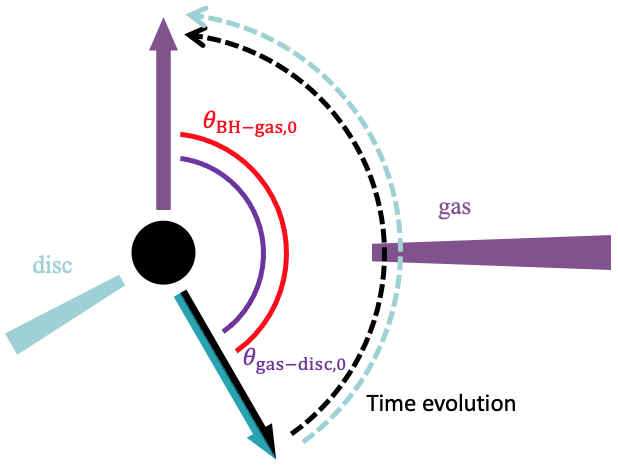}}}\\
        \rule{0pt}{3ex}Q-VH & $10^6$ & $1$ & $0.1$ & 0.8 & $5\pi/6$ & \textbf{4} & \multicolumn{1}{c!{\vrule width 0.5pt}}{0}\\ 
        Q-H & $10^6$ & $1$ & $0.1$ & 0.8 & $5\pi/6$ & \textbf{2} & \multicolumn{1}{c!{\vrule width 0.5pt}}{0}\\
        Q-L & $10^6$ & $1$ & $0.1$ & 0.8 & $5\pi/6$ & \textbf{0.5} & \multicolumn{1}{c!{\vrule width 0.5pt}}{0}\\
        \rule{0pt}{3ex}MBH-H    & $\bm{10^7}$ & $1$ & $0.1$ & 0.8 & $5\pi/6$ & 1 & \multicolumn{1}{c!{\vrule width 0.5pt}}{0}\\
        MBH-L    & $\bm{10^5}$ & $1$ & $0.1$ & 0.8 & $5\pi/6$ & 1 & \multicolumn{1}{c!{\vrule width 0.5pt}}{0} \\ 
        \rule{0pt}{3ex}\WB{Counter-align} & \WB{$\bm{10^7}$} & \WB{\bm{$0.1$}} & \WB{$0.1$} & \WB{0.8} & \WB{$\bm{\pi}$} & \WB{1} & \multicolumn{1}{c!{\vrule width 0.5pt}}{\WB{0}}\B \\ 
        \cline{1-8}
        \rule{0pt}{3ex}GD-Misalign-VH & $10^6$ & $1$ & $0.1$ & 0.8 & $0$ & 1 & $\bm{5\pi/6}$\\
        GD-Misalign-H & $10^6$ & $1$ & $0.1$ & 0.8 & $0$ & 1 & $\bm{3\pi/4}$\\
        GD-Misalign-L & $10^6$ & $1$ & $0.1$ & 0.8 & $0$ & 1 & \WB{$\bm{7\pi/12}$} \\
        \rule{0pt}{3ex}\WB{GD-Misalign-05} & \WB{$10^6$} & \WB{$\bm{0.5}$} & \WB{$0.1$} & \WB{0.8} & \WB{$0$} & \WB{1} & \WB{$\bm{5\pi/6}$}\B\\
        \hline
    \end{tabular}
    \label{tab:simulation_runs}
\end{table*}

The vertical density profile and velocity field are computed using the publicly available code \textsc{gd\_basic}\footnote{\href{https://alessandrolupiastro.wordpress.com/software/}{https://alessandrolupiastro.wordpress.com/software/}} to ensure vertical hydrostatic equilibrium under the combined potential of the bulge, CND, and BH \citep{Lupi_et_al_2015}. The gas is initially set up assuming an ideal-gas equation of state with a uniform temperature of $T_0 = 2 \times 10^4$~K and a polytropic index of $\gamma = 5/3$. The system is first evolved in isolation for 20~Myr, with only gravity and hydrodynamics active, to allow for relaxation. Subsequently, during the simulations, a lower polytropic index of $\gamma = 7/5$ is adopted to mimic mild cooling and drive gas accretion onto BH without using a dedicated cooling model (\citealt{Dotti_et_al_2009}; \citetalias{Cenci_et_al_2021}; \citealt{Sala_et_al_2021}).

Both the number of stellar particles, $N_{\rm stars}$, and the number of gas particles, $N_{\rm gas}$, are set to $10^5$ ($N_{\rm stars} = N_{\rm gas} = 10^5$). For the stellar particles, the Plummer-equivalent gravitational softening length is set to $\epsilon_{\rm stars} = 0.16$~pc, whereas for the BH particle, it is $\epsilon_{\rm BH} = 1$~pc. For the gas particles, a fully adaptive softening scheme is employed. Their resolution is defined by the size of the smoothing kernel, which is chosen to include an effective number of neighbours, $N_{\text{ngb}} = 32$. The minimum gravitational softening length for the gas, which sets the maximum spatial resolution, is set to $\epsilon_{\rm gas} = 0.16$~pc. 

The mass transfer rate from resolved scales onto the accretion disc of the BH particle, $\dot{M}_{\text{in}}$, can be determined using different prescriptions. 
\WB{In contrast to \citet{Fiacconi_et_al_2018}, who directly compute the mass flux onto the BH particle using the surrounding mesh, for the tests in this work, we set the inflow rate equal to the BHL formula (but we stress that any recipe can be chosen in its place):}

\begin{equation}
    \dot{M}_{\text{in}} = \frac{4 \pi \alpha_{\text{acc}} G^2 M_{\text{BH}}^2 \rho_{\rm gas}}{(c_{\rm s, gas}^2 + v^2)^{3/2}}~,
    \label{eq:Bondi_accretion}
\end{equation}

\noindent where $\rho_{\rm gas}$ is the density of the surrounding gas, $v$ represents the relative velocity between the gas and the BH, $c_{\rm s, gas}$ is the sound speed of the surrounding gas, and $\alpha_{\rm acc}$ is a dimensionless parameter to boost the mass accretion rate for unresolved dense gas in kpc-scale simulations \citep[e.g.][]{Springel_et_al_2005,Booth_and_Schaye_2009}. At high-enough resolution, as in isolated simulations of galaxies \citep[e.g.][]{Tamburello_et_al_2017} and their central regions \citep[e.g. CNDs;][]{SouzaLima_et_al_2017,SouzaLima_et_al_2020}, this boost parameter is not needed (but see \citealt{Negri_and_Volonteri_2017} for a discussion): therefore, we set $\alpha_{\rm acc} = 1$. \WB{An underlying assumption of the BHL prescription is that the inflowing gas has negligible angular momentum at the \citet{Bondi_1952} radius, $R_{\rm B} = 2 G M_{\rm BH}/c_{\rm s,gas}^2$. We verified that the angular momentum of the inflowing gas is much smaller than the Keplerian angular momentum at $R_{\rm B}$. Nevertheless, we note that alternative prescriptions for $\dot M_{\rm in}$ are also valid \citep[e.g.][]{Bleuler_and_Teyssier_2014, Angles-Alcazar_et_al_2015}. Furthermore, here $\mbh$ should be replaced by $\mbh + \mdisc$, since the disc mass also contributes to the gravitational force on the surrounding gas, but we ignore its contribution since $\mdisc \ll \mbh$.}

\begin{figure*}
    \centering
    \includegraphics[width=1.0\linewidth]{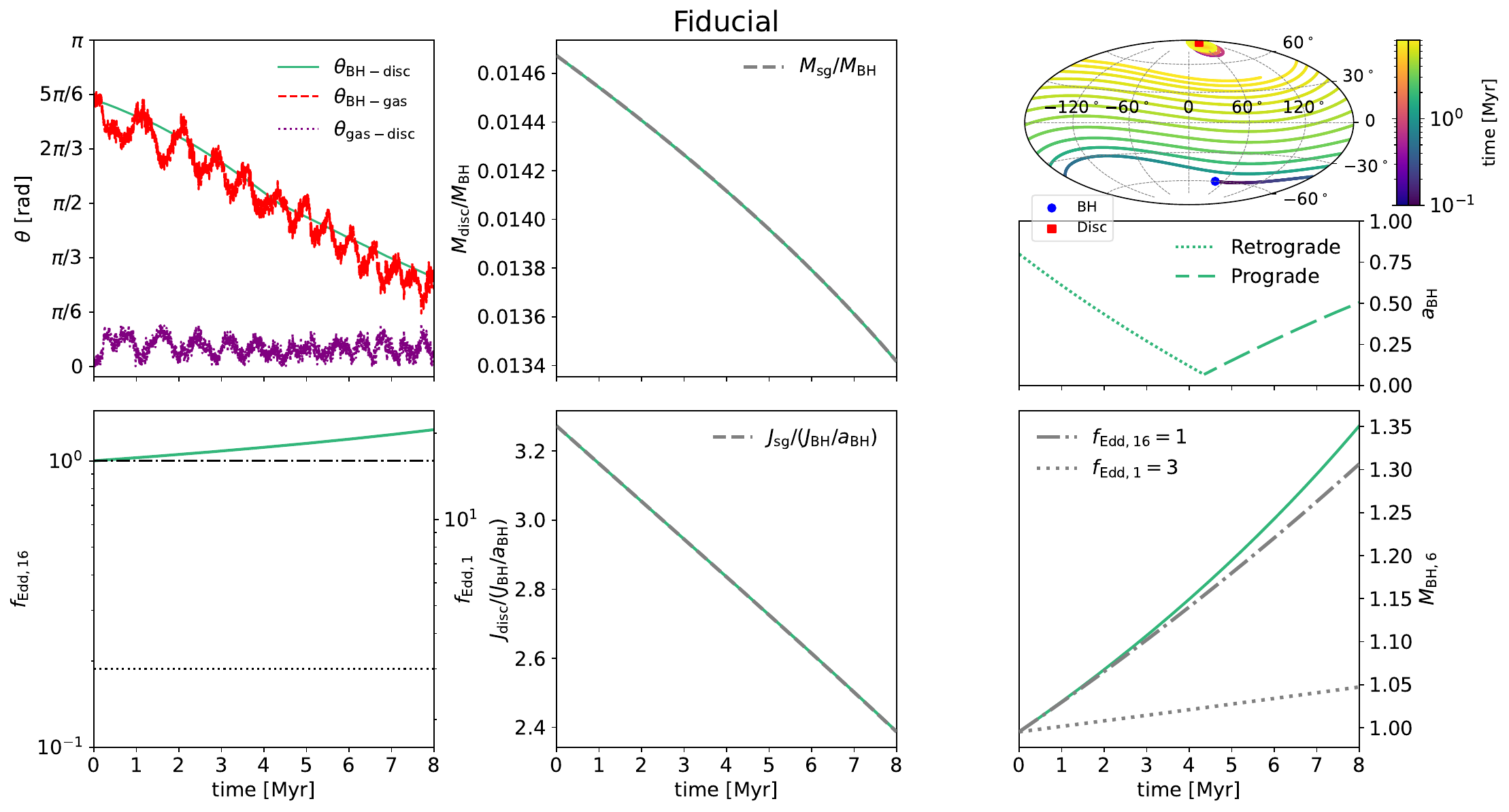}
    \caption{\WB{Time evolution of key quantities for the Fiducial run. \textit{Top-left panel}: misalignment angles $\theta_{\rm BH-disc}$ (solid green line), $\theta_{\rm BH-gas}$ (dashed red line), and $\theta_{\rm gas-disc}$ (dotted purple line). \textit{Top-central panel}: disc mass, $M_{\rm disc}$, in units of $M_{\rm BH}$. The self-gravitating mass, $M_{\rm sg}$, is also shown as a grey dashed line for comparison. \textit{Top-right panel}: BH (blue circle) and disc (red square) angular momentum unit vectors displayed using Hammer projections, wherein the equatorial plane is defined by the angular momentum of the gas within the BH kernel in the initial conditions (before relaxation). The colour coding indicates time evolution from 0.1 to 8~Myr. The direction of $\bm{j}_{\rm gas}$ is nearly aligned with the $z$-axis and thus not shown. \textit{Middle-right panel}: BH spin parameter, $a_{\rm BH}$. Dotted and dashed lines correspond to a retrograde and prograde disc, respectively. \textit{Bottom-left panel}: Eddington ratio, $\feddsixteen$, with $f_{\rm Edd, 1}$ shown on the right axis for comparison. For reference, the black dash-dotted line marks $\feddsixteen = 1$ and the black dotted line indicates $f_{\rm Edd, 1} = 3$. \textit{Bottom-central panel}: disc angular momentum, $J_{\rm disc}$, in units of $J_{\rm BH}/a_{\rm BH} = G M_{\rm BH}^2/c$ (i.e. the maximum angular momentum of a BH). \textit{Bottom-right panel}: BH mass in units of $10^6 \msun$, $M_{\rm BH,6}$. The black dash-dotted and dotted lines represent reference tracks for constant specific accretion rates of $\feddsixteen = 1$ and $f_{\rm Edd, 1} = 3$, respectively. Because of the sufficient gas inflow onto the accretion disc, the disc mass and angular momentum remain at the self-gravity limit, resulting in a slow increase in $\feddsixteen$. The figure also shows strong precession and slow alignment between the BH and the disc due to the Lense-Thirring torque at high mass accretion rates (the thick-disc precession model). The change in $a_{\rm BH}$ is significant because of the extended period of retrograde accretion.}}
    \label{fig:result_fiducial}
\end{figure*}

When computing $\dot M_{\rm in}$, the gas properties are determined using a mass-weighted average over the nearest $N_{\rm ngb, BH}$ neighbour gas particles around the BH. To ensure a sufficiently large sampling of the gas properties surrounding the BH, we set $N_{\rm ngb, BH} = 3 N_{\rm ngb}$. The BH particle kernel is defined as the region encompassing $N_{\rm ngb, BH}$ neighbours, up to a maximum accretion radius, $r_{\rm max, acc} = 10$~pc. 

There are also many different prescriptions to determine $\bm{\dot{J}}_{\text{in}}$. In this paper, we consider coherent accretion, in which the angular momentum of the inflow is aligned to that of the surrounding gas \citep[][]{Volonteri_et_al_2007}. Consequently, we use the relation $\bm{\dot{J}}_{\text{in}} = \Lambda_{\rm in} {\dot{M}}_{\text{in}} \, \bm j_{\rm gas}$, where $\Lambda_{\rm in}$ is the specific angular momentum of the inflowing material \citepalias[as in][]{Cenci_et_al_2021}, and $\bm j_{\rm gas}$ is the unit vector of the total angular momentum of the gas in the BH kernel, $\bm J_{\rm gas}$ (of magnitude $J_{\rm gas}$). In the simulation, we define the $z$-axis as the direction of the initial $\bm j_{\rm gas}$ (i.e. the initial CND lies in the $x$-$y$ plane before relaxation).

Since angular momentum transport is not fully resolved for the inflowing gas, we assume that it circularises at the circularisation radius, $R_{\rm circ}$, which reduces the angular momentum influx and prevents the formation of a self-gravitating disc with excessive angular momentum. We further define the ratio of the circularisation radius to the self-gravitating radius as $W_{\rm circ} = R_{\rm circ} / R_{\rm sg}$. The value of $W_{\rm circ}$ can be specified in the code. We impose an upper limit on ${\Lambda}_{\text{in}}$ by comparing two possible values:

\begin{equation}
    {\Lambda}_{\text{in}} = \min \left( \frac{J_{\rm disc}(R_{\rm circ})}{M_{\rm disc}(R_{\rm circ})} , \, \frac{J_{\rm{gas}}}{M_{\rm{gas}}} \right) ~,
\end{equation}

\noindent where $M_{\rm{gas}}$ is the total mass of the gas particles in the BH particle kernel.

For the properties of the BH and disc, we perform simulations by initialising the BH mass, spin, and Eddington ratio as follows: $\mbh = M_{\rm BH, 0}$, $a_{\rm BH} = a_{\rm BH, 0}$, and $f_{\rm Edd, 16} = f_{\rm Edd, 16, 0}$. To prevent an initial rapid increase in $M_{\rm disc}$ due to a large inflow onto the accretion disc (as was the case in \citetalias{Cenci_et_al_2021}, wherein the disc was initially less massive than its own self-gravitating mass), we set the initial disc mass to $M_{\rm disc,0} = M_{\rm sg}$. The initial disc angular momentum, $J_{\rm disc, 0}$, is determined based on other initial given values.

To describe the initial angle between the BH, disc, and surrounding gas, we introduce the following angles: $\theta_{\rm BH-disc}\equiv \arccos ( \bm{j}_{\rm BH} \cdot \bm{j}_{\rm disc})$, $\theta_{\rm gas-disc} \equiv \arccos ( \bm{j}_{\rm gas} \cdot \bm{j}_{\rm disc})$, and $\theta_{\rm BH - gas} \equiv \arccos ( \bm{j}_{\rm BH} \cdot \bm{j}_{\rm gas})$. We remind the reader that $\bm{j}_{\rm disc}$ refers to the outer part of the accretion disc. We initialise $\theta_{\rm BH-disc} = \theta_{\rm BH-disc,0}$ and $\theta_{\rm gas-disc} = \theta_{\rm gas-disc, 0}$. Two types of initial misalignment configurations are considered to explore different scenarios of BH-disc-gas alignment:

\begin{enumerate}

\item Gas-disc alignment: we set $\theta_{\rm gas-disc, 0} = 0$, meaning that $\bm J_{\rm gas}$ is parallel to $\bm J_{\rm{disc}}$,  while the BH is misaligned with the disc ($\theta_{\rm BH-disc, 0} \neq 0$). In this case, $\theta_{\rm BH-gas, 0} = \theta_{\rm BH-disc, 0}$.
    
\item BH-disc alignment: we set $\theta_{\rm BH-disc, 0} = 0$, meaning that $\bm J_{\rm BH}$ is parallel to $\bm J_{\rm{disc}}$, while both are misaligned with the angular momentum of the surrounding gas ($\theta_{\rm gas-disc, 0} \neq 0$). In this case, $\theta_{\rm BH-gas, 0} = \theta_{\rm gas-disc, 0}$.
    
\end{enumerate}

These two types of initial conditions represent two different scenarios wherein a new stream of gas with a different angular momentum direction flows into the centre of the galaxy. If the BH initially lacks a disc or hosts only a small accretion disc with negligible mass and angular momentum, the incoming gas stream is expected to establish a configuration similar to the gas-disc alignment case. Conversely, if the BH already possesses a large accretion disc aligned with its own spin, the initial condition would resemble the BH-disc alignment scenario.

$Q_{\rm min}$ and $W_{\rm circ}$ are constants that can be set in the code. The choices of parameters for the simulation runs are summarised in Table~\ref{tab:simulation_runs}. The fixed parameters used in all simulations are $\alpha_{0.1} = 1$, $\xi = 0.7$, and $\dot M_{\rm out} = 0$.


\section{Results}\label{sec:results}

In this section, we analyse our simulation results using the new sub-grid model presented in this paper. 
We first discuss the results of the Fiducial run in Section~\ref{sec:fiducial_run}. In Section~\ref{sec:result_gas_disc_alignment}, we explore variations in model parameters for the initial gas-disc alignment configuration. Section~\ref{sec:result_BH_disc_alignment} presents the results for the 
\WB{simulations with a BH-disc} alignment initial condition, which can lead to higher mass accretion rates.

\subsection{The Fiducial run}\label{sec:fiducial_run}

\WB{Figure~\ref{fig:result_fiducial} shows the evolution of $\theta_{\rm BH-disc}$, $\theta_{\rm BH-gas}$, $\theta_{\rm gas-disc}$, $\feddsixteen$, $\mdisc$, $\jdisc$, $\bm{j}_{\rm BH}$, $\bm{j}_{\rm disc}$, $a_{\rm BH}$, and $\mbh$ for one of our initial gas-disc alignment simulations, the Fiducial run. In this run, the initial conditions are $M_{\rm BH,0}=10^6\msun$, $f_{\rm Edd,16,0}=1$, $W_{\rm circ}=0.1$, $a_{\rm BH,0}=0.8$, $\theta_{\rm BH-disc,0}=5\pi/6$, $Q_{\rm min}=1$, and $\theta_{\rm gas-disc,0}=0$.}

\WB{Although the BHL accretion rate always exceeds the BH mass accretion rate by at least one order of magnitude, this does not translate into rapid BH growth. This is because the mass inflow rate onto the accretion disc is constrained by the condition $\mdisc \leq M_{\rm sg}$ (see Section~\ref{sec:self_gravitating}), which limits how much of the surrounding gas can be accreted onto the disc at any given time.  As a result, $\mdisc = M_{\rm sg}$ throughout the run due to sufficient inflow from the surrounding gas (see the top-central panel).} 
\WB{Because of the continuous inflow of angular momentum from the surrounding gas disc, $J_{\rm disc}$ also remains equal to $J_{\rm sg}$ throughout the simulation (see the bottom-central panel). The values of $M_{\rm sg}$ and $J_{\rm sg}$ evolve over time because they depend on $\mbh$.}

\WB{In the bottom-left panel, we show the mass accretion rate onto the BH, using the Eddington ratio as a proxy. For an easier comparison with alternative definitions of Eddington ratio, we also show the values of $f_{\rm Edd, 1}$ in the panel. It is important to note that the definition of super-Eddington accretion varies across the literature. Above $\WB{\mathcal L_{\rm BH}}/\WB{\mathcal L_{\rm Edd}} = 0.3$, the thin $\alpha$-disc model ceases to properly describe the inner region of the accretion disc \citep[][]{Koratkar_and_Blaes_1999}; hence, some studies classified observed sources as super-Eddington for $f_{\rm Edd, 1} > 3$ \citep[assuming a typical value of $\eta=0.1$; e.g.][]{Du_et_al_2018, Liu_et_al_2021}, while it is common in theoretical studies to adopt $\feddsixteen = 1$ as the super-Eddington threshold for simplicity (e.g. \citealt{Madau_et_al_2014}). Following these definitions, we include horizontal lines for $\feddsixteen = 1$ \citep[the threshold adopted in, e.g.][]{Sassano_et_al_2023,Lupi_et_al_2024} and $f_{\rm Edd, 1} = 3$ \citep[as done in, e.g.][]{Capelo_et_al_2015,Capelo_et_al_2017} in the panel.} \WB{The evolution of $\feddsixteen$ is set by the evolution of $\mdisc$ and $\jdisc$, which are governed by the self-gravity limit, leading to a gradual increase in $\feddsixteen$.}

\WB{The top-right panel} shows the evolution of the directions of the disc and BH angular momentum vectors, projected onto the full-sky sphere using Hammer projections. Initially, the directions of both $\bm{j}_{\rm disc}$ and $\bm{j}_{\rm gas}$ are approximately aligned with the $z$-axis [i.e. the longitude and latitude are around ($0^\circ$, $90^\circ$)], although not perfectly, due to relaxation effects. The initial longitude and latitude of $\bm{j}_{\rm BH}$ are approximately ($45^\circ$, $-60^\circ$). We fix the initial longitude to $45^\circ$ for all runs in this work, noting that this value does not affect the results due to symmetry. The initial latitude is set by $\theta_{\rm BH-disc, 0}$. Because $\bm{j}_{\rm gas}$ is roughly aligned with the $z$-axis, the initial latitude is therefore around $-60^\circ$.

\WB{The value of $\hat f_{\rm Edd,16}$ varies within the range $\sim$$0.2$--$0.8$ within this run, while $\feddsixteen$ is slightly larger than unity, thus $\feddsixteen > \hat f_{\rm Edd,16}$ throughout the entire run (Section~\ref{sec:angular_momentum}). As a result, the Lense-Thirring torque is dominated by the thick-disc precession model, leading to a long alignment time-scale of a few Myr between the BH and the disc (Equation~\ref{eq:LT_highedd}). Therefore, the BH spin experiences a combination of strong precession and gradual alignment towards the disc. The value of $\theta_{\rm BH-disc}$ decreases slowly and continues to decline until the end of the simulation (see the top-left panel). The BH undergoes pronounced precession around the disc, as clearly shown in the top-right panel. Due to this precession, the evolution of $\theta_{\rm BH-gas}$ fluctuates around $\theta_{\rm BH-disc}$. During this period, $\bm j_{\rm disc}$ and $\theta_{\rm gas-disc}$ undergo slight precession from their initial direction as a result of the Lense-Thirring torque. Since $\jdisc$ is several times larger than $J_{\rm BH}$, its direction evolves less significantly.}

\WB{The sustained high accretion rate results in significant evolution of $a_{\rm BH}$ in this run. Initially, the disc is retrograde, causing $a_{\rm BH}$ to decrease from it initial value of 0.8 to $\sim$0.1. Once the BH aligns with the disc, $a_{\rm BH}$ gradually increases due to prograde accretion. This demonstrates that the BH spin can vary substantially over just a few Myr during a period of sustained mildly super-Eddington accretion.}

\WB{The bottom-right panel of Figure~\ref{fig:result_fiducial} shows the evolution of the BH mass in the simulation, along with reference cases of a BH accreting at constant (specific) rates of $\feddsixteen = 1$ and $f_{\rm Edd, 1} = 3$. In this run, $\mbh$ increases by around 35 per~cent in 8~Myr. This growth is slightly larger than that obtained with a constant $\feddsixteen=1$, which yields a $30$ per-cent increase, and the $e$-folding time-scale \citep[Salpeter time-scale;][]{Salpeter_1964} for $\mbh$ is about 29~Myr (assuming $\eta=1/16$ for simplicity). In contrast, $\mbh$ can only increase by approximately 5 per~cent for $f_{\rm Edd, 1} = 3$.}

\subsection{Initial gas-disc alignment}\label{sec:result_gas_disc_alignment}

\begin{figure*}
    \centering
    \includegraphics[width=1.0\linewidth]{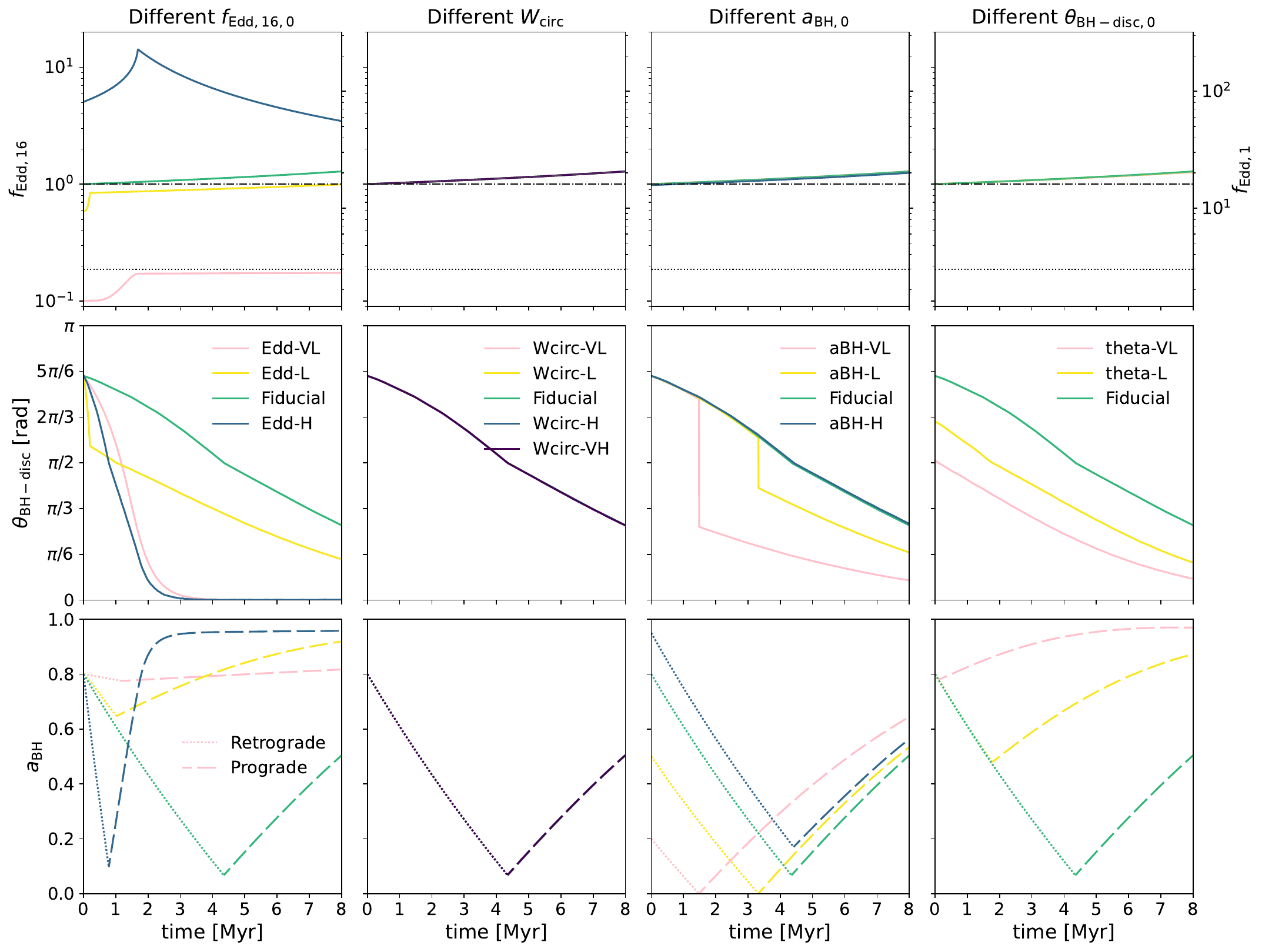}
    \caption{\WB{Time evolution of $\feddsixteen$ (top row; the corresponding values of $f_{\rm Edd, 1}$ are also shown for comparison), $\theta_{\rm BH-disc}$ (middle row), and $a_{\rm BH}$ (bottom row) for a series of runs in which a single parameter of the Fiducial run is varied at a time. In the top row, the black dash-dotted line marks $\feddsixteen = 1$, whereas the black dotted line indicates $f_{\rm Edd, 1} = 3$. In the bottom row, dotted and dashed lines correspond to a retrograde and a prograde disc, respectively. In all runs shown, $M_{\rm BH,0} = 10^6$~M$_{\sun}$, $Q_{\rm min} = 1$, and $\theta_{\rm gas-disc,0} = 0$. From left to right, the panels show the effects of varying the initial Eddington ratio, $f_{\rm Edd,16,0}$, the ratio of the circularisation radius to the self-gravitating radius, $W_{\rm circ}$, the initial BH spin, $a_{\rm BH,0}$, and the initial misalignment angle between the BH and the disc, $\theta_{\rm BH-disc, 0}$ (see the values of these parameters in Table~\ref{tab:simulation_runs}). The Fiducial run is shown in green for reference. Most initial conditions yields a similar evolution of $\feddsixteen$, because sufficient inflow keeps the disc in the self-gravity state. Exceptions are the Edd-L and Edd-VL runs, where the Lense-Thirring torque in the low accretion rates regime (Bardeen-Petterson configuration) reduces the disc angular momentum, thereby increasing $\feddsixteen$. The Edd-H run shows a decrease in $\feddsixteen$ after $t \sim 2$~Myr because it reaches $f_{\rm Edd,16,max}$, which decreases as $\mbh$ increases.} }
    \label{fig:result_fedd_wcirc_a_theta}
\end{figure*}

\begin{figure*}
    \centering
    \includegraphics[width=0.7\linewidth]{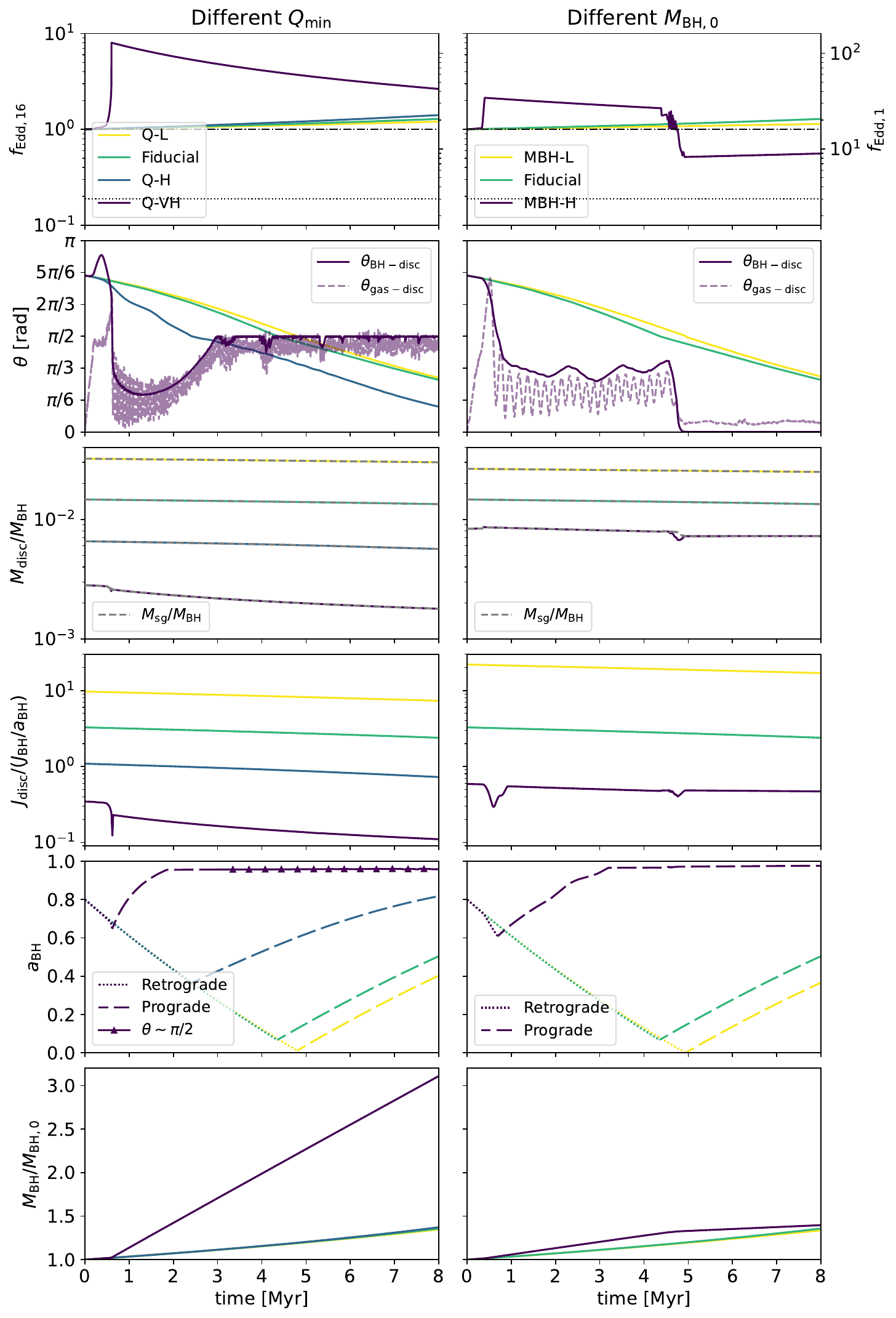}
    \caption{\WB{Time evolution of BH-disc quantities for different values of $Q_{\rm min}$ (left-hand column) and $M_{\rm BH,0}$ (right-hand column). From top to bottom: $\feddsixteen$ (with the corresponding $f_{\rm Edd,1}$ on the right axis for comparison), $\theta_{\rm BH-disc}$ (solid line), $M_{\rm disc}/M_{\rm BH}$, $J_{\rm disc}/(J_{\rm BH}/a_{\rm BH})$, $a_{\rm BH}$, and $\mbh/M_{\rm BH,0}$. In the top row, the black dash-dotted line marks $\feddsixteen = 1$, whereas the black dotted line indicates $f_{\rm Edd, 1} = 3$. The Fiducial run is shown in green for reference. The evolution of $\theta_{\rm gas-disc}$ is additionally shown in the second row only for the Q-VH and MBH-H runs (purple dashed lines). In the third row, $M_{\rm sg}/\mbh$ is shown as a grey dashed line for comparison. We note that the evolution of $\mdisc$ and $\jdisc$ shown in the third and fourth rows is plotted on a logarithmic scale to better compare the results amongst different runs. In the fifth row, the dotted and dashed lines represent retrograde and prograde discs, respectively. For the Q-VH run, $\theta_{\rm BH-disc}$ remains around $\pi / 2$ from $t \sim 3$~Myr to the end of the simulation, and is highlighted with a line and triangle marker in this row. The Q-VH and MBH-H runs have smaller discs with lower $\jdisc$, resulting in a misalignment between the disc and the gas. This configuration enables efficient removal of disc angular momentum through the inflow of misaligned gas, allowing the system to reach $\feddsixteen = f_{\rm Edd, 16, max}$. For the MBH-H run, $\feddsixteen$ drops between $t=4$ and $t=5$ Myr due to insufficient gas inflow onto the accretion disc, which causes $\mdisc$ to fall below $M_{\rm sg}$.}
   }
    \label{fig:result_Q_Mbh}
\end{figure*}

In this section, \WB{we consider and compare all runs (except for the Counter-align one, discussed separately in Appendix~\ref{app:extra_run}) with the gas-disc alignment initial condition listed in Table~\ref{tab:simulation_runs} (including the Fiducial run), wherein the gas surrounding the BH is aligned with the accretion disc, while the BH is misaligned with the disc, i.e. $\theta_{\rm BH-disc, 0} \neq 0$ and $\theta_{\rm gas-disc, 0} = 0$. Figure~\ref{fig:result_fedd_wcirc_a_theta} compares the runs with varying $f_{\rm Edd,16,0}$, $W_{\rm circ}$, $a_{\rm BH}$, and $\theta_{\rm BH-disc,0}$.}

In the first column of Figure~\ref{fig:result_fedd_wcirc_a_theta}, we present the evolution of $\feddsixteen$, $\theta_{\rm BH-disc}$\WB{, and $a_{\rm BH}$} for simulations with different initial Eddington ratios. We compare the runs Edd-VL, Edd-L, Fiducial, and Edd-H, corresponding to $f_{\rm Edd,16,0} = 0.1$, $0.6$, $1$, and $5$, respectively. 

\WB{For the Edd-VL run, $\feddsixteen$ initially increases from 0.1 to approximately 0.2 during the first 2~Myr, after which it increases slightly throughout the simulation. In this case, since $\feddsixteen < \hat f_{\rm Edd,16}$ (which is more or less constant at $\sim$0.8 throughout the run), the Lense-Thirring torque is governed by the Bardeen-Petterson configuration (Section~\ref{sec:LT_low}), which exerts a strong alignment torque between the BH and the disc. This torque efficiently reduces $\jdisc$, resulting in a more compact disc and thereby increasing $\feddsixteen$, and also drives a rapid decrease in $\theta_{\rm BH-disc}$. After the first 2~Myr, the disc is nearly aligned with the BH (i.e. $\theta_{\rm BH-disc} \lesssim \pi/6$). The Lense-Thirring torque can no longer efficiently counterbalance the angular momentum supplied by the inflowing gas. Because of the sufficient inflow, $\mdisc = M_{\rm sg}$ and $\jdisc = J_{\rm sg}$ as well, and the disc is therefore governed by the self-gravity limit, leading to a slow increase in $\feddsixteen$. Because of the low accretion rate, the accreted angular momentum onto the BH is smaller, and $a_{\rm BH}$ therefore changes mildly.}

\WB{For the Edd-L run, at the start of the simulation, $\feddsixteen < \hat f_{\rm Edd,16}$. The Lense-Thirring torque rapidly reduces the disc angular momentum, causing a sharp decrease in $\theta_{\rm BH-disc}$ and an increase in $\feddsixteen$ towards $\hat f_{\rm Edd,16}$. After $\feddsixteen = \hat f_{\rm Edd,16}$, the alignment process becomes much more slowly because the alignment torque becomes much weaker for the thick-disc precession model at high mass accretion rates, and the disc properties are then governed by the self-gravity limit. Thus, $\feddsixteen$ continues to increase slowly until the end of the simulation. This run demonstrates that the Lense-Thirring torque can reduce the disc angular momentum and increase the BH accretion rate only up to $\feddsixteen \sim \hat f_{\rm Edd, 16} \sim 1$.}

\WB{For the Edd-H run, the system is also governed by the self-gravity limit, which leads to an initial increase in $\feddsixteen$. At $t\sim$2~Myr, $\feddsixteen = f_{\rm Edd,16,max} \sim 15$. After this point, $\feddsixteen$ remains at $f_{\rm Edd,16,max}$, which gradually decreases as $\mbh$ increases. It starts with a higher initial $\feddsixteen$, leading to a faster alignment between the BH and the disc, and exhibits a more pronounced evolution of $a_{\rm BH}$, reaching $a_{\rm BH} \sim a_{\rm BH,max} = 0.957$ by the end of the simulation.} 

\WB{Interestingly, the Fiducial run shows the longest alignment time amongst the runs with different $f_{\rm Edd,16,0}$. This is because in the Fiducial run, $\feddsixteen$ is only slightly larger than $\hat f_{\rm Edd, 16}$, resulting in a longer alignment time-scale due to the weaker torque. In contrast, the Edd-H run maintains a higher $\feddsixteen$, which leads to faster alignment even though $\feddsixteen > \hat f_{\rm Edd,16}$. This difference in alignment time-scales leads to different evolutions in $f_{\rm Edd,16}$.}

\WB{In the second column of Figure~\ref{fig:result_fedd_wcirc_a_theta}, we show the lack of impact of different values of $W_{\rm circ}$ on the evolution of $\feddsixteen$, $\theta_{\rm BH-disc}$, and $a_{\rm BH}$ by comparing the runs Wcirc-VH, Wcirc-H, Fiducial, Wcirc-L, and Wcirc-VL, with $W_{\rm circ} =$ 0.9, 0.5, 0.1, 0.05, and 0.01, respectively. All runs show a similar evolution because the inflow rates are extremely high. As a result, in this setup, the value of $W_{\rm circ}$ does not affect the outcome, except for extremely small values of $W_{\rm circ}$ (not shown here). However, for lower inflow rates, the value of $W_{\rm circ}$ could influence the results (see \citetalias{Cenci_et_al_2021}).}

In the third column of Figure~\ref{fig:result_fedd_wcirc_a_theta}, we present the effects of varying $a_{\rm BH, 0}$. We include the runs aBH-VL, aBH-L, Fiducial, and aBH-H, with $a_{\rm BH, 0} =$ 0.2, 0.5, 0.8, and 0.99, respectively. \WB{All these runs show similar $\feddsixteen$ evolution due to the self-gravity limit. For the aBH-VL and aBH-L runs, $a_{\rm BH}$ spins down to zero during retrograde accretion, leading to a spin flip and a corresponding drop in $\theta_{\rm BH-disc}$. For the aBH-H run, the evolution of $\theta_{\rm BH-disc}$ is similar to that in the Fiducial run, since the alignment time-scale in the precessing thick-disc model only depends on $\feddsixteen$ rather than on $a_{\rm BH}$.}

In the fourth column of Figure~\ref{fig:result_fedd_wcirc_a_theta}, we display the results of varying $\theta_{\rm BH-disc, 0}$ by comparing the runs theta-VL, theta-L, and Fiducial, with $\theta_{\rm BH-disc, 0} = \pi/2, \ 3\pi/2, \text{ and } 5\pi/6$, respectively. \WB{All these runs show similar evolution of $\feddsixteen$ because they are dominated by the self-gravity limit. A smaller initial misalignment angle between the BH and the disc results in a shorter duration spent in the retrograde-disc phase, leading to a higher final $a_{\rm BH}$ due to a longer phase of prograde accretion.} 

\WB{Figure~\ref{fig:result_Q_Mbh} compares the runs with varying $Q_{\rm min}$ and $\mbh$.}
The left-hand column of Figure~\ref{fig:result_Q_Mbh} presents the results of varying the Toomre parameter by comparing the runs Q-VH, Q-H, Fiducial, and Q-L, with $Q_{\rm min} =$ 4, 2, 1, and 0.5, respectively. The evolution of $\feddsixteen$, $\theta_{\rm BH-disc}$, $\mdisc$, $\jdisc$, \WB{$a_{\rm BH}$, and $\mbh$} is shown in the figure. 
$\mdisc$ remains equal to $M_{\rm sg}$, since the inflow of mass is sufficient to maintain this condition. Consequently, the value of $Q_{\rm min}$ sets the size of the accretion disc, with a higher $Q_{\rm min}$ producing a smaller disc. Although values such as $Q_{\rm min} = 0.5$ or $Q_{\rm min} = 4$ may be physically unrealistic, they are included to explore how different disc sizes may influence the results.

\WB{For the Q-H, Fiducial, and Q-L runs, a larger $Q_{\rm min}$ leads to a slightly larger $\feddsixteen$ by the end of the simulation. A larger $Q_{\rm min}$ corresponds to a smaller $\mdisc$ and $\jdisc$. Due to the conservation of angular momentum, a smaller disc is more strongly influenced by the Lense-Thirring torque. Consequently, a smaller disc exhibits stronger precession motion. When combined with the inflow of gas onto the disc, this leads to a faster alignment between the BH and the disc, and therefore to a longer phase of prograde accretion and a correspondingly higher final $a_{\rm BH}$.}

For the Q-VH run, in which $Q_{\rm min} = 4$, we also plot the evolution of $\theta_{\rm gas - disc}$ for comparison. The disc angular momentum is slightly smaller than the BH angular momentum, i.e. 
\WB{$\jdisc < J_{\rm BH}$}, due to the reduced disc size. When $\theta_{\rm BH-disc} > \pi/2$, the alignment torque acting on the BH pulls the BH angular momentum towards the disc, while the corresponding torque on the disc pushes the disc angular momentum further away from the BH, owing to conservation of angular momentum. \WB{Since $J_{\rm disc} < J_{\rm BH}$,} the Lense-Thirring torque has a stronger effect on the disc than on the BH. Together with the influence of the precession torque, this causes $\theta_{\rm gas-disc}$ to increase significantly from its initial value of zero, because the disc deviates from its original orientation. 

\WB{Soon after $\theta_{\rm gas - disc}$ exceeds $\pi / 2$, the continued accretion of misaligned gas significantly decreases $\jdisc$, leading to a substantial drop in $\theta_{\rm gas-disc}$ and $\theta_{\rm BH-disc}$, thereby driving an increase in $\feddsixteen$. In this simulation, $\feddsixteen$ rapidly reaches $f_{\rm Edd,16,max}$ at $t \sim 1$ Myr and remains at $f_{\rm Edd,16,max}$ until the end of the simulation because the disc is limited by self-gravity under sufficient inflow. We note that the value of $f_{\rm Edd,16,max}$ gradually decreases, as $f_{\rm Edd,16,max} \propto 1/\mbh$. The prolonged phase of prograde, super-Eddington accretion results in $a_{\rm BH} = a_{\rm BH,max} \sim 0.96$ by the end of the simulation, yielding a final BH mass of approximately $3 \times 10^6 \msun$.} 

\begin{figure*}
    \centering
    \includegraphics[width=1.0\linewidth]{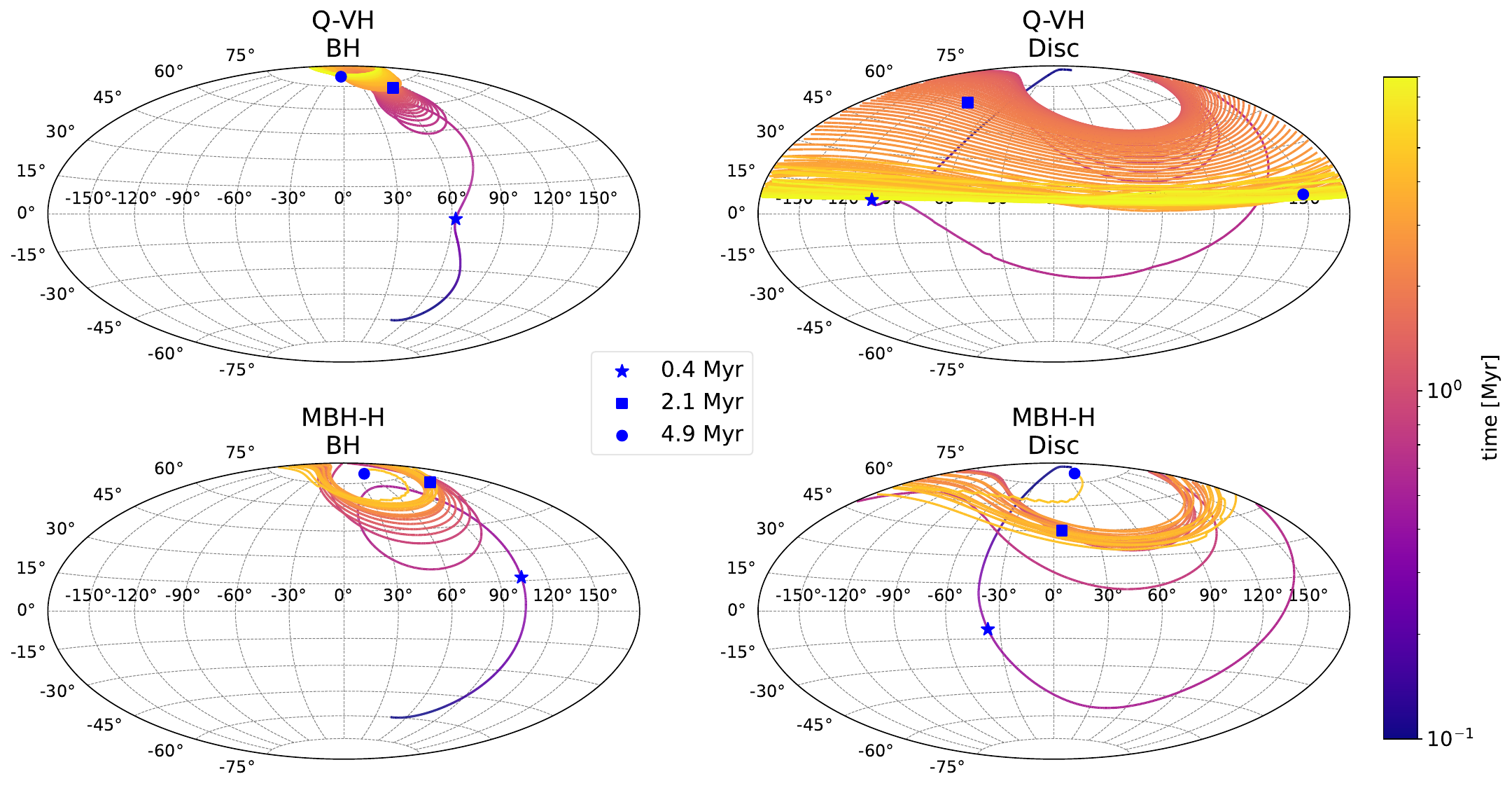}
    \caption{\WB{Evolution of the BH (left-hand panels) and disc (right-hand panels) angular momentum unit vectors for the Q-VH (top row) and MBH-H (bottom row) runs, shown via Hammer projections, wherein the equatorial plane is defined by the angular momentum of the gas within the BH kernel of the initial conditions (before relaxation). Time is colour-coded from 0.1 to 8~Myr. The blue markers highlight specific time stamps at 0.4 (star), 2.1 (square), and 4.9 (circle) Myr. In both runs, the direction of $\bm{j}_{\rm gas}$ is nearly aligned with the $z$-axis and is therefore omitted. Both runs feature discs with angular momentum smaller than that of the BH ($J_{\rm disc} < J_{\rm BH}$), which causes the disc's angular momentum vector to be pushed away from its initial direction and results in pronounced precession between the BH and the disc.} }
    \label{fig:spin_evolution_Mbh_Q}
\end{figure*}

\WB{Interestingly, $\theta_{\rm BH-disc}$ remains around $\pi/2$ from $t=3$ Myr to the end of the simulation (see also Figure~\ref{fig:spin_evolution_Mbh_Q}). This occurs because, when $\theta_{\rm BH-disc} < \pi/2 $, the accretion of angular momentum from the disc to the BH tends to push the disc's angular momentum vector away from that of the BH, whereas when $\theta_{\rm BH-disc} > \pi/2$, it instead pushes the disc and BH angular momentum vectors towards each other. Thus, $\theta_{\rm BH-disc} \sim \pi/2$ throughout the remainder of the run.} \WB{The reason why it does not reach this equilibrium state at $t\sim1$~Myr, when $\theta_{\rm BH-disc}$ first reaches $\pi/2$, is because there is still sufficient misaligned inflow of angular momentum onto the accretion disc at that time, which leads to a reduction in both $\theta_{\rm gas-disc}$ and $\theta_{\rm BH-disc}$.}

In the right-hand column of Figure~\ref{fig:result_Q_Mbh}, we present the results of varying $M_{\rm BH, 0}$ by comparing the MBH-L, Fiducial, and MBH-H runs, with $M_{\rm BH,6} = $ 0.1, 1, and 10, respectively. Since $R_{\rm sg}/ \rg$ decreases as $\mbh$ increases (Section~\ref{sec:self_gravitating}), a more massive BH leads to an initially smaller $\mdisc/\mbh$ and $\jdisc/J_{\rm BH}$. 
\WB{Consequently, the MBH-L run shows results similar to those of the Q-L run owing to the small disc. $\feddsixteen$ slowly increases from its initial value because of the self-gravity limit, and the alignment between the BH and the disc is also slightly slower than in the Fiducial run.}

\WB{For the MBH-H run, we also plot the evolution of $\theta_{\rm gas-disc}$. As in the Q-VH run, $\feddsixteen$ increases and reaches $f_{\rm Edd, 16, max} \sim 2$ at the beginning of the simulation because of the smaller disc. From $t=0.5$ to 4 Myr, $\feddsixteen$ remains at $f_{\rm Edd, 16, max}$ and the evolution of $\theta_{\rm gas-disc}$ shows a clear precessional motion between the BH and the disc (see also Figure~\ref{fig:spin_evolution_Mbh_Q}). At $t \sim 4$ Myr, $\feddsixteen$ drops because the inflow rate (determined by the BHL accretion rate) is not sufficient to sustain $\feddsixteen = f_{\rm Edd,16,max}$. After this, the combined effect of the fluctuating inflow and the interplay between the two torque regimes causes strong oscillations around $\hat f_{\rm Edd,16} \sim 1$. Because the alignment torque is stronger when $\feddsixteen < \hat f_{\rm Edd, 16}$ under the Bardeen-Peterson configuration, the value of $\theta_{\rm BH-disc}$ decreases rapidly. During this time, $\mdisc$ is also clearly smaller than $M_{\rm sg}$ because of the insufficient inflow (see Figure~\ref{fig:nonSG_noinflow_compare_C21} for an example in which $\mdisc$ is always smaller than $M_{\rm sg}$). The system eventually settles into a state with $\feddsixteen \approx 0.6$, $\mdisc = M_{\rm sg}$, and an aligned configuration. Owing to the extended period of prograde, super-Eddington accretion, $a_{\rm BH}$ reaches $a_{\rm BH,max} \sim 0.97$ by the end of the simulation.}

\WB{Interestingly, in this setup, the MBH-H run, which corresponds to the most massive BH, exhibits a higher mass accretion rate at the beginning. This result is somewhat counter-intuitive, and the main reason is that $\jdisc/J_{\rm BH}$ decreases with increasing $\mbh$, so that a smaller $\jdisc/J_{\rm BH}$ enhances the influence of the Lense-Thirring torque. Due to the insufficient gas inflow rates, the mass accretion rates drops significantly after $t \sim 5$ Myr. Thus, the BH growth slows down and results in a similar final $\Delta \mbh / \mbh \sim 40$ per cent by the end of the simulation for all three runs shown in the right-hand column of Figure~\ref{fig:result_Q_Mbh}.}

In Figure~\ref{fig:spin_evolution_Mbh_Q}, we present the evolution of $\bm j_{\rm BH}$ and $\bm j_{\rm disc}$ for the Q-VH and MBH-H runs via Hammer projections. In both cases, $\jdisc \lesssim J_{\rm BH}$. Due to conservation of angular momentum, the Lense-Thirring torque pushes the disc away from its initial direction, leading to an increase in $\theta_{\rm gas-disc}$. As a result of both the Lense-Thirring torque and gas inflow, the BH gradually aligns with the disc and surrounding gas, while exhibiting strong precessional motion. \WB{For the Q-VH run, the disc remains stable at $\theta_{\rm gas-disc} = \pi/2$ by the end of the simulation.}

\WB{For these two runs, the initial conditions do not satisfy the alignment criterion proposed by \citeauthor{King_et_al_2005} (\citeyear{King_et_al_2005}; Equation~\ref{eq:counter_align_King_2005}). However, the BH and disc do not evolve towards counter-alignment, as predicted by the criterion.} 
This outcome arises because $R_{\rm disc} > R_{\rm warp}$, meaning that the time-scale to reach counter-alignment is not negligible. During this period, the angular momentum inflow can significantly modify the disc angular momentum, thereby preventing the system from reaching counter-alignment. \WB{By contrast, in the Counter-align run (Figure~\ref{fig:result_plot_counteralign_1e7}), the BH and disc remain stably counter-aligned throughout most of the run.}

\subsection{Initial BH-disc alignment}\label{sec:result_BH_disc_alignment}

\begin{figure*}
    \centering
    \includegraphics[width=1.0\linewidth]{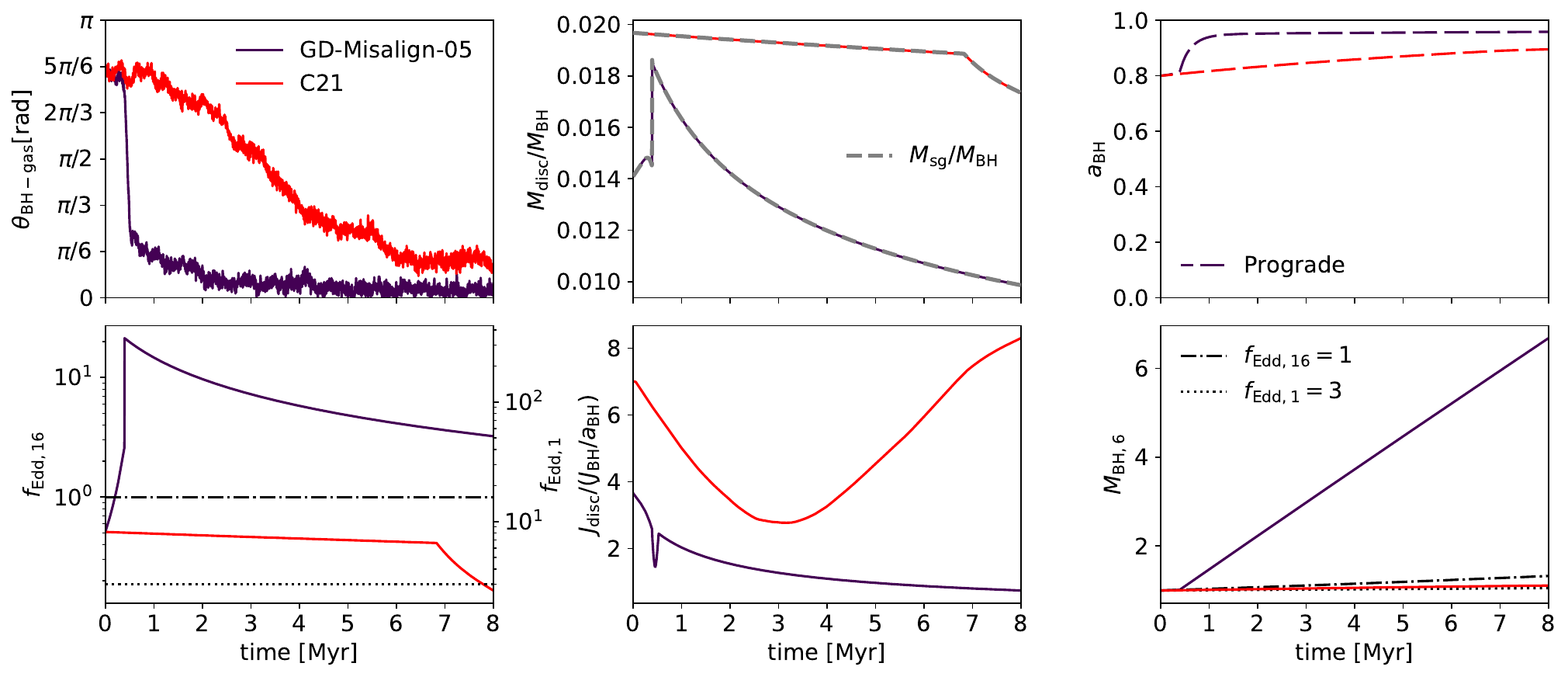}
    \caption{\WB{Comparison of the GD-Misalign-05 run using our model (purple) and the model by \citetalias{Cenci_et_al_2021} (red). Each panel shows the time evolution of key quantities. \textit{Top-left panel}: misalignment angle between the BH and the gas, $\theta_{\rm BH-gas}$. \textit{Top-central panel}: disc mass, $M_{\rm disc}$, in units of $M_{\rm BH}$. The self-gravitating mass, $M_{\rm sg}$, is also shown as a grey dashed line for comparison. \textit{Top-right panel}: BH spin parameter, $a_{\rm BH}$. The dashed line indicates that the disc remains prograde during the entire simulation. \textit{Bottom-left panel}: Eddington ratio, $\feddsixteen$, with $f_{\rm Edd, 1}$ shown on the right axis for comparison. For reference, the black dash-dotted line marks $\feddsixteen = 1$ and the black dotted line indicates $f_{\rm Edd, 1} = 3$. \textit{Bottom-central panel}: disc angular momentum, $J_{\rm disc}$, in units of $J_{\rm BH}/a_{\rm BH} = G M_{\rm BH}^2/c$ (i.e. the maximum angular momentum of a BH).  \textit{Bottom-right panel}: BH mass in units of $10^6 \msun$, $M_{\rm BH,6}$. The black dash-dotted and dotted lines represent reference tracks for constant (specific) accretion rates of $\feddsixteen = 1$ and $f_{\rm Edd, 1} = 3$, respectively. The two models yield markedly different results. In both runs, $\jdisc$ decreases substantially due to the inflow of misaligned gas onto the disc. In our model, we capture the rise of $\feddsixteen$ to $f_{\rm Edd,16,max} \sim 20$ as a result of this decrease, whereas \citetalias{Cenci_et_al_2021} impose a (much lower) cap at $f_{\rm Edd, \eta} = 1$ and therefore cannot model this increase in BH accretion rate.}}
    \label{fig:result_misalign_disc_gas}
\end{figure*}
 
\begin{figure*}
    \centering
    \includegraphics[width=1.0\linewidth]{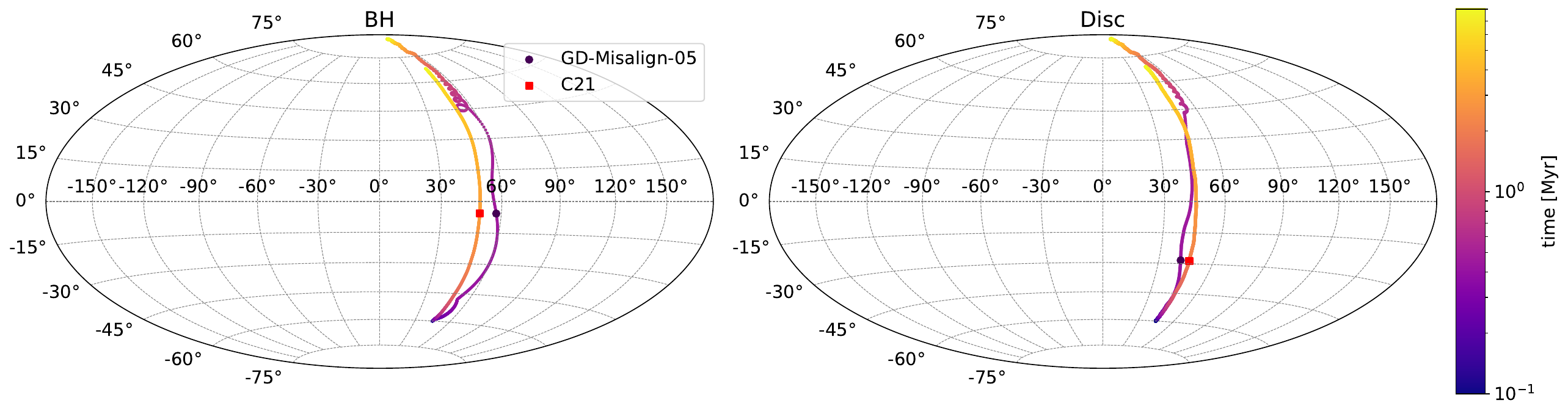}
    \caption{\WB{Evolution of the BH (left-hand panel) and disc (right-hand panel) angular momentum unit vectors for the GD-Misalign-05 run using our model (purple) and the model by \citetalias{Cenci_et_al_2021} (red), shown via Hammer projections, wherein the equatorial plane is defined by the angular momentum of the gas within the BH kernel of the initial conditions (before relaxation). The colour bar indicates time evolution from 0.1 to 8~Myr. In both models, the direction of $\bm{j}_{\rm gas}$ is nearly aligned with the $z$-axis and is therefore omitted. Our model shows faster alignment and slightly stronger precession motion of both the BH and disc angular momentum vectors compared to \citetalias{Cenci_et_al_2021}, due to the higher $\feddsixteen$ at early times, driven by the misaligned gas inflow.} }
    \label{fig:spin_evolution_misalign}
\end{figure*}

In this section, we consider the four runs with the BH-disc alignment initial condition listed in Table~\ref{tab:simulation_runs}, when the gas surrounding the BH is misaligned with the BH-disc system, and the BH is aligned with the disc, i.e. $\theta_{\rm BH-disc, 0} = 0$ and $\theta_{\rm gas-disc, 0} \neq 0$. \WB{Three of these simulations have the same initial conditions of the Fiducial run, except for $\theta_{\rm BH-disc,0}$, which is zero here, and for $\theta_{\rm gas-disc,0}$, which is equal to $5\pi/6$ (GD-Misalign-VH), $3\pi/4$ (GD-Misalign-H), and $7\pi/12$ (GD-Misalign-L). The fourth run, GD-Misalign-05, is instead initially identical to the GD-Misalign-VH case, except for having a smaller initial accretion rate ($f_{\rm Edd,16,0} = 0.5$).}\footnote{\WB{The choice of $f_{\rm Edd,16,0}$ for this simulation was dictated by the wish to compare our model to that of \citetalias{Cenci_et_al_2021}. Since the initial BH spin is $a_{\rm BH,0} = 0.8$, an initial value of $f_{\rm Edd,16} = 1$ would have implied $f_{\rm Edd,\eta,0} \sim 1.1$, thus above their limit of $f_{\rm Edd,\eta} \leq 1$.}}

\WB{For the GD-Misalign-05 run, we also apply the model from \citetalias{Cenci_et_al_2021}, using similar initial conditions. In both models, we set $M_{\rm disc, 0} = M_{\rm sg}$ and compute $J_{\rm disc, 0}$ accordingly. However, since $M_{\rm sg}$ differs between the two models, the resulting $J_{\rm disc, 0}$ varies. This discrepancy arises because our model adopts the thin $\alpha$-disc structure from \citetalias{Kato_et_al_2008}, whereas \citetalias{Cenci_et_al_2021} refer to \citet{Frank_et_al_2002}. Due to the lower surface density in \citeauthor{Frank_et_al_2002} ({\citeyear{Frank_et_al_2002}}; see Figure~\ref{fig:profile_continuous} and Appendix~\ref{app:difference_alpha_disc}), the results of \citetalias{Cenci_et_al_2021} yield a larger $R_{\rm sg}$. As a result of the combined effects of lower surface density and larger $R_{\rm sg}$, the values of $M_{\rm disc}$ and corresponding $\jdisc$ in \citetalias{Cenci_et_al_2021} are typically 1.5--2 times larger than those obtained with our model for the same $\feddsixteen$ and $\mbh$. However, this difference has only a minor impact on the evolution of the BH properties, while all other initial conditions remain identical.}

In Figure~\ref{fig:result_misalign_disc_gas}, we show the results of the 
\WB{GD-Misalign-05} run using our model and that of \citetalias{Cenci_et_al_2021}. These two models produce significantly different outcomes.

In our model, $\feddsixteen$ rises rapidly at the beginning and reaches $f_{\rm Edd, 16, max} \sim 20$. This occurs because the inflow of angular momentum from the surrounding gas onto the accretion disc significantly reduces $\jdisc$, due to the misalignment between the gas and the disc. The decrease of $\jdisc$ leads to a high BH mass accretion rate. The strong mass inflow onto the disc at $\feddsixteen \sim 20$ results in the rapid alignment of the disc and BH with the surrounding gas within the first $0.5$~Myr. 
\WB{After $t > 0.5$~Myr, $\feddsixteen$ remains at $f_{\rm Edd, 16, max}$ until the end of the simulation, because both $\mdisc$ and $\jdisc$ are constrained by self-gravity.}

By contrast, in the \citetalias{Cenci_et_al_2021} model, the imposed limit of $f_{\rm Edd, \eta} \leq 1$ prevents the model from capturing the effects of high mass accretion rates. The strong reduction in $\jdisc$ due to the gas inflow onto the disc does not translate into an increased Eddington ratio because of this cap. As a result, $\feddsixteen$ remains at approximately 0.5 \WB{($f_{\rm Edd,\eta} = 1$)} for around 7~Myr, and the BH aligns with the surrounding gas much more slowly than in our model.

\WB{In our model, $\mdisc$ shows a rapid increase at $t \sim 0.5$~Myr. This occurs because the value of $M_{\rm sg}$ is larger at $f_{\rm Edd, 16, max}$.}
In the model by \citetalias{Cenci_et_al_2021}, $\mdisc$ varies only slightly over the first 7~Myr, as $\feddsixteen$ remains nearly constant during this period. Both models show a significant reduction in $\jdisc$ due to the misaligned gas inflow, although the time-scale of this decrease differs considerably because of the different mass accretion rates.

\WB{Our model shows $a_{\rm BH} \sim a_{\rm BH,max} \sim 0.95$ at the end of the simulation, due to BH mass growth through a prograde super-Eddington accreting disc. On the other hand, the model from \citetalias{Cenci_et_al_2021} shows a much slower increase in $a_{\rm BH}$. Furthermore, our model produces a much larger final BH mass of approximately $6 \times 10^6 \msun$}

In Figure~\ref{fig:spin_evolution_misalign}, we illustrate the evolution of the direction of $\bm j_{\rm BH}$ and $\bm j_{\rm disc}$ for both models, projected onto the full-sky sphere using Hammer projections. In both models, the BH and disc progressively align with the gas direction (which is approximately aligned with the $z$-axis). In our model, the time-scale for alignment is significantly shorter, as a larger $\feddsixteen$ drives a stronger inflow of angular momentum onto the disc. Moreover, our model also exhibits a more pronounced precession between the BH and the disc, although it is relatively mild in this scenario, as the misalignment angle between the BH and the disc is generally small.

This \WB{comparison} clearly highlights the importance of incorporating the super-Eddington model when the angular momentum of the surrounding gas is misaligned with the accretion flow. 

\begin{figure}
    \centering
    \includegraphics[width=0.9\linewidth]{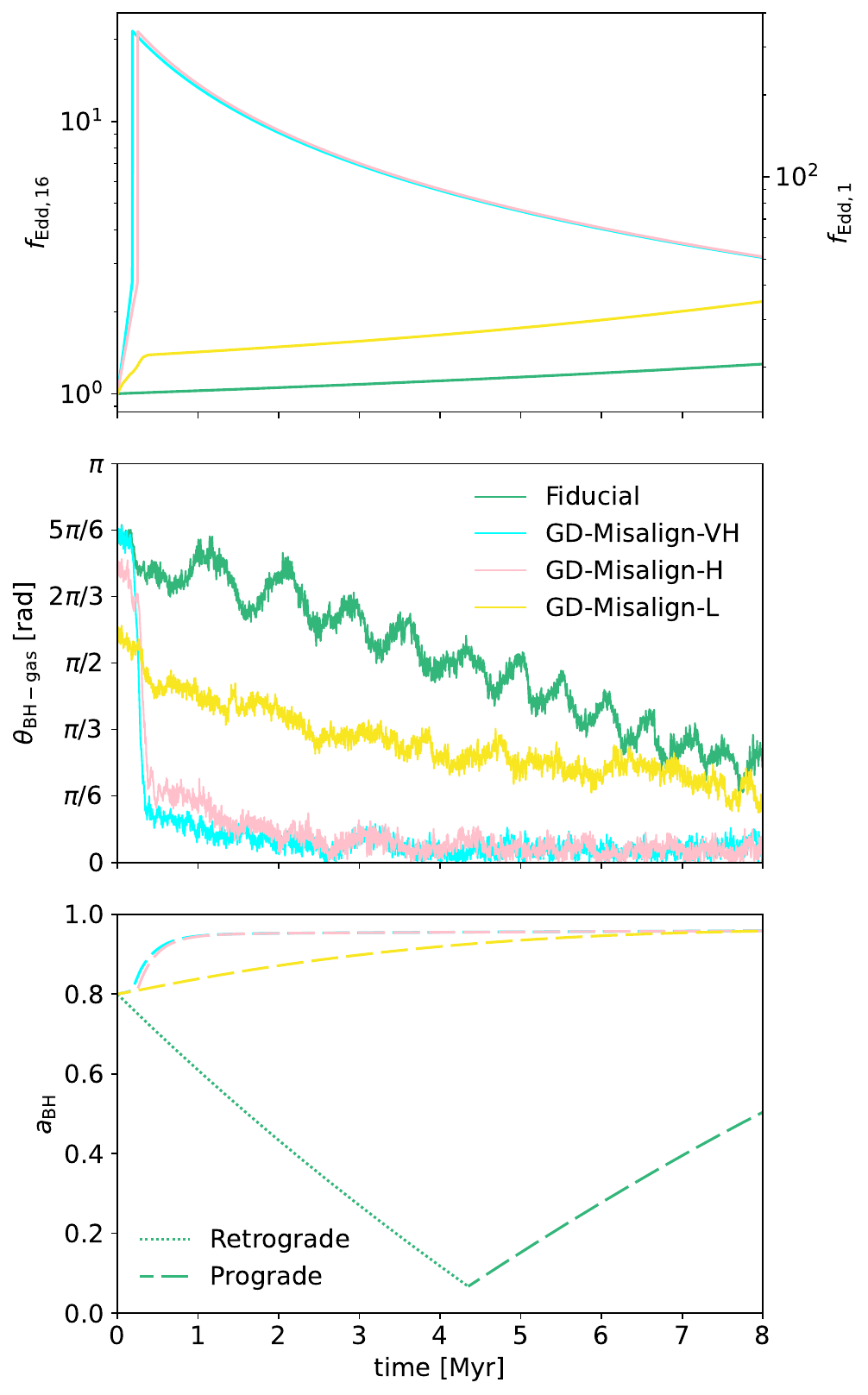}
    \caption{\WB{Time evolution of $\feddsixteen$ (top panel, with $f_{\rm Edd,1}$ shown on the right axis for comparison), misalignment angle between the BH and the surrounding gas, $\theta_{\rm BH-gas}$ (middle panel), and $a_{\rm BH}$ (bottom panel) for the Fiducial, GD-Misalign-VH, GD-Misalign-H, and GD-Misalign-L runs. All GD-Misalign runs adopt the gas-disc alignment initial conditions and exhibit higher peaks in $\feddsixteen$ compared to the Fiducial case. This is because the misaligned gas inflow efficiently reduces the disc angular momentum, thereby enabling higher mass accretion rates. In particular, the GD-Misalign-VH and GD-Misalign-H runs reach $\feddsixteen = f_{\rm Edd,16,max} \sim 20$, which drives faster alignment between the BH and the gas.}}
    \label{fig:result_misalign_theta}
\end{figure}

Figure~\ref{fig:result_misalign_theta} compares the runs Fiducial, GD-Misalign-L, GD-Misalign-H, and GD-Misalign-VH. The Fiducial run adopts the gas-disc alignment initial condition \WB{(i.e. $\theta_{\rm gas-disc,0} = 0$, with $\theta_{\rm BH-disc} = 5\pi/6$)}, whereas the other three runs adopt the BH-disc alignment initial condition \WB{(i.e. $\theta_{\rm BH-disc} = 0$)}, with $\theta_{\rm gas-disc,0} = \WB{7\pi/12}$, $3\pi/4$, and $5\pi/6$, respectively.

\WB{Comparing these three GD-Misalign runs, we observe that the GD-Misalign-VH and GD-Misalign-H runs show results similar to those of the GD-Misalign-05 run. The GD-Misalign-L run exhibits a much slower alignment and a slower evolution of $a_{\rm BH}$. In addition, the maximum Eddington ratio in the GD-Misalign-L run is approximately one order of magnitude lower than in the other two runs. This difference arises because $M_{\rm sg}$ is not a strictly monotonic function of $\feddsixteen$, due to the complex structure of the accretion disc structure. In particular, for $2 \lesssim \feddsixteen \lesssim 2.5$, $M_{\rm sg}$ slightly decreases with increasing $\feddsixteen$. However, once $\feddsixteen \gtrsim 2.5$, $M_{\rm sg}$ increases rapidly with $\feddsixteen$. As a result, once $\feddsixteen$ exceeds this threshold, the corresponding rise in $M_{\rm sg}$ drives a stronger inflow onto the accretion disc. This enhanced inflow further reduces the disc angular momentum, leading to a significantly higher mass accretion rate.}

\WB{Comparing the GD-Misalign runs with the Fiducial run, we can see that the Fiducial run consistently shows a smaller $\feddsixteen$, because it cannot effectively reduce $\jdisc$. On the contrary, the GD-Misalign runs are able to reduce $\jdisc$ effectively through the inflow of misaligned gas onto the disc. As a result, the GD-Misalign runs exhibit a larger $\feddsixteen$, particularly for a larger $\theta_{\rm gas-disc, 0}$. For the GD-Misalign-VH and GD-Misalign-H runs, the alignment between the BH and the gas proceeds much more rapidly than in the Fiducial run due to the significantly higher $\feddsixteen$ and misaligned inflow. It is also worth noting that the final value of $a_{\rm BH}$ is larger for the GD-Misalign runs, because they remain in the state of prograde accretion throughout the entire simulation.}


\section{Discussion}\label{sec:discussion}

We developed a new sub-grid model for the evolution of BH mass and spin that incorporates super-Eddington accretion, regulated via an accretion disc-mediated growth. Our work builds upon the model proposed by \citetalias{Cenci_et_al_2021}. We conducted a suite of simulations in idealised setups to test the model and investigate the evolution of the BH. In Section~\ref{sec:relevance_super_edd}, we discuss the relevance of super-Eddington accretion based on our simulation results. In Section~\ref{sec:caveat}, we discuss a few potential caveats of our model. 

\subsection{Relevance of super-Eddington accretion}\label{sec:relevance_super_edd}

We begin by examining the simulations in which the surrounding gas is initially aligned with the accretion disc but the BH is initially misaligned with both (i.e. gas-disc alignment initial conditions: $\theta_{\rm gas-disc, 0} = 0$ and $\theta_{\rm BH-disc, 0} \neq 0$). \WB{In this setup, the BH typically shows a slow increase in mass accretion rate due to sufficient inflow that maintains a disc governed by the self-gravity limit (Figure~\ref{fig:result_fiducial}). For lower initial accretion rates (Edd-L and Edd-VL runs, with $f_{\rm Edd,16,0} = 0.6$ and 0.1, respectively), the value of $\feddsixteen$ increases, as the Lense-Thirring torque efficiently reduces the disc angular momentum. However, the maximum value of $\feddsixteen$ is around $\hat f_{\rm Edd,16} \sim 1$ for the Edd-L run, because the torque model at higher mass accretion rates (thick precessing disc configuration) is less efficient at reducing $\jdisc$ (Figure~\ref{fig:result_fedd_wcirc_a_theta}). If the disc is smaller, due to a larger $Q_{\rm min}$ or a larger $\mbh$, $\feddsixteen$ can reach $f_{\rm Edd,16,max}$, as a lower $\jdisc/J_{\rm BH}$ enhances the misalignment between the disc and the surrounding gas, allowing inflows to further reduce the disc angular momentum (Figure~\ref{fig:result_Q_Mbh}).}

If we consider a different initial condition, in which the BH is aligned with the disc but both are misaligned with the surrounding gas (i.e. BH-disc alignment initial conditions: $\theta_{\rm gas-disc, 0} \neq 0$ and $\theta_{\rm BH-disc, 0} = 0$), the misaligned inflow can significantly reduce $\jdisc$. Under these conditions, our simulations show that the system can reach $\feddsixteen \sim 20$ with $M_{\rm BH, 6} = 1$ (Figure~\ref{fig:result_misalign_disc_gas}). \WB{If sufficient inflow continues to feed the accretion disc, the system can remain at $\feddsixteen = f_{\rm Edd,16,max}$. This results in a substantial increase in BH mass. For instance, $\mbh$ grows from $10^6 \msun$ to $6 \times 10^6 \msun$ within 8~Myr in the GD-Misalign-05 run.} 

In this case, the direction of the angular momentum inflow onto the accretion disc plays a crucial role in determining the BH mass growth rate. If a new inflow of misaligned gas enters the centre of the galaxy, a configuration similar to the BH-disc alignment initial condition might emerge, provided that the BH and disc were originally aligned (which is 
\WB{typically found} at the end of our simulations). A misaligned inflow can enhance the mass accretion rate, \WB{potentially leading to a phase of super-Eddington accretion.} However, this process is highly non-linear, making it challenging to predict without detailed simulations. This highlights the importance of considering the direction of the angular momentum of the gas surrounding the accretion disc in order to better estimate the rate of BH growth. 

In general, reducing the disc angular momentum is crucial for 
\WB{reaching} a high mass accretion rate. Otherwise, the mass accretion rate onto the BH can remain low, even if the BHL accretion rate is 
\WB{sufficient to provide adequate inflow to the disc}. One mechanism for reducing $\jdisc$ is via the Lense-Thirring torque, but in our simulations this effect can bring $\feddsixteen$ to at most $\sim$1 \WB{(see, e.g. the Edd-L run in Figure~\ref{fig:result_fedd_wcirc_a_theta})}. Another approach is through the inflow of low-angular-momentum and/or misaligned gas, which could lead to a larger mass accretion rate (e.g. $\feddsixteen = f_{\rm Edd, 16, max} \sim 20$ for $M_{\rm BH,6} = 1$ \WB{in the GD-Misalign-05 run in Figure~\ref{fig:result_misalign_disc_gas}}). Such inflows may arise from chaotic cosmic accretion, disturbed morphologies in proto-galaxies, merger events, starbursts, gravitational instabilities, or non-axisymmetric gravitational torques (e.g. \citealt{Shlosman_et_al_1989, Hopkins_and_Quataert_2010, Capelo_and_Dotti_2017, Blumenthal_and_Barnes_2018, Yu_et_al_2022, Shin_et_al_2025}; see \citealt{Capelo_et_al_2023} for a recent review). Additionally, turbulence on scales slightly larger than those of the accretion disc can induce chaotic accretion, contributing to the efficient removal of disc angular momentum and thereby promoting 
high accretion rates \citep[][]{Dotti_et_al_2013, Fiacconi_et_al_2018}.

In high-redshift environments, it is more frequent to have chaotic gas accretion onto the (proto-)galaxy, a highly turbulent gas kinematics, and galaxy-galaxy interactions, leading to more frequent strong and misaligned inflows onto the accretion disc \citep[][]{Gabor_and_Bournaud_2014, Danhaive_et_al_2025, Duan_et_al_2025}. In such cases, super-Eddington accretion might be (re-)triggered when a misaligned inflow occurs, making episodic super-Eddington accretion in the early Universe a plausible mechanism for facilitating rapid BH growth \citep[][]{Volonteri_et_al_2015, Khoperskov_et_al_2021}. However, the detailed exploration of this scenario is left for future work. 

\citet{Sassano_et_al_2023} modelled BH growth and feedback in the super-Eddington regime in simulations of the central region of a proto-galaxy, starting from an initial BH mass of $\mbh = 10^3 \msun$. The BH in their simulations accretes up to $\sim$$10^4 \msun$ within 
\WB{1~Myr}, corresponding to a growth of more than an order of magnitude -- substantially higher than our results. 
\WB{\citet{Zana_et_al_2025}}, using an improved version of the code used by \citet{Sassano_et_al_2023} and different initial conditions, obtain even larger growth rates (of the order of $\lesssim$$10^5$~M$_{\sun}$ in less than \WB{0.}1~Myr). Part of this discrepancy may arise from their use of BHL accretion rates for the BH accretion rate, which neglects the angular momentum of the inflowing gas and thus overestimates the accretion rate. \WB{We posit, however, that their scenario is reproducible by our model in the cases where $\mbh$ is smaller than the values adopted in the simulation runs presented in this paper and the angular momenta of the external gas and of the accretion disc have opposite directions. This is not an uncommon situation at high redshift, if we assume an isotropic reservoir of the gas angular momentum.}

\begin{figure}
    \centering
    \includegraphics[width=1.0\linewidth]{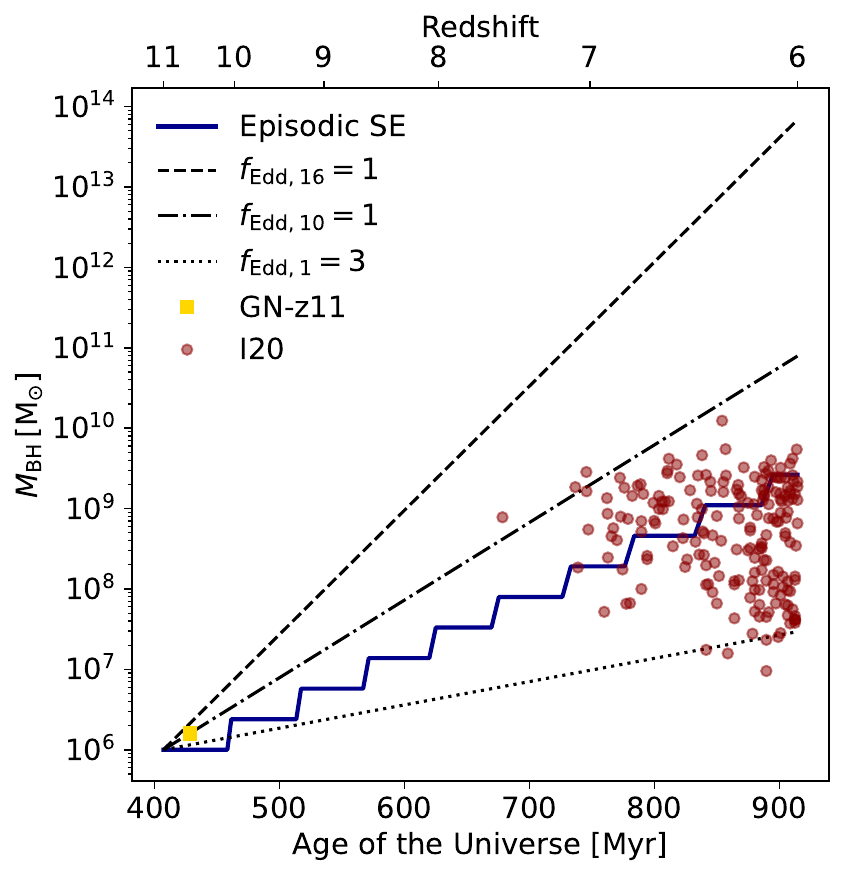}
    \caption{Time evolution of BH mass from $z = 11$ to $z = 6$, from an initial value of $\mbh = 10^6 \msun$. The blue line shows BH growth via episodic super-Eddington accretion (Episodic SE), assuming a dynamic time-scale of $t_{\rm dyn} = 2$~Myr, a misaligned inflow fraction of $p_{\rm mis} = \WB{0.37}$, \WB{a fraction of these misaligned inflows capable of triggering super-Eddington accretion, $p_{\rm SE} = 0.1$, and a mass increase of $\Delta \mbh/\mbh = 1.4$} during each \WB{super-Eddington event}. The dashed, dash-dotted, and dotted lines correspond to constant specific accretion rates of $\feddsixteen = 1$, $f_{\rm Edd,10} = 1$, and $f_{\rm Edd,1} = 3$, respectively. The yellow square denotes GN-z11 \citep[][]{Maiolino_et_al_2024}, whereas the dark red circles represent 196 quasars observed at $z \geq 6$, taken from the supplemental table in \citeauthor{Inayoshi_2020} (\citeyear{Inayoshi_2020}; I20). This figure shows that our model can potentially reproduce the most massive BHs observed at $z = 6$.
    }
    \label{fig:BH_mass_growth}
\end{figure}

We extrapolate our results to estimate the growth of a BH from $z=11$ down to $z=6$, using a simple calculation that consider\WB{s} a sequence of episodic super-Eddington accretion events\WB{, similar to those in the GD-Misalign-05 run (Figure~\ref{fig:result_misalign_disc_gas})}, following the approach of \citet{Sassano_et_al_2023}. We consider a BH at $z=11$ with an initial mass of $M_{\rm BH, z=11} = 10^6 \msun$, consistent with a massive BH seed formed via direct collapse in \citet{Mayer_et_al_2024} and with GN-z11, observed by \citet{Maiolino_et_al_2024}. \citet{Sassano_et_al_2023} showed that the dynamical time-scale at the centre of a galaxy is $t_{\rm dyn} = 2$ Myr, corresponding to the typical time required for gravitational torques and dynamical instabilities to replenish the gas within the central 2~pc.

We assume that the minimum time between new gas inflows is $t_{\rm dyn}$, yielding a maximum of $N_{\rm max} = 253$ inflow events between $z = 11$ and $z = 6$. We define $p_{\rm mis}$ as the fraction of these events involving misaligned gas inflows, \WB{and $p_{\rm SE}$ as the fraction of these misaligned inflows} capable of triggering super-Eddington accretion. \WB{Based on Figure~\ref{fig:result_misalign_theta}, we define misaligned gas inflow as one with $\theta_{\rm gas-disc} > 7\pi/12$, since such configurations can lead to rapid BH growth. For simplicity, we assume that the angular momentum of the inflowing gas is isotropically distributed during each event, which yields $p_{\rm mis} = 0.37$, and we adopt $p_{\rm SE} = 0.1$. We further assume that BH growth ceases after one dynamical time-scale, resulting in $\Delta M_{\rm BH}/M_{\rm BH} = 1.4$ within 2~Myr, as illustrated in the GD-Misalign-05 run.\footnote{One important difference between this calculation and that of \citet{Sassano_et_al_2023} is that we assume a fixed $\Delta \mbh / \mbh$ (i.e. a fixed $\feddsixteen$; as in \citealt{Regan_et_al_2019}) for each event, whereas they assume a fixed $\Delta \mbh$. We adopt this approach because our simulations always feature sufficient gas inflow onto the accretion disc. Therefore, we expect that, for a more massive BH, similar growth rates can be sustained due to the available gas reservoir. However, both a fixed $\Delta \mbh/\mbh$ and a fixed $\Delta \mbh$ are strong assumptions, as the actual BH growth rates should also depend on several other factors, such as the surrounding gas mass, angular momentum, dynamics, temperature, and the BH mass and spin.} Thus, the final BH mass at $z=6$ is}

\begin{equation}
    \begin{split}
        M_{\rm BH, z=6} &\sim M_{\rm BH, z=11} \left(1 + \frac{\Delta M_{\rm BH}}{M_{\rm BH}} \right)^{N_{\rm max}p_{\rm mis}\WB{p_{\rm SE}}} \\
        &= \WB{3.8} \times 10^9 \msun~,
    \end{split}
\end{equation}

\noindent which is consistent with the most massive BHs observed at $z \sim 6$ \citep[e.g.][]{Inayoshi_2020, Fan_et_al_2023}. The corresponding BH growth rates are equivalent to an overall effective Eddington ratio of $\feddsixteen = \WB{0.46}$ (i.e. $f_{\rm Edd, \eta} / \eta_{0.1} = f_{\rm Edd, 10} = \WB{0.74}$, $f_{\rm Edd, 1} = \WB{7.4}$).
\WB{The evolution of $\mbh$ from $z=11$ to $z=6$ is} shown in Figure~\ref{fig:BH_mass_growth}. The GN-z11 source \citep[][]{Maiolino_et_al_2024} and 196 quasars observed at $z \sim 7.5$--6, listed in \citet{Inayoshi_2020}, are also included for comparison. 
Additionally, the figure displays the BH mass growth under a constant specific accretion rate of $\feddsixteen = 1$, $f_{\rm Edd, 10} = 1$, and $f_{\rm Edd, 1} = 3$, which correspond to three common definitions of super-Eddington accretion in the literature: we note that the choice of definition for super-Eddington accretion substantially impacts the final $\mbh$, as illustrated in the figure. 

\WB{Our results successfully reproduce the most massive BHs observed at $z \sim 6$. We also note that, if we begin with a low-mass seed \citep[e.g. $10^3$~M$_{\sun}$;][]{Sassano_et_al_2021}, the final mass would be much lower, even when starting at a much earlier redshift. However, this discrepancy can be mitigated if episodes of super-Eddington accretion are more frequent at higher redshift, which is plausible given the more chaotic environments in the early Universe. We note that only about ten super-Eddington episodes are required to grow a BH from $10^6 \msun$ to $10^{10} \msun$, indicating that misaligned gas inflows can be highly efficient in driving BH growth. Nevertheless, we emphasize that this remains a simplified, order-of-magnitude calculation based on idealized assumptions. Determining how long such super-Eddington phases can persist, how frequently misaligned accretion occurs in realistic environments, and how BH feedback processes regulate BH growth will require more detailed hydrodynamical simulations to obtain a more robust estimate of BH mass evolution.}

Both \citet{Regan_et_al_2019} and  \citet{Massonneau_et_al_2023} conduct high-resolution hydrodynamics simulations to investigate BH growth, allowing for super-Eddington accretion in the presence of BH feedback. In both studies, episodic super-Eddington accretion is found, as BH feedback expels gas from the surrounding region, which is later replenished via inflows. \citet{Regan_et_al_2019} reported an effective Eddington ratio of $f_{\rm Edd,10} \sim 0.1$--0.5 (i.e. $\feddsixteen \sim 0.06$--0.3), whereas \citet{Massonneau_et_al_2023} found that an effective Eddington ratio of $f_{\rm Edd,10} \sim 1$ (i.e. $\feddsixteen \sim 0.6$) can be achieved when jet feedback efficiencies are weak. Although both studies yield mass accretion rates comparable to ours, the cumulative growth of $\mbh$ from $z=11$ to $z=6$ can differ by more than an order of magnitude. The discrepancy in effective Eddington ratios may arise from the inclusion of BH feedback, as well as differences in BH accretion prescriptions and initial BH mass. 

\subsection{Caveats}\label{sec:caveat}

\subsubsection{Accretion disc structure}\label{sec:caveat_accretion_model}

In this paper, we modelled the accretion disc structure by combining the fitting formulae for the photon-trapping region from simulations by \citetalias{Kitaki_et_al_2018} with the three regions of the thin $\alpha$-disc. We selected \citetalias{Kitaki_et_al_2018} because they provide fitting formulae for the disc structure across a wide range of values of $\feddsixteen$ and $\mbh$. However, we note that different simulations and theoretical models predict varying structures \WB{and scale heights} for the photon-trapping region. We compared the results from \citetalias{Kitaki_et_al_2018} to those of \citet{Watarai_2006} and \citet{Sadowski_2011} in Section~\ref{sec:accretion_disc_structure}. Simulations that include magnetic fields and/or GR effects may yield more significant differences \citep[e.g.][]{Sadowski_and_Narayan_2016, Jiang_et_al_2019, Hu_et_al_2022_RHDsimulaion}. Furthermore, we renormalise the surface density and specific angular momentum profiles of this inner region to ensure continuity. While these differences may not strongly affect the calculation of the BH mass accretion rate as a function of $\mdisc$ and $\jdisc$, they could influence the magnitude of the Lense-Thirring torque exerted by the inner precessing thick\WB{-}disc in the photon-trapping region. \WB{In particular, the value of $H/R$ affects both $R_{\rm bw}$ and $R_{\rm trap}$, and thereby the precession frequency.} 

Furthermore, the simulations in \citetalias{Kitaki_et_al_2018} assumed a constant $\alpha = 0.1$. Even though many of our equations are formulated for a generic $\alpha$, using different values of $\alpha$ in our model may lead to an inaccurate representation of the photon-trapping region structure. In addition, GRMHD simulations predict significant radial variations in  $\alpha$ \citep[e.g.][]{Jiang_et_al_2019, Liska_et_al_2021}. Moreover, we adopt $\xi=0.7$ in this study, consistent with $\alpha = 0.1$ \citep[][]{Lodato_and_Pringle_2007}: a different value of $\alpha$ would imply a different value of $\xi$ \citep[][]{Lodato_and_Pringle_2007, Perego_et_al_2009}.\footnote{\WB{We note that \citet{Fiacconi_et_al_2018} write $\xi \sim O(1)$ instead.}} The parameter $\xi$ affects the Lense-Thirring torque through $R_{\rm warp}$, $t_{\rm gm}$, and $\hat f_{\rm Edd, 16}$.

The surface density and specific angular momentum profiles of the disc used in this work are all derived assuming a non-spinning BH. If the effects of BH spin are taken into consideration, the disc structure might be altered in the innermost regions near the event horizon. We neglected this effect, as the integral properties of the disc are dominated by the outer parts, and thus it does not significantly affect our results. 

The thin $\alpha$-disc model is widely used to describe the structure of accretion discs due to its simplicity and success at explaining many observed properties of such systems \citep[][]{Abramowicz_et_al_2013}. Accordingly, we adopt it to describe the disc where $\rdisc \geq R_{\rm trap}$. However, several fundamental questions remain unsolved. For instance, it has difficulties accounting for the variability commonly observed in AGN \citep[e.g.][]{Burke_et_al_2021}, the discrepancy between observed disc sizes and theoretical predictions \citep[e.g.][]{Dai_et_al_2010}\WB{, and the influence of strong magnetic fields \citep[][]{Narayan_et_al_2003, Hopkins_2024}}. 

In our model, we impose $\feddsixteen \leq f_{\rm Edd,16,max}$ to avoid the ``degeneracy'' behaviour in the $(\mdisc,\ \jdisc)$ parameter space discussed in Appendix~\ref{app:Newton_Raphson_method}, which arises when region~(a) dominates the accretion disc structure. By doing this, we effectively ignore region~(a) for the mass accretion rate calculations, but we note that we still include this region for the torque computations. This constraint prevents us from studying the evolution of the BH when the Eddington ratio exceeds this limit. We note, however, that this limit is much higher than the usually applied Eddington limit for a wide range of BH and disc masses, and it becomes low only for very massive BHs and/or the extreme case of a nearly depleted accretion disc. Furthermore, region~(a) of the disc is both viscously and thermally unstable in the thin $\alpha$-disc theory \citep[][]{Lightman_and_Eardley_1974}. These instabilities might produce a different disc structure than what assumed in this work, and the degeneracy might disappear. However, these instabilities are not observed in the simulations conducted by \citet{Kitaki_et_al_2021}, and the underlying reasons remain unclear \citep[][]{King_et_al_2023}. 

Another caveat is that our model is not applicable at very low mass accretion rates ($\feddsixteen \lesssim 10^{-2}$), when the innermost part ($R < R_{\rm tran}$ in Figure~\ref{fig:characteristic_radius}) of the accretion disc transitions into an ADAF that cannot be adequately described by the thin $\alpha$-disc model. In contrast to our approach, \citet{Koudmani_et_al_2024} extends the sub-grid model of \citet{Fiacconi_et_al_2018} to include this regime by coupling an inner ADIOS flow with an outer thin $\alpha$-disc. \WB{We neglect this effect since the aim of this study is super-Eddington accretion, and we note that none of the runs listed in Table~\ref{tab:simulation_runs} enter this regime.} 

Furthermore, the viscous time-scale, $t_{\rm visc}$, for the thin $\alpha$-disc at $R_{\rm disc}$ is approximately 0.1--1~Myr in our model. This implies that if the disc structure changes rapidly on a time-scale shorter than $t_{\rm visc}$, the results may be inaccurate, as the system does not have sufficient time to reach a stable and stationary accretion disc configuration. In contrast, \citet{Tartenas_and_Zubovas_2022} explicitly model the viscous evolution of the disc surface density to account for this situation. They find that including viscous evolution leads to a reduced and more extended BH accretion rate and alters the impact of BH feedback on the surrounding gas, compared to the BHL prescription, in simulations of collisions between a gas ring and a molecular cloud in the galactic centre. 

As discussed in Section~\ref{sec:self_gravitating}, we impose the limit that $R_{\rm disc} \leq R_{\rm sg}$ to prevent the disc from becoming self-gravitating. However, if the local cooling time is not sufficiently short at $R \gtrsim R_{\rm sg}$, the disc might not collapse and a gravitoturbulent disc could instead form, as the spiral shocks can heat the gas and increase $Q$ \citep[][]{Durisen_et_al_2007, Deng_et_al_2020}. Several models have also been proposed to describe this region \citep[e.g.][]{Goodman_2003, Sirko_and_Goodman_2003, Thompson_et_al_2005}. For simplicity, our model does not include this potential gravitoturbulent regime. \WB{When combined with the opacity effects discussed in Section~\ref{sec:caveat_opacity}, this regime could lead to some modifications of our results. A detailed treatment, however, lies beyond the scope of the present study and may be explored in future work.}

\subsubsection{Torque modelling and angular momentum inflow}\label{sec:caveat_angular_momentum}

In our torque evolution model, we consider two scenarios based on the mass accretion rate. For $\feddsixteen > \hat f_{\rm Edd, 16}$, we apply the Lense-Thirring effect derived from an inner precessing thick\WB{-}disc. For $\feddsixteen < \hat f_{\rm Edd, 16}$, we adopt the Lense-Thirring effect from the Bardeen-Petterson configuration. 
\WB{T}he transition between the two torque models in our framework is quite abrupt, leading to \WB{an abrupt change in the alignment time-scale between BH and disc.}
In a more realistic model, the change between the two regimes should be smoother, yielding a more gradual evolution 
when 
$\feddsixteen$ is close to $\hat f_{\rm Edd, 16}$. 

\WB{Unlike \citet{Shin_et_al_2025}, who fully resolve the self-gravity radius of the accretion disc to accurately measure the angular momentum of the inflow onto it, our simulations do not have sufficient resolution to achieve this. Instead, we introduced the $W_{\rm circ}$ parameter to set an upper limit on the specific angular momentum of the inflow onto the accretion disc.} However, determining its exact value (and whether it remains constant) requires a detailed study that resolves spatial scales from the accretion disc to the larger environments, providing insights into how angular momentum evolves in this region. 
\WB{Nonetheless, in the simulations presented here, we find that the value of $W_{\rm circ}$ does not influence the results, since all runs are governed by the self-gravity limit (Figure~\ref{fig:result_fedd_wcirc_a_theta}).}
Furthermore, we do not account for the possibility of chaotic accretion with varying angular momentum direction of the inflow in this model \citep[e.g.][]{Dotti_et_al_2013, Fiacconi_et_al_2018, Bustamante_et_al_2019}.

Smoothed particle hydrodynamics simulations and GRMHD simulations indicate that a thin disc might experience a sharp change in density and tilt angle, or even fragment into rings, when the inclination angle between the BH and the disc exceeds $45^\circ$ \citep[][]{Nealon_et_al_2015, Liska_et_al_2021, Speicher_and_Blaes_2025}. This could significantly impact the structure of the accretion disc, the BH mass accretion rate, and the strength of the Lense-Thirring torque. However, quantifying how this effect influences the evolution of the SMBH is beyond the scope of this study.

\subsubsection{Opacity effect}\label{sec:caveat_opacity}

It is well known that an accurate opacity table is crucial for correctly modelling the structure of accretion discs in protoplanetary discs \citep[e.g.][]{Bell_and_Lin_1994, Semenov_et_al_2003}. This is often neglected for AGN accretion disc models, due to their typically high temperature, which justifies the use of a simplified opacity law. In the original thin $\alpha$-disc model developed by \citet{Shakura_Sunyaev_1973}, the dominant opacity source is assumed to be either electron scattering opacity or free-free absorption opacity, a treatment that is retained in, e.g. \citet{Frank_et_al_2002} and \citetalias{Kato_et_al_2008}. Consequently, our model adopts the same assumption for the opacity source when computing the structure of the thin $\alpha$-disc. 

However, as pointed out by, e.g. \citet{Hure_et_al_1994b}, \citet{Frank_et_al_2002}, and \citet{Perego_et_al_2009}, this assumption breaks down when $T \lesssim 10^4$~K. At such temperatures, hydrogen recombination significantly reduces the opacity (known as the opacity gap), meaning that it no longer follows the simple electron scattering or free-free absorption prescriptions. In the outermost regions of the accretion disc, the temperature can drop below $10^4$~K when the mass accretion rate is low, thereby invalidating this opacity assumption.

For regions where $T \lesssim 10^4 ~\rm K$, the opacity gap leads to a lower optical depth, allowing photons to escape more easily from the accretion disc. This further lowers the disc temperature, which in turn causes a significant drop in $Q$. \citet{Hure_et_al_1994a, Hure_et_al_1994b} and \citet{Derdzinski_and_Mayer_2023} demonstrate that $Q$ drops rapidly by one to two orders of magnitude when $T$ enters the opacity gap, potentially rendering the low-temperature regions of the disc gravitationally unstable and triggering fragmentation and a reduction in disc size.

We can calculate the accretion disc temperature at $R_{\rm sg}$ to assess the validity of this assumption in our model. Assuming that $R_{\rm sg}$ lies in region~(c) of the accretion disc (i.e. $R_{\rm sg} = R_{\rm sg, c}$), we use the temperature profile from \citetalias{Kato_et_al_2008} to calculate $T(R=R_{\rm sg}) \equiv T_{\rm sg}$:

\begin{equation}
    T_{\rm sg} = 5.0 \times 10^3 \, \alpha_{0.1}^{-2/3} \, M_{\rm BH, 6}^{2/3} \, f_{\rm Edd, 16}^{2/3} \, Q_{\rm min}^{2/3} ~ \rm K~.  
\end{equation}

\noindent If we impose the condition $T_{\rm sg} \geq 10^4$~K to ensure that the opacity assumption remains valid, then we obtain the following criterion (see also \citealt{Perego_et_al_2009}; their numerical value slightly differs from ours because we employ the \citetalias{Kato_et_al_2008} solution and they use the solution by \citealt{Frank_et_al_2002}): 

\begin{equation}
    M_{\rm BH, 6} \ f_{\rm Edd, 16} / \alpha_{0.1} \geq 2.84/Q_{\rm min}~.
    \label{eq:perego_criteria}
\end{equation}

In our test runs, this criterion is not always satisfied, indicating that, in those instances, we likely overestimate the disc size. A smaller disc could potentially result in a slightly higher mass accretion rate initially, followed by a more rapid decline (Figure~\ref{fig:result_Q_Mbh}). We note that all previous accretion disc-based sub-grid models, which assume that the outer disc follows the thin $\alpha$-disc and enforce $R_{\rm disc} \leq R_{\rm sg}$, adopt similar assumptions and may therefore be subject to the same limitations (e.g. \citealt{Perego_et_al_2009, Dubois_et_al_2014, Fiacconi_et_al_2018, Bustamante_et_al_2019}; \citetalias{Cenci_et_al_2021}; \citealt{Koudmani_et_al_2024}).\footnote{\citet{Perego_et_al_2009} also noted this limitation, but they neglected its effect, as they assume a constant $f_{\rm Edd}$. Consequently, they do not require a relation between $\mdisc$, $\jdisc$, and $f_{\rm Edd}$ to update the BH mass accretion rate, which is strongly influenced by the outer parts of the disc.}

To accurately model the low-temperature regions of the accretion disc ($T < 10^4$~K), a detailed computation using a comprehensive opacity table would be required \citep[e.g.][]{Gangardt_et_al_2024, Rozner_et_al_2025}. However, such an approach is beyond the scope of this paper. Moreover, \citet{Jiang_and_Blaes_2020} show that the ``iron opacity bump'' in the detailed opacity table can cause the disc to become convectively unstable, leading to large fluctuations in surface density and temperature. 

\subsubsection{Other caveats}
 
We assume that the radiative efficiency follows the expressions from \citet{Madau_et_al_2014}, which was derived only for aligned discs ($\theta_{\rm BH-disc} = 0$) and extended in this work to include counter-aligned discs ($\theta_{\rm BH-disc} = \pi$). For a misaligned disc (0 < $\theta_{\rm BH-disc} < \pi$), we still use Equation~\eqref{eq:radiative_efficiency}, using the minus (plus) sign for $\theta_{\rm BH-disc} \leq \pi/2$ ($\theta_{\rm BH-disc} > \pi/2$) in Equations~\eqref{eq:radiative_efficiency_A}--\eqref{eq:radiative_efficiency_C}, although this expression becomes less and less accurate as we approach $\theta_{\rm BH-disc} = \pi/2$. \citet{Hughes_and_Blandford_2003} provide a fitting function of $\eta$ as a function of $\theta_{\rm BH-disc}$. However, we do not adopt it in this work, as the precise value of $\eta$ only slightly influences our results, since it only affects the BH growth rate through the $1-\eta$ term. Nonetheless, if BH feedback is included, the precise value of $\eta$ becomes significantly more important. In addition, these changes are only relevant if the disc remains misaligned in the innermost region, i.e. $\feddsixteen > \hat f_{\rm Edd, 16}$, as most of the energy conversion occurs in the inner region close to the ISCO. 

Within our model, we only model gas accretion in the absence of AGN feedback mechanisms, which are crucial to the evolution of SMBHs and their host galaxies. If feedback effects are included, the BH mass accretion rate may be suppressed due to interactions between AGN-driven outflows and the surrounding gas. \WB{It is also worth noting that jet launching can rapidly reduce the BH spin in the presence of a strongly-magnetized super-Eddington accretion disc \citep[e.g.][]{Ricarte_et_al_2023}}. Coupling our model with a feedback 
\WB{prescription} \WB{\citep[e.g.][]{Dubois_et_al_2012, Sala_et_al_2021, Talbot_et_al_2021, Bollati_et_al_2024}} and exploring the influence of BH feedback will be addressed in future work. 


\section{Conclusions}\label{sec:conclusions}

We have developed a new sub-grid model for super-Eddington accretion which is based on structural and thermodynamical properties of realistic accretion discs in RHD simulations. We have implemented this model in the \textsc{gizmo} code using a BH-accretion disc particle that evolves the BH mass and spin through a disc-mediated accretion rate. We have extended the models by \citet{Fiacconi_et_al_2018} and \citetalias{Cenci_et_al_2021} to incorporate super-Eddington accretion, by combining simulation results of super-Eddington flows from \citetalias{Kitaki_et_al_2018} with the three-region structure of the thin $\alpha$-disc model to describe the disc structure. The evolution of the BH spin is determined by two models for calculating the Lense-Thirring torque: the Bardeen-Petterson configuration for low mass accretion rates and the inner precessing thick\WB{-}disc model for high mass accretion rates. An overview of the model is provided in Table~\ref{tab:overview}.

This model enables us to explore the relevance of super-Eddington accretion under different boundary conditions around the BH and the accretion disc. We ran a suite of simulations in idealised scenarios with a spinning SMBH surrounded by an accretion disc embedded in a gaseous disc \WB{and a spherically distributed stellar component}. We considered two types of initial conditions: (i) gas-disc alignment, in which initially the surrounding gas is aligned with the accretion disc but both are misaligned with the BH; (ii) BH-disc alignment, wherein initially the BH is aligned with the accretion disc but both are misaligned with the surrounding gas. We summarise our findings as follows: 

\begin{itemize}

    \item 
    \WB{To achieve high mass accretion rates, not only a sufficient gas inflow is required, but the efficient removal of disc angular momentum is also essential. Without efficient removal, the disc becomes limited by the self-gravity limit, leading to a slow increase in $\feddsixteen$ (Figure~\ref{fig:result_fiducial}). One mechanism responsible for this is the Lense-Thirring torque between a misaligned BH and disc. However, when the gas is nearly aligned with the disc ($\theta_{\rm gas-disc} \sim 0$), this mechanism can increase the mass accretion rate only up to $\feddsixteen \sim 1$ (Figure~\ref{fig:result_fedd_wcirc_a_theta}). We stress that this level of mass accretion corresponds to $f_{\rm Edd, 1} \sim 16$ and thus is considered super-Eddington in many works. This behaviour arises from the interplay between the two torque prescriptions: the torque model for high mass accretion rates has a longer alignment time-scale, whereas the model for low mass accretion rates leads to a much faster alignment.}
    
    \item Another way to achieve a high mass accretion rate is to reduce the disc angular momentum through the misaligned inflow of gas onto the accretion disc. We tested this scenario using BH-disc alignment initial conditions ($\theta_{\rm BH-disc, 0} =0$).  
    \WB{We demonstrated that, in this case, the system can clearly achieve super-Eddington accretion, with $\feddsixteen$ reaching as high as 20 (Figures~\ref{fig:result_misalign_disc_gas} and \ref{fig:result_misalign_theta}), resulting in a rapid increase in BH mass from $10^6 \msun$ to $6 \times 10^6 \msun$ within approximately 8~Myr. By contrast, under the same initial conditions, the model by \citetalias{Cenci_et_al_2021} reaches a much smaller final BH mass, because it does not  allow super-Eddington accretion.} This highlights the relevance of episodic super-Eddington accretion triggered by new inflows of gas with varying angular momentum into the galactic centre. 

    \item In our simulations, the BH spin typically aligns with the direction of the surrounding gas within 
    \WB{approximately 1--10 Myr}. For the gas-disc alignment initial conditions, this alignment occurs due to the Lense-Thirring torque between the BH and the disc. The magnitude of $a_{\rm BH}$ initially decreases when the disc is retrograde and then 
    \WB{increases} when the disc becomes prograde \WB{(Figure~\ref{fig:result_fedd_wcirc_a_theta})}. For the BH-disc alignment initial conditions, alignment results from the combined effects of gas accretion and the Lense-Thirring torque. The magnitude of $a_{\rm BH}$ increases steadily due to prograde accretion \WB{(Figure~\ref{fig:result_misalign_theta})}. If the BH continues to experience coherent accretion over an extended period, $a_{\rm BH}$ will continue to increase due to sustained prograde accretion. This trend is consistent with the findings in \citet[][]{Dotti_et_al_2013}. 

    \item We extrapolate our results and explore episodic super-Eddington accretion, triggered by successive misaligned gas inflows into the galactic centre. We find that the BH can grow from an initial mass of $\mbh = 10^6 \msun$ at $z=11$ to $\mbh \sim \WB{4} \times 10^9 \msun$ at $z=6$ (Figure~\ref{fig:BH_mass_growth}), which is consistent with the most massive SMBHs observed at such redshift. This suggests that episodic super-Eddington accretion may provide a viable mechanism for reaching the mass of the most massive BHs at $z \sim 6$, when starting at $z \sim 11$ from a BH seed of $\sim$$10^6$~M$_{\sun}$ (typical of direct-collapse scenarios).
    
\end{itemize}

This model offers a new method for modelling the growth of SMBHs in the super-Eddington regime, based on a detailed physics framework. After combining our model with a suitable BH feedback model, the relevance of rapid BH growth in the early Universe can be studied in greater detail within a cosmological context. This can provide further insight into the massive SMBHs observed in the early Universe. The model could also be applied to other astrophysical systems to investigate super-Eddington accretion onto BHs, such as binary BHs 
in galaxy-scale simulations.


\section*{Acknowledgements}
\WB{We thank the anonymous referee for useful comments that helped us improve the manuscript.} We \WB{also} thank \WB{Nicholas Choustikov, Julien Devriendt,} Robert Feldmann, Mudit Garg, Alexandre Refregier, \WB{Debora Sijacki, Adrianne Slyz,} and Alessandro Trani for useful and inspiring discussions. PRC acknowledges support from the Swiss National Science Foundation under the Sinergia Grant CRSII5\_213497 (GW-Learn). AL acknowledges support by the PRIN MUR ``2022935STW'' funded by European Union-Next Generation EU, Missione 4 Componente 2, CUP C53D23000950006. \WB{This work made use of infrastructure services provided by S3IT (\url{www.s3it.uzh.ch}), the Service and Support for Science IT team at the University of Zurich. All plots were created with the matplotlib library for visualisation with Python \citep[][]{Hunter2007}.}

\section*{Data Availability}

The data underlying this article will be shared on reasonable request to the corresponding author.


\bibliographystyle{mnras}
\bibliography{mnras}


\appendix

\section{Differences between our model and Cenci et al. (2021)}\label{app:difference_between_cenci_2021}

Our model is built upon that of \citetalias{Cenci_et_al_2021}. We first note that \citetalias{Cenci_et_al_2021} use $f_{\rm Edd, \eta}$ to represent the Eddington ratio (see also Footnote~\ref{foot:fEdd}). In contrast, we adopt $\feddsixteen$ throughout this paper, ensuring that the Eddington ratio remains independent of the radiative efficiency. Below, we summarise the differences between our model and that of \citetalias{Cenci_et_al_2021}:

\begin{enumerate}

    \item In \citetalias{Cenci_et_al_2021}, only region~(c) of the thin $\alpha$-disc is used to describe the accretion disc structure. By contrast, our model incorporates all three regions -- (a), (b), and (c) -- of the thin $\alpha$-disc, together with a fitting formula for the photon-trapping region from \citetalias{Kitaki_et_al_2018}. This results in differences in the surface density and specific angular momentum profiles, which in turn lead to differences in the disc mass and angular momentum. Consequently, the relationship between the Eddington ratio, $\mdisc$, and $\jdisc$ is no longer a simple analytic expression like in \citetalias{Cenci_et_al_2021}. Instead, we determine $\feddsixteen$ numerically using the Newton-Raphson method. 
    
    \item \citetalias{Cenci_et_al_2021} adopt the solution in \citet{Frank_et_al_2002} for the thin $\alpha$-disc structure. In contrast, we use that of \citetalias{Kato_et_al_2008}, as that work provides detailed expressions for all three regions of the thin $\alpha$-disc. As described in Appendix~\ref{app:difference_alpha_disc}, there are several differences between \citet{Frank_et_al_2002} and \citetalias{Kato_et_al_2008} \citep[and the original description by][]{Shakura_Sunyaev_1973}, including the treatment of opacity, the definitions adopted, and the value of the mean molecular weight. Consequently, the surface density in \citetalias{Kato_et_al_2008} is approximately 1.7 times higher than that in \citet{Frank_et_al_2002}, which leads to different values of $M_{\rm sg}$ for a fixed $\mbh$, $\feddsixteen$, and $Q_{\rm min}$. 
    
    \item As we adopt a different accretion disc structure, we have re-derived the expressions for $R_{\rm warp}$ (Equation~\ref{eq:r_warp}), $t_{\rm gm}$ (Equation~\ref{eq:t_gm}), and $R_{\rm sg}$ (Equation~\ref{eq:r_sg_abc}) to ensure consistency.
    
    \item \citetalias{Cenci_et_al_2021} impose $f_{\rm Edd, \eta} \leq 1$ to prevent super-Eddington accretion. In our model, we do not have this limit. Instead, we introduce a new upper limit for BH accretion, denoted as $f_{\rm Edd, 16, max}$ (Equation~\ref{eq:f_edd_max}), which is much higher than 1 for a wide range of BH masses. For example, $f_{\rm Edd, 16, max} \sim 20$ for $M_{\rm BH, 6} = 1$ (and $M_{\rm disc} = M_{\rm sg}$). 
    
    \item In \citetalias{Cenci_et_al_2021}, the Lense-Thirring torque is consistently calculated using the Bardeen-Petterson configuration, which is  valid only at relatively low mass accretion rates ($\feddsixteen < \hat f_{\rm Edd, 16} \sim 1$ for $a_{\rm BH} \gtrsim 0.5$) in our model. For higher mass accretion rates, we adopt a different torque prescription based on the inner precessing thick\WB{-}disc model (Section~\ref{sec:LT_high}).
    
    \item We consider the dependence on $Q_{\rm min}$ in our model, exploring cases with varying $Q_{\rm min}$ values, which set the maximum size of the accretion disc, to examine their impact on the SMBH evolution.
    
    \item \citetalias{Cenci_et_al_2021} use the Kerr metric to compute the radiative efficiency \citep[][]{Bardeen_et_al_1972}, which does not account for the photon-trapping effect in super-Eddington flows. In contrast, we employ an improved version of the original fitting function by \citeauthor{Madau_et_al_2014} (\citeyear{Madau_et_al_2014}; Equations~\ref{eq:radiative_efficiency}--\ref{eq:radiative_efficiency_C}) to include this effect in the calculation of the radiative efficiency.
    
    \item In our \WB{simulations}, we initialise $M_{\rm disc, 0} = M_{\rm sg}$, whereas in \citetalias{Cenci_et_al_2021}, $M_{\rm disc, 0}$ is treated as a free parameter. We adopt this approach because a substantial inflow onto the disc typically occurs at the beginning of the simulation if $M_{\rm disc, 0} < M_{\rm sg}$. This strong inflow would lead to a discontinuous increase in both $\mdisc$ and $\jdisc$, causing $\feddsixteen$ to deviate abruptly from its given initial value $f_{\rm Edd, 16, 0}$.
    
    \item \WB{When $\mdisc = M_{\rm sg}$, we additionally require that $\jdisc \leq J_{\rm sg}$. This condition is imposed to ensure that the disc remains constrained by the self-gravity limit.} 

    \item \WB{In \citetalias{Cenci_et_al_2021}, the maximum value of $a_{\rm BH}$ is set to 0.998 \citep[][]{Thorne_1974}. In contrast, we derive a new upper limit for $a_{\rm BH}$ at high accretion rates, accounting for the effect of reduced radiative efficiency, as discussed in Section~\ref{sec:max_a_BH}. Accordingly, the maximum value of $a_{\rm BH}$ in our model is given by $a_{\rm BH} \leq \min(0.998,  a_{\rm BH,max})$. }

\end{enumerate}

\section{Differences between thin $\alpha$-disc descriptions}\label{app:difference_alpha_disc}

\begin{table*}
    \centering
    \caption{Comparison of three references used to compute the structure of a thin $\alpha$-disc structure.}
    \begin{tabular}{lccccc}
        \hline 
        \rule{0pt}{2.5ex}{Reference} & $\nu_1$ & $\Sigma$ & $\mu$ & $\tau$ & $\kappa_{\rm ff}$ (in cgs units)\T \B  \\
        \hline 
        \rule{0pt}{3ex}\citet{Shakura_Sunyaev_1973} & $\alpha c_{\rm s} H$ & $2 \rho H$ & $1$ & $ 2 \kappa \rho H$ & $0.64 \times 10^{23} \rho T^{-7/2} $ \\
        \citet{Frank_et_al_2002} & $\alpha c_{\rm s} H$ & $\rho  H$ & $0.615$ & $ \kappa \rho H$ & $5 \times 10^{24} \rho T^{-7/2}$ \\
        \citetalias{Kato_et_al_2008}  & $ 2\alpha c_{\rm s} H/3$ & $2 \rho H$ & $0.5$ & $ \kappa \rho  H$ & $0.64 \times 10^{23} \rho T^{-7/2}$\B \\
        \hline
    \end{tabular}
    \label{tab:difference_alpha_disc}
\end{table*}

The thin-$\alpha$ disc model was originally developed by \citet{Shakura_Sunyaev_1973} and later extended to include GR effects by \citet{Novikov_and_Thorne_1973}. In this work, we adopt the formulation of \citetalias{Kato_et_al_2008} for describing such a structure. In contrast, \citet{Perego_et_al_2009}, \citet{Fiacconi_et_al_2018}, \citetalias{Cenci_et_al_2021}, and \citet{Koudmani_et_al_2024} refer to \citet{Frank_et_al_2002}. All three non-GR references (\citealt{Shakura_Sunyaev_1973}; \citealt{Frank_et_al_2002}; \citetalias{Kato_et_al_2008}) provide similar descriptions for the thin $\alpha$-disc structure, with only minor differences in normalisation and characteristic radii (e.g. $R_{\rm ab}$ and $R_{\rm bc}$). These discrepancies arise from differences in the definitions and opacity assumptions adopted in each reference, which are summarised in Table~\ref{tab:difference_alpha_disc}.

We also note that the expression for $\nu_1$ differs slightly in \citetalias{Kato_et_al_2008}, as they define the $\alpha$ parameter via the $r\phi$-component of the shear stress tensor, i.e. $t_{r \phi} = -\alpha P$, as opposed to \citet{Shakura_Sunyaev_1973} and \citet{Frank_et_al_2002}, who use $\nu_1 = \alpha c_{\rm s} H$. In addition, \citet{Frank_et_al_2002} adopt a significantly larger value for $\kappa_{\rm ff}$ compared to the other two references, likely due to a higher metallicity. However, this has only a weak impact on the overall results, as most quantities depend on the opacity through very low power-law exponents (typically $\sim$0.1).

\section{Applying the Newton-Raphson method to calculate the mass accretion rate}  \label{app:Newton_Raphson_method}

Considering the accretion disc structure discussed in Section~\ref{sec:accretion_disc_structure}, we can calculate $\mdisc$ and $\jdisc$ (Equations~\ref{eq:M_disc_integral} and \ref{eq:J_disc_integral}) by providing $R_{\rm disc}$ and $\feddsixteen$, along with given values of $\mbh$, $\alpha$, and $a_{\rm BH}$. We then use the Newton-Raphson method to calculate $\feddsixteen$ (Section~\ref{sec:mass_accretion}). However, it is necessary to carefully explore the parameter space of $(\mdisc, \ \jdisc)$, as degeneracies in this space can cause the Newton-Raphson method to fail.

\begin{figure}
    \centering
    \includegraphics[width=0.8\linewidth]{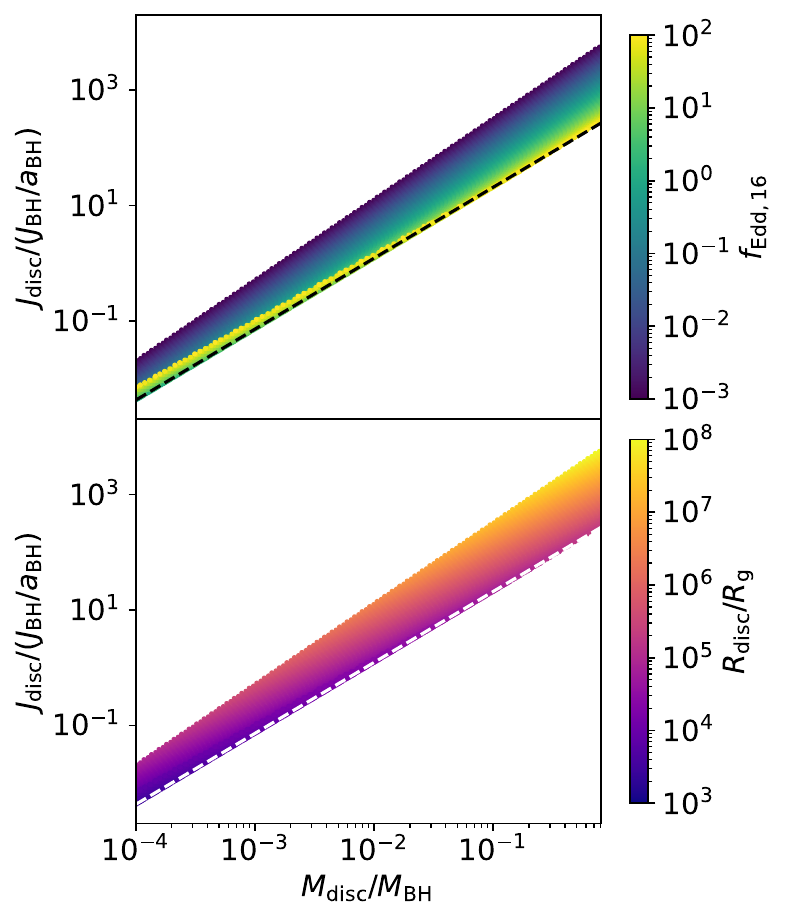}
    \caption{A section of the ($M_{\rm disc}, J_{\rm disc}$) parameter space for $M_{\rm BH, 6} = 1$ and $\alpha_{0.1} = 1$, for different values of $f_{\rm Edd,16}$ (top panel) and $R_{\rm disc}$ (bottom panel). The black (white) dashed line represents the lower boundary of the parameter space in the top (bottom) panel, determined by $f_{\rm Edd, 16, max}$. We consider $\feddsixteen$ values ranging from $10^{-3}$ to $10^2$, with $R_{\rm disc}$ varying from $10^3 \, \rg$ to $10^8 \, \rg$. 
    }
    \label{fig:parameter_space_M_J}
\end{figure}

In Figure~\ref{fig:parameter_space_M_J}, we set $M_{\rm BH, 6} = 1$ and $\alpha_{0.1} = 1$ (the value of $a_{\rm BH}$ is not relevant, since it only affects the value of the ISCO radius, which is not used in the integration) and explore the parameter space of $\mdisc$ and $\jdisc$. This parameter space exhibits a clear lower boundary in $\jdisc$ at any given $M_{\rm disc}$. This behaviour arises because $\Sigma_{\rm b}$ and $\Sigma_{\rm c}$ increase with $\feddsixteen$, whereas $\Sigma_{\rm a} \propto f_{\rm Edd, 16}^{-1}$. To demonstrate how this behaviour can lead to the degeneracy, we first define 

\begin{equation}
    \mathcal P = \left(\frac{\partial \jdisc}{\partial \feddsixteen}\right)_{\mdisc}~.
\end{equation}

When $\feddsixteen \ll 1$, the disc is dominated by regions~(b) and (c). In this regime, for a given $\mdisc$, if $\feddsixteen$ increases, $\Sigma$ also increases, resulting in a more compact disc and hence a smaller $\jdisc$, hence $\mathcal P$ is negative. However, when $f_{\rm Edd, 16 }$ is large, the disc becomes dominated by region~(a), provided that $\rdisc \lesssim R_{\rm ab}$. In this case, as $\feddsixteen$ increases, $\Sigma$ decreases, resulting in a larger $\jdisc$, hence $\mathcal P$ is positive. 

Consequently, $\mathcal P$ transitions from  negative to positive when $\rdisc \sim R_{\rm ab}$. This results in a degeneracy in the $(\mdisc, \ \jdisc)$ parameter space. We can estimate the boundary of the parameter space by identifying when the condition $M_{\rm disc,a} \ll M_{\rm disc}$ is not satisfied, where $M_{\rm disc,a}$ is the disc mass in region~(a). We estimate that this condition breaks down at $R_{\rm disc} = 2.7 R_{\rm ab}$, where $M_{\rm disc, a} \sim 0.1\mdisc$. We can estimate the minimum disc angular momentum $J_{\rm disc, min}$ for a given $\mdisc$ by setting $R_{\rm disc} = 2.7 R_{\rm ab}$. The Eddington ratio at $(\mdisc, \ J_{\rm disc, min})$ is then defined as $f_{\rm Edd, 16, max}$ (Equation~\ref{eq:f_edd_max}). Cases where $\feddsixteen > f_{\rm Edd, 16, max}$ may result in degeneracy or no solution for a given $(\mdisc, J_{\rm disc})$. To avoid such failures in the Newton-Raphson method, we impose the constraint $\feddsixteen \leq f_{\rm Edd, 16, max}$.

\vspace*{-0.5cm}
\section{Radiative efficiency}

\begin{figure}
    \centering
    \includegraphics[width=0.8\linewidth]{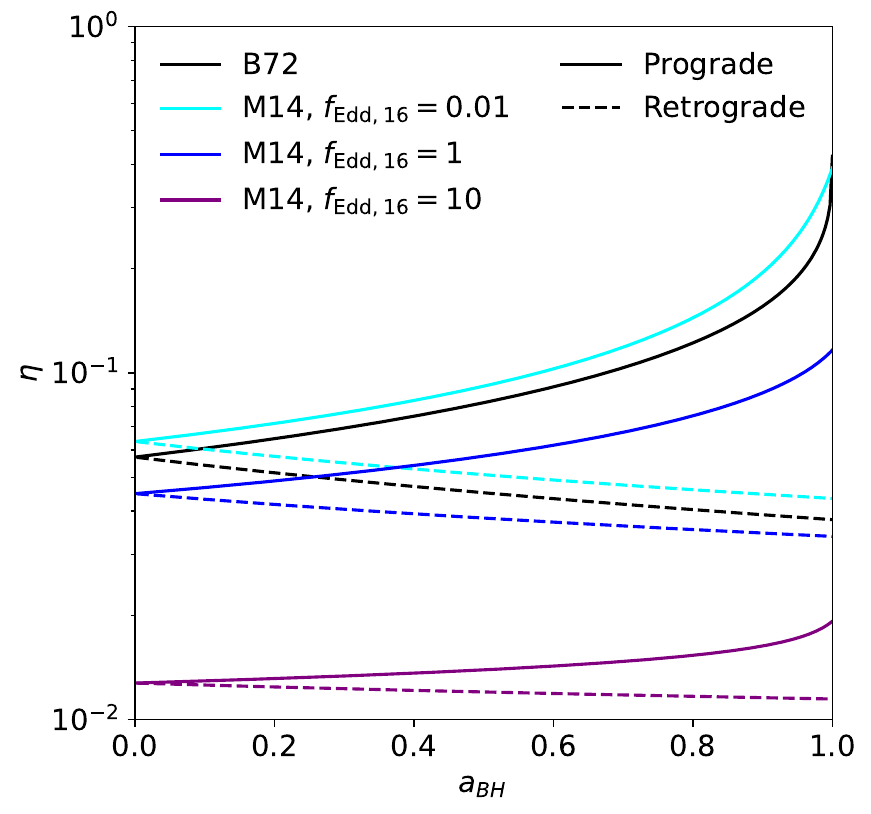}
    \caption{Radiative efficiency as a function of BH spin. The solid and dashed lines represent prograde and retrograde discs, respectively. The black lines show the relation from \citeauthor{Bardeen_et_al_1972} (\citeyear{Bardeen_et_al_1972}: B72), whereas the coloured lines indicate an improved version of the original fitting function by \citeauthor{Madau_et_al_2014} (\citeyear{Madau_et_al_2014}: M14; Equations~\ref{eq:radiative_efficiency}--\ref{eq:radiative_efficiency_C}) for different Eddington ratios: $\feddsixteen = 0.01$ (cyan), 1 (blue), and 10 (purple).}
    \label{fig:eta}
\end{figure}

In Figure~\ref{fig:eta}, we show the value of $\eta$ as a function of $a_{\rm BH}$. We compare the results from \citet{Bardeen_et_al_1972} to those from an improved version of the original fitting function by \citeauthor{Madau_et_al_2014} (\citeyear{Madau_et_al_2014}; Equations~\ref{eq:radiative_efficiency}--\ref{eq:radiative_efficiency_C}) with $\feddsixteen = 0.01$, 1, and 10. It is clear that, at low mass accretion rates, both models show similar values of $\eta$. However, at high mass accretion rates, Equations~\eqref{eq:radiative_efficiency}--\eqref{eq:radiative_efficiency_C} yield a significantly lower value than that of \citet{Bardeen_et_al_1972}, due to the photon-trapping effect.

\section{\WB{Extra runs}}  \label{app:extra_run}

\WB{In this section, we present the results of two additional runs, to show how the model behaves when (i) the BH is counter-aligned with the accretion disc and when (ii) the inflow onto the accretion disc is set to zero.}

\begin{figure*}
    \centering
    \includegraphics[width=1.0\linewidth]{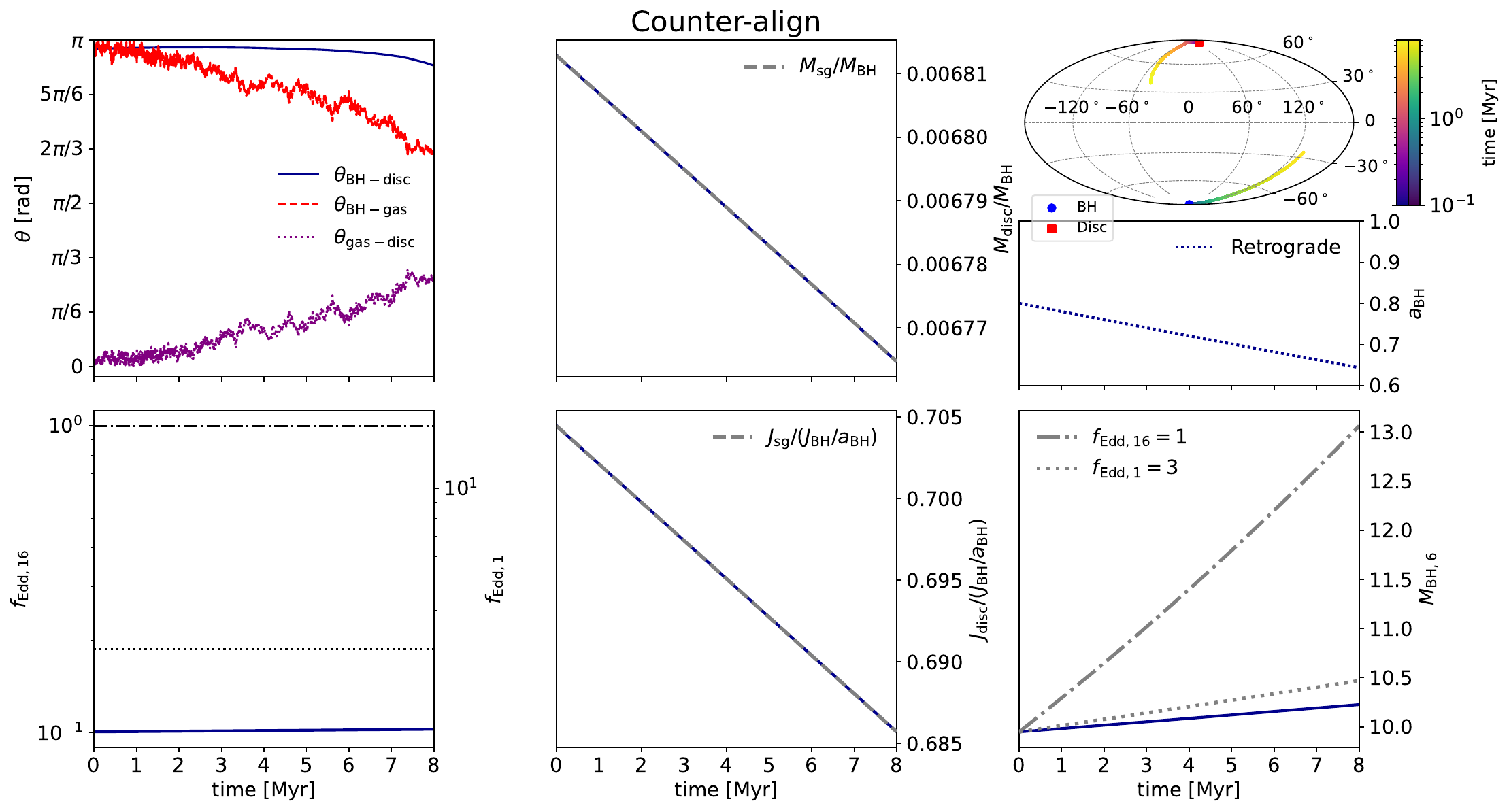}
    \caption{\WB{Time evolution of key quantities for the Counter-align run. Panels from top-left to bottom-right show: misalignment angles $\theta_{\rm BH-disc}$ (solid blue line), $\theta_{\rm BH-gas}$ (dashed red line), and $\theta_{\rm gas-disc}$ (dotted purple line); disc mass, $M_{\rm disc}$, in units of $M_{\rm BH}$ along with the self-gravitating mass, $M_{\rm sg}$ (grey line); BH (blue circle) and disc (red square) angular momenta shown via Hammer projections, colour-coded by time; $a_{\rm BH}$ (which remains retrograde, shown as a dotted line); Eddington ratio $\feddsixteen$, with $f_{\rm Edd,1}$ indicated on the right axis; disc angular momentum, $J_{\rm disc}$, in units of $J_{\rm BH}/a_{\rm BH}$; and BH mass $M_{\rm BH,6}$, with black lines showing constant (specific) accretion at $\feddsixteen = 1$ and $f_{\rm Edd, 1} = 3$. In this setup, the disc is initially counter-aligned with the BH while aligned with the gas. The BH and disc remain counter-aligned throughout most of the run, gradually drifting from their initial orientation. As time progresses, $\theta_{\rm gas-disc}$ increases, leading to an increasingly misaligned inflow onto the accretion disc. Consequently, $\theta_{\rm BH-disc}$ deviates from $\pi$ by the end of the simulation.}}
    \label{fig:result_plot_counteralign_1e7}
\end{figure*}

\WB{Figure~\ref{fig:result_plot_counteralign_1e7} shows the result of the Counter-align run, whose parameters are listed in Table~\ref{tab:simulation_runs}. The accretion disc is initially counter-aligned with the BH and is aligned with the gas (i.e. $\theta_{\rm BH-disc,0}=\pi$ and $\theta_{\rm gas-disc} = 0$). We note that the the initial BH mass is $10^7 \msun$, which yields $J_{\rm disc}/J_{\rm BH} \sim 0.7$; therefore, the criterion proposed by \citet{King_et_al_2005} is not satisfied (Equation~\ref{eq:counter_align_King_2005}). The BH and disc remain counter-aligned throughout most of the run. Both $\theta_{\rm BH-gas}$ and $\theta_{\rm gas-disc}$ gradually increase due to the Lense-Thirring torque exerted between the BH and the disc. This interaction drives an increasingly misaligned inflow onto the accretion disc, causing $\theta_{\rm BH-disc}$ to deviate slightly from $\pi$. We also note that $\feddsixteen$ shows a slight increase, as it is constrained by the self-gravity limit.}

\begin{figure*}
    \centering
    \includegraphics[width=1.0\linewidth]{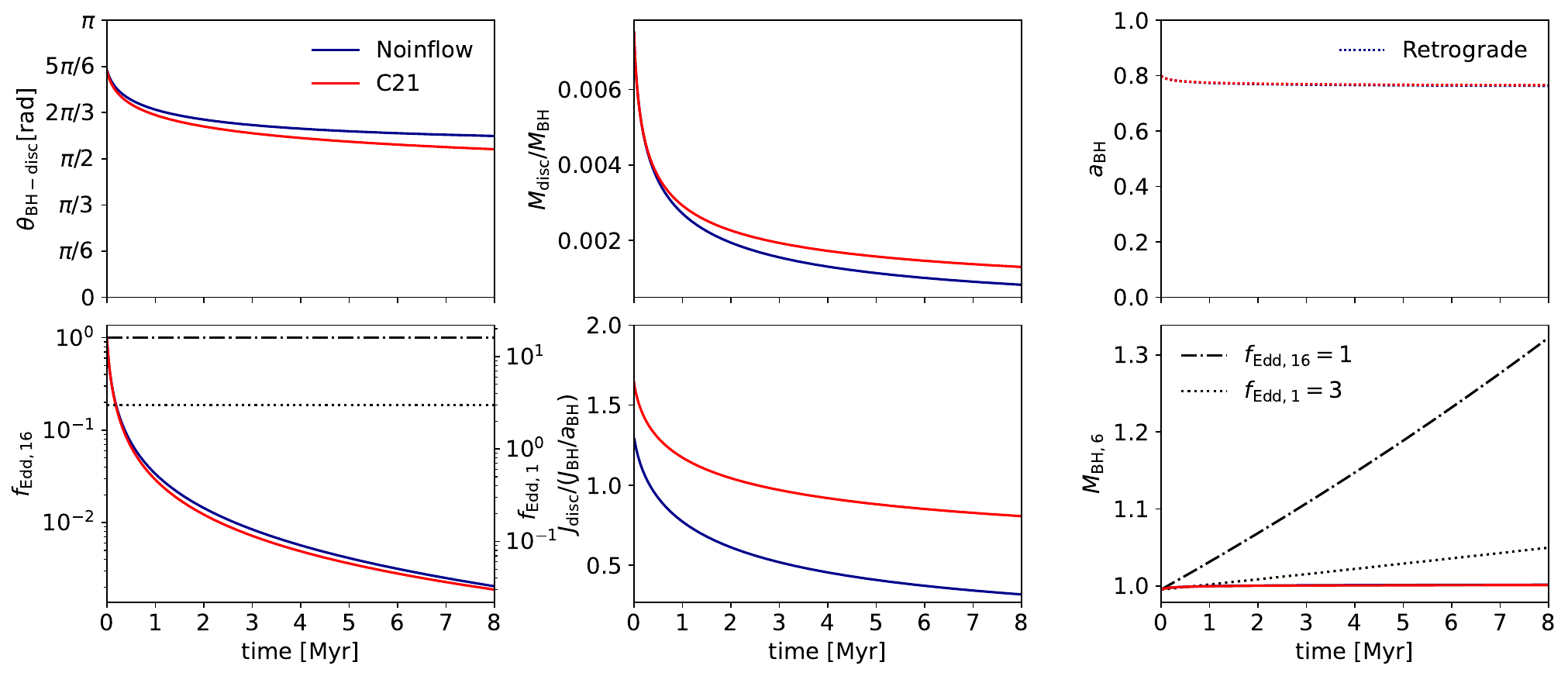}
    \caption{\WB{Time evolution of key quantities for the No-inflow run using our model (blue lines) and that of \citetalias{Cenci_et_al_2021} (red lines). Both runs start from $\feddsixteen = 1$ and $M_{\rm disc, 0} = 7.5 \times 10^{-3} \mbh$, while the inflow rate is (artificially) set to zero. This figure is similar to Figure~\ref{fig:result_plot_counteralign_1e7}, except that we only show the misalignment angle between the BH and the disc, and do not include $M_{\rm sg}$ or the evolution of angular momenta via Hammer projections. Both models show qualitatively similar behaviours, as expected, since they should converge at low accretion rates.}}
    \label{fig:nonSG_noinflow_compare_C21}
\end{figure*}

\WB{Figure~\ref{fig:nonSG_noinflow_compare_C21} shows the results of the No-inflow run, obtained using both our model and that of \citetalias{Cenci_et_al_2021}. In this setup, both runs start from $\feddsixteen = 1$ ($f_{\rm Edd, \eta} \sim 0.6$ for the \citetalias{Cenci_et_al_2021} model) and $M_{\rm disc, 0} = 7.5 \times 10^{-3} \mbh$. $J_{\rm disc, 0}$ is set accordingly and differs between the two runs because of the different accretion disc structures adopted in each model. The inflow rate onto the disc is (artificially) set to zero for both runs, to test the model behaviour and to prevent the disc from becoming self-gravitating, given that the value of $M_{\rm sg}$ differs between the two models. For both runs, $\theta_{\rm BH-disc, 0} = 5\pi/6$, $\theta_{\rm gas-disc, 0} = 0$, $a_{\rm BH, 0} = 0.8$, $M_{\rm BH, 0} = 10^6 \msun$, $W_{\rm circ}= 0.1$, and $Q_{\rm min} = 1$. Due to the absence of an inflow, $\feddsixteen$, $\mdisc$, and $\jdisc$ decrease steadily in both cases. The alignment between the BH and the disc also becomes less efficient as $\feddsixteen$ decreases. Both runs show qualitatively similar behaviour, as expected, since at low accretion rates the two models should yield similar results. The slight differences between the two models arise from their distinct accretion-disc structures.} 

\bsp 
\label{lastpage}
\end{document}